\documentclass[10pt,paper,aps,prstab,twocolumn,amsmath,superscriptaddress,preprintnumbers,floatfix,showpacs]{revtex4-2}

\usepackage{siunitx}
\usepackage{graphicx}
\usepackage[colorlinks]{hyperref}
\hypersetup{
	colorlinks=true, 
	linktoc=all,     
	linkcolor=blue,  
	citecolor=magenta,
	urlcolor=blue	
	}

\usepackage[dvipsnames]{xcolor}
\usepackage{tikz}
\usepackage{booktabs}
\usepackage{dcolumn}
\usepackage{mathtools}
\usepackage{braket}
\usepackage[utf8]{inputenc}
\usepackage{placeins}
\usepackage[T1]{fontenc}
\usepackage{natbib}
\usepackage{amssymb,latexsym}
\usepackage{wasysym}
\usepackage{soul}
\usepackage{xfrac}

\newcommand{\matr}[1]{\mathbf{#1}}
\newcommand{\dd}{\text{d}}

\DeclareMathOperator{\sgn}{sgn}
\begin{document}
	
\author{N.N.\,Nikolaev}
\affiliation{L.D. Landau Institute for Theoretical Physics, 142432 Chernogolovka, Russia}
	
\author{F.\,Rathmann}
\email[Corresponding author: ]{frathmann@bnl.gov}
\thanks{Present address: Brookhaven National Laboratory, Upton, NY 11973, USA}
\affiliation{Institut f\"ur Kernphysik, Forschungszentrum J\"ulich, 52425 		J\"ulich, Germany}
	
\author{J.\,Slim}
\thanks{Present address: Deutsches Elektronen-Synchrotron, 22607 Hamburg, Germany}
\affiliation{III. Physikalisches Institut B, RWTH Aachen University, 52056 Aachen, Germany}
	
\author{A.\,Andres}
\affiliation{Institut f\"ur Kernphysik, Forschungszentrum J\"ulich, 52425 		J\"ulich, Germany}
\affiliation{III. Physikalisches Institut B, RWTH Aachen University, 52056 Aachen, Germany}

\author{V.\,Hejny}
\affiliation{Institut f\"ur Kernphysik, Forschungszentrum J\"ulich, 52425 J\"ulich, Germany}
	
\author{A.\,Nass}
\affiliation{Institut f\"ur Kernphysik, Forschungszentrum J\"ulich, 52425 J\"ulich, Germany}

\author{A.\,Kacharava}
\affiliation{Institut f\"ur Kernphysik, Forschungszentrum J\"ulich, 52425 J\"ulich, Germany}

\author{P.\,Lenisa}
\affiliation{University of Ferrara and Istituto Nazionale di Fisica Nucleare, 44100 Ferrara, Italy}

\author{J.\,Pretz}
\affiliation{Institut f\"ur Kernphysik, Forschungszentrum J\"ulich, 52425 J\"ulich, Germany}	
\affiliation{III. Physikalisches Institut B, RWTH Aachen University, 52056 Aachen, Germany}
	
\author{A.\,Saleev}
\thanks{Present address: GSI Helmholtz Centre for Heavy Ion Research, 64291 Darmstadt, Germany}
\affiliation{University of Ferrara and Istituto Nazionale di Fisica Nucleare, 44100 Ferrara, Italy}
	
\author{V.\,Shmakova}
\thanks{Present address: Brookhaven National Laboratory, Upton, NY 11973, USA}
\affiliation{University of Ferrara and Istituto Nazionale di Fisica Nucleare, 44100 Ferrara, Italy}
		
\author{H.\,Soltner}
\affiliation{Zentralinstitut f\"ur Engineering, Elektronik und Analytik, Forschungszentrum J\"ulich, 52425 J\"ulich, Germany}
		
\author{F.\,Abusaif}
\thanks{Present  address: Karlsruhe Institute of Technology, 76344 Eggenstein-Leopoldshafen, Germany}
\affiliation{Institut f\"ur Kernphysik, Forschungszentrum J\"ulich, 52425 J\"ulich, Germany}
\affiliation{III. Physikalisches Institut B, RWTH Aachen University, 52056 Aachen, Germany}

\author{A.\,Aggarwal}
\affiliation{Marian Smoluchowski Institute of Physics, Jagiellonian University, 30348 Cracow, Poland}
		
\author{A.\,Aksentev}
\affiliation{Institute for Nuclear Research, Russian Academy of Sciences, 117312 Moscow, Russia}
		
\author{B.\,Alberdi}
\thanks{Present address: Humboldt-Universität zu Berlin, Institut für Physik, 12489 Berlin, Germany}
\affiliation{Institut f\"ur Kernphysik, Forschungszentrum J\"ulich, 52425 J\"ulich, Germany}
\affiliation{III. Physikalisches Institut B, RWTH Aachen University, 52056 Aachen, Germany}
	
\author{L.\,Barion}
\affiliation{University of Ferrara and Istituto Nazionale di Fisica Nucleare, 44100 Ferrara, Italy}
		
\author{I.\,Bekman}
\thanks{Present address: Zentralinstitut für Engineering, Elektronik und Analytik, Forschungszentrum Jülich, Jülich, Germany.}
\affiliation{Institut f\"ur Kernphysik, Forschungszentrum J\"ulich, 52425 J\"ulich, Germany}
		
\author{M.\,Bey\ss}
\affiliation{Institut f\"ur Kernphysik, Forschungszentrum J\"ulich, 52425 J\"ulich, Germany}
\affiliation{III. Physikalisches Institut B, RWTH Aachen University, 52056 Aachen, Germany}
		
\author{C.\,B\"ohme}
\affiliation{Institut f\"ur Kernphysik, Forschungszentrum J\"ulich, 52425 J\"ulich, Germany}
		
\author{B.\,Breitkreutz}
\thanks{Present address: GSI Helmholtz Centre for Heavy Ion Research, 64291 Darmstadt, Germany}
\affiliation{Institut f\"ur Kernphysik, Forschungszentrum J\"ulich, 52425 J\"ulich, Germany}
		
\author{N.\,Canale}
\affiliation{University of Ferrara and Istituto Nazionale di Fisica Nucleare, 44100 Ferrara, Italy}
		
\author{G.\,Ciullo}
\affiliation{University of Ferrara and Istituto Nazionale di Fisica Nucleare, 44100 Ferrara, Italy}
		
\author{S.\,Dymov}
\affiliation{University of Ferrara and Istituto Nazionale di Fisica Nucleare, 44100 Ferrara, Italy}
		
\author{N.-O.\,Fr\"ohlich}
\thanks{Present address: Deutsches Elektronen-Synchrotron, 22607 Hamburg, Germany}
\affiliation{Institut f\"ur Kernphysik, Forschungszentrum J\"ulich, 52425 J\"ulich, Germany}
		
\author{R.\,Gebel}
\affiliation{Institut f\"ur Kernphysik, Forschungszentrum J\"ulich, 52425 J\"ulich, Germany}
		
\author{M.\,Gaisser}
\affiliation{III. Physikalisches Institut B, RWTH Aachen University, 52056 Aachen, Germany}
		
\author{K.\,Grigoryev}
\thanks{Present address: GSI Helmholtz Centre for Heavy Ion Research, 64291 Darmstadt, Germany}
\affiliation{Institut f\"ur Kernphysik, Forschungszentrum J\"ulich, 52425 J\"ulich, Germany}
		
\author{D.\,Grzonka}
\affiliation{Institut f\"ur Kernphysik, Forschungszentrum J\"ulich, 52425 J\"ulich, Germany}
		
\author{J.\,Hetzel}
\affiliation{Institut f\"ur Kernphysik, Forschungszentrum J\"ulich, 52425 J\"ulich, Germany}
		
\author{O.\,Javakhishvili}
\affiliation{Department of Electrical and Computer Engineering, Agricultural University of Georgia, 0159 Tbilisi, Georgia}
		
\author{V.\,Kamerdzhiev}
\thanks{Present address: GSI Helmholtz Centre for Heavy Ion Research, 64291 Darmstadt, Germany}
\affiliation{Institut f\"ur Kernphysik, Forschungszentrum J\"ulich, 52425 J\"ulich, Germany}
			
\author{S.\,Karanth}
\affiliation{Marian Smoluchowski Institute of Physics, Jagiellonian University, 30348 Cracow, Poland}
		
\author{I.\,Keshelashvili}
\thanks{Present address: GSI Helmholtz Centre for Heavy Ion Research, 64291 Darmstadt, Germany}
\affiliation{Institut f\"ur Kernphysik, Forschungszentrum J\"ulich, 52425 J\"ulich, Germany}
		
\author{A.\,Kononov}
\affiliation{University of Ferrara and Istituto Nazionale di Fisica Nucleare, 44100 Ferrara, Italy}
		
\author{K.\,Laihem}
\thanks{Present address: GSI Helmholtz Centre for Heavy Ion Research, 64291 Darmstadt, Germany}
\affiliation{III. Physikalisches Institut B, RWTH Aachen University, 52056 Aachen, Germany}
	
\author{A.\,Lehrach}
\affiliation{Institut f\"ur Kernphysik, Forschungszentrum J\"ulich, 52425 J\"ulich, Germany}
\affiliation{III. Physikalisches Institut B, RWTH Aachen University, 52056 Aachen, Germany}
		
\author{N.\,Lomidze}
\affiliation{High Energy Physics Institute, Tbilisi State University, 0186 Tbilisi, Georgia}
		
\author{B.\,Lorentz}
\affiliation{GSI Helmholtzzentrum für Schwerionenforschung, 64291 Darmstadt, Germany}
		
\author{G.\,Macharashvili}
\affiliation{High Energy Physics Institute, Tbilisi State University, 0186 Tbilisi, Georgia}
		
\author{A.\,Magiera}
\affiliation{Marian Smoluchowski Institute of Physics, Jagiellonian University, 30348 Cracow, Poland}
		
\author{D.\,Mchedlishvili}
\affiliation{High Energy Physics Institute, Tbilisi State University, 0186 Tbilisi, Georgia}
		
\author{A.\,Melnikov}
\affiliation{Institute for Nuclear Research, Russian Academy of Sciences, 117312 Moscow, Russia}
		
\author{F.\,Müller}
\affiliation{Institut f\"ur Kernphysik, Forschungszentrum J\"ulich, 52425 J\"ulich, Germany}
\affiliation{III. Physikalisches Institut B, RWTH Aachen University, 52056 Aachen, Germany}
		
\author{A. Pesce}
\affiliation{Institut f\"ur Kernphysik, Forschungszentrum J\"ulich, 52425 J\"ulich, Germany}
		
\author{V.\,Poncza}
\affiliation{Institut f\"ur Kernphysik, Forschungszentrum J\"ulich, 52425 J\"ulich, Germany}
		
\author{D.\,Prasuhn}
\affiliation{Institut f\"ur Kernphysik, Forschungszentrum J\"ulich, 52425 J\"ulich, Germany}
		
\author{D.\,Shergelashvili}
\affiliation{High Energy Physics Institute, Tbilisi State University, 0186 Tbilisi, Georgia}	
		
\author{N.\,Shurkhno}
\thanks{Present address: GSI Helmholtz Centre for Heavy Ion Research, 64291 Darmstadt, Germany}
\affiliation{Institut f\"ur Kernphysik, Forschungszentrum J\"ulich, 52425 J\"ulich, Germany}
		
\author{S.\,Siddique}
\thanks{Present address: GSI Helmholtz Centre for Heavy Ion Research, 64291 Darmstadt, Germany}
\affiliation{Institut f\"ur Kernphysik, Forschungszentrum J\"ulich, 52425 J\"ulich, Germany}
\affiliation{III. Physikalisches Institut B, RWTH Aachen University, 52056 Aachen, Germany}
		
\author{A.\,Silenko}
\affiliation{Bogoliubov Laboratory of Theoretical Physics, Joint Institute for Nuclear Research, 141980 Dubna, Russia}
		
\author{S.\,Stassen}
\affiliation{Institut f\"ur Kernphysik, Forschungszentrum J\"ulich, 52425 J\"ulich, Germany}
		
\author{E.J.\,Stephenson}		
\affiliation{Indiana University, Department of Physics, Bloomington, Indiana 47405, USA}
		
\author{H.\,Ströher}
\affiliation{Institut f\"ur Kernphysik, Forschungszentrum J\"ulich, 52425 J\"ulich, Germany}
		
\author{M.\,Tabidze}
\affiliation{High Energy Physics Institute, Tbilisi State University, 0186 Tbilisi, Georgia}
		
\author{G.\,Tagliente}
\affiliation{Istituto Nazionale di Fisica Nucleare sez.\ Bari, 70125 Bari, Italy}
		
\author{Y.\,Valdau}
\thanks{Present address: GSI Helmholtz Centre for Heavy Ion Research, 64291 Darmstadt, Germany}
\affiliation{Institut f\"ur Kernphysik, Forschungszentrum J\"ulich, 52425 J\"ulich, Germany}
		
\author{M.\,Vitz}
\affiliation{Institut f\"ur Kernphysik, Forschungszentrum J\"ulich, 52425 J\"ulich, Germany}
\affiliation{III. Physikalisches Institut B, RWTH Aachen University, 52056 Aachen, Germany}
		
\author{T.\,Wagner}
\thanks{Present address: GSI Helmholtz Centre for Heavy Ion Research, 64291 Darmstadt, Germany}
\affiliation{Institut f\"ur Kernphysik, Forschungszentrum J\"ulich, 52425 J\"ulich, Germany}
\affiliation{III. Physikalisches Institut B, RWTH Aachen University, 52056 Aachen, Germany}
		
\author{A.\,Wirzba}
\affiliation{Institut f\"ur Kernphysik, Forschungszentrum J\"ulich,  52425 J\"ulich, Germany}
\affiliation{Institute for Advanced Simulation, Forschungszentrum J\"ulich, 52425 J\"ulich, Germany}
			
\author{A.\,Wro\'{n}ska}
\affiliation{Marian Smoluchowski Institute of Physics, Jagiellonian University, 30348 Cracow, Poland}
		
\author{P.\,W\"ustner}
\affiliation{Zentralinstitut f\"ur Engineering, Elektronik und Analytik, Forschungszentrum J\"ulich, 52425 J\"ulich, Germany}
		
\author{M.\,\.{Z}urek}
\thanks{Present address: Argonne National Laboratory, Lemont, IL 60439, USA}
\affiliation{Institut f\"ur Kernphysik, Forschungszentrum J\"ulich, 52425 J\"ulich, Germany}
		
\collaboration{JEDI collaboration}

	
\date {\today}
	
\title{Spin decoherence and off-resonance behavior of radiofrequency-driven spin rotations \\in storage rings}
	
\begin{abstract}
	Radiofrequency-driven resonant spin rotators are routinely used as standard instruments in polarization experiments in particle and nuclear physics. Maintaining the continuous exact parametric spin-resonance condition of the equality of the spin rotator and the spin precession frequency during operation constitutes one of the challenges. We present a detailed analytic description of the impact of detuning the exact spin resonance on the vertical and the in-plane precessing components of the polarization. An important part of the formalism presented here is the consideration of experimentally relevant spin-decoherence effects.  We discuss applications of the developed formalism to the interpretation of the experimental data on the novel pilot bunch approach to control the spin-resonance condition during the operation of the radiofrequency-driven Wien filter that is used as a spin rotator in the first direct deuteron electric dipole moment measurement at COSY. We emphasize the potential importance of the hitherto unexplored phase of the envelope of the horizontal polarization as an indicator of the stability of the radiofrequency-driven spin rotations in storage rings. The work presented here serves as a satellite publication to the work published concurrently on the proof of principle experiment about the so-called pilot bunch approach that was  developed to provide co-magnetometry for the deuteron electric dipole moment experiment at COSY.
\end{abstract}
\maketitle
\tableofcontents
	
\section{Introduction} 
\label{sec:introduction}
Controlled spin rotations, notably the spin flips (SF), are imperative for particle and nuclear physics experiments that involve polarized particles (see \textit{e.g.}, \cite{PhysRevC.58.658}, for extensive reviews, see\,\cite{SYLee, Yokoya}). In storage rings, the radiofrequency (RF) magnetic field resonant to the idle spin  precession acts as a spin flipper, resembling the familiar case of nuclear magnetic resonance (NMR). In an ideal magnetic ring, one stores beam particles with on average vertically oriented polarization, and the spin precession frequency is given by $f_\text{s}   = G\gamma f_\text{c}$, where $f_\text{c}$ denotes the cyclotron frequency of the ring, and $G$ and $\gamma$ denote magnetic anomaly and relativistic $\gamma$-factor of the stored particles\,\footnote{In accelerator physics, $f_\text{s}  $ usually defines the spin rotation with respect to the particle momentum, {\it i.e.,} $f_\text{s}  $ denotes the spin precession frequency in the laboratory frame with cyclotron frequency \textit{subtracted}.}.
	
In practice, the  magnetic field imperfections in the machine, especially the ones tangential to the beam orbit, bring about a substantial and often poorly known correction to the above simple formula for $f_\text{s}  $\,\cite{SYLee,Yokoya,SpinTuneMapping}. There are other complications that contribute, such as spin decoherence due to beam momentum spread $\Delta p/p$ from synchrotron oscillations and from orbit lengthening due to betatron oscillations, which require chromaticity tuning\,\cite{SCT1000sJEDI2,SCTchromaticityJEDI,KoopShatunov}. A more fundamental obstacle is	that the beam energy is so poorly known that, rather conversely, the spin precession frequency can be used to calibrate the beam energy\,\cite{Skrinsky:1989ie}. For instance, this problem of $f_\text{s}  $ being uncertain can be overcome with the Froissart-Stora scan approach, where the particle spin is subjected to a magnetic field of slowly varying frequency\,\cite{Froissart}. When the scanned frequency range is sufficiently broad to cover the not so well-known spin precession frequency $f_\text{s}  $, then during the scan, the nuclear magnetic resonance condition will be encountered. 
	
There are important spin-physics experiments in storage rings being conducted or anticipated, where it is imperative to maintain the exact spin-resonance condition for a long time, including a large number of SFs under continuous operation of an RF spin rotator. As part of the program of studies of systematic effects in electric dipole moment (EDM) searches of charged particles in storage rings, the JEDI collaboration\,\cite{jedi-collaboration2} at the Cooler Synchrotron (COSY) storage ring in Forschungszentrum J\"ulich \,\cite{MaierCOSY,PhysRevSTAB.18.020101} has developed a technique of measuring the  idle spin precession frequency to $\num{e-10}$  precision within a \SI{100}{s} time interval \,\cite{PhysRevLett.115.094801,JEDIspintune2}.  When brought to interaction with an internal polarimeter target, the precessing horizontal polarization component of the beam gives rise to an up-down asymmetry 	oscillating with the spin precession frequency. The Fourier analysis of the time-stamped events in the polarimeter (see Ref.\,\cite{PhysRevLett.115.094801}	for details) makes it possible to determine the oscillation frequency and also the envelope of the precessing polarization. The measurement of the spin precession frequency relies on the  oscillating \textit{horizontal} polarization component. Thus when during a single or multiple spin flips\,\cite{JEDIphase} the spins are closely aligned along the \textit{vertical} axis in the machine, the control of the spin precession frequency fails,  because in that case the horizontal polarization component is either too small or vanishes.
	
Recently, the JEDI collaboration proposed a solution to this issue based on the so-called pilot bunch approach, applicable to a situation with multiple bunches stored in the ring.  The spin manipulations applied to the orbiting particles are organized in three stages:
\begin{itemize}
		\item[I:] In the first stage 
		the initial vertical spins of multiple bunches of the stored deuterons are rotated into the horizontal plane by the radiofrequency solenoid, operated as a fixed-frequency spin rotator like in previous JEDI experiments. 
	    \item[II:] In the second stage, the frequency of the idle spin precession $f_\text{s}$  of the in-plane polarization is measured. 
	    \item[III:] In the third stage, the radiofrequency Wien filter (WF) is used as a spin rotator in a special mode where it is switched off once per beam revolution for a short period of time when one of several stored bunches passes through the spin rotator. The operation of the WF starts at the frequency $f_\text{WF}=f_\text{s}$ as measured in the second stage, and is kept locked to the continuously measured idle spin precession frequency $f_\text{s}$ of the \textit{unperturbed} (pilot) bunch. Thus, the pilot bunch acts as a co-magnetometer, providing crucial information about $f_\text{s}$, which can be used in the interpretation of the spin dynamics of signal bunches exposed to the RF fields in the WF which operates at frequency $f_\text{WF}$.
\end{itemize}

\begin{table*}[t]
	\caption{\label{tab:hierarchy-frequencies}  Hierarchy of typical frequencies.  }
	\begin{ruledtabular}
		\renewcommand{\arraystretch}{1.2} 
		\begin{tabular}{lcd}
			System 							& Frequency  	& \text{Value}\,\text{ [Hz]}\\\hline
			Cyclotron motion				& $f_\text{c}  $			& 750000 \\
			Spin precession	with respect to particle momentum				& $f_\text{s}  $			& 120000 \\
			Synchrotron motion				& $f_\text{sy}$	& 200 \\
			RF-driven spin flip				& $f_\text{SF}$	& 1 \\
			Feedback system induced spin precession spread  	& $\Delta f_\text{s}  ^\text{fb}$	& 0.005
		\end{tabular}
	\end{ruledtabular}
\end{table*}		

The JEDI collaboration reports in Ref.\,\cite{Slim:2023lpd} the first successful application of the pilot-bunch technique using two bunches stored in COSY with the radiofrequency Wien filter employed as a spin rotator. The experiment was carried out with deuterons of momentum $p = \SI{970}{MeV/c}$. The sophisticated technical details of the development of the fast radiofrequency switches, operating at the ring frequency $f_\text{c}\simeq \SI{750}{kHz}$, which allowed us to turn off the radiofrequency of the Wien filter when one of the two orbiting beams passed the Wien filter, are discussed in Ref.\,\cite{Slim:2023lpd}. While the polarization of the bunch exposed to the radiofrequency  fields undergoes continuous SFs, the pilot bunch is immune to the radiofrequency of the Wien filter, and it provides a continuous determination of the idle spin precession frequency. The spin precession frequency is then employed to lock frequency and phase of the Wien filter. The pilot-bunch technique was proposed primarily in connection to the precision spin experiments on tests of fundamental symmetries such as a search for the parity and time-reversal-invariance violating permanent EDMs of charged particles\,\cite{srEDM,YannisHybrid,AbusaifCYR}, but it may find other applications in spin physics at storage rings.
	
In practice, a certain amount of detuning is an indispensable feature of the RF-driven spin dynamics in storage rings. The frequency of  radiofrequency  power supplies can only be controlled with finite accuracy, leaving room for residual detuning of the Wien filter and spin precession frequencies. Moreover, the betatron and synchrotron oscillation-induced spin tune spread is endemic in ensembles of stored particles. Finally, the process of  feedback to lock the Wien filter and spin precession phases is nothing more than a continuous compensation of the detuning caused by the instabilities of the storage ring. It is important to assess the impact of constant or time-varying detuning of individual particles in the ensemble on various aspects of the long-time continuous spin flips, ranging from the amplitude and tunes of the vertical spin oscillations to the time dependence of the envelope and phase of the precessing horizontal polarization. A very different effect of synchrotron oscillations, namely their impact on single Froissart-Stora crossings of the spin resonance\,\cite{Froissart} and the behavior of the polarization in the relatively short time periods thereafter, was studied earlier at COSY\,\cite{PhysRevSTAB.15.124202}.	

Yet another closely related issue is the role of the finite spin-coherence time. For instance,  damping is known to shift the frequency of the classical harmonic oscillator. In the case of  a parametric spin resonance, involving non-commuting spin rotations, this requires a dedicated treatment of the impact of spin decoherence on the spin precessions and its dependence on the mechanism leading to spin decoherence. 

Considering the JEDI spin experiments with polarized deuterons at a beam momentum of  $p = \SI{0.97}{GeV/c}$, a hierarchy of typical frequencies  as listed in Table\,\ref{tab:hierarchy-frequencies} is involved that defines the small parameters in the problem. The typical time scales involved are the spin observation times (cycle times) $t_\text{exp} \approx \SI{100}{s}$ and the in-plane (horizontal) spin-coherence time $\tau_\text{SCT} \sim \SI{1000}{s}$. 

Still another time scale results from the feedback system (fb) used to synchronize the radiofrequency  Wien filter     with the spin precession frequency. The JEDI studies revealed a non-negligible variation of the idle spin precession frequency on the order of about $10^{-8}$ from one fill to another and during each fill\,\cite{PhysRevLett.115.094801}. In practice, about 5 consecutive measurements of $1-\SI{2}{s}$ duration are required to obtain a trend of the spin-phase response with a spread of the order of $\sigma_\text{fb} \sim \SI{0.2}{rad}$ to obtain a feedback to correct the  Wien filter     frequency\,\cite{PhaseLock}.  It can be assumed that this phase response is smooth during the feedback time interval of $t_\text{fb} = 5 - \SI{10}{s}$, and one can speak of a corresponding non-negligible detuning of the radiofrequency  Wien filter     with respect to the spin precession,
		\begin{equation}
		\Delta f_\text{s}^\text{fb} \sim \frac{\sigma_\text{fb} }{2\pi t_\text{fb}} \sim 5\,\text{mHz} \,.
		\label{FeedBackDetuning}	
		\end{equation} 

A similar hierarchy was observed for polarized protons at a beam kinetic energy of \SI{49.3}{MeV} in COSY, where 99 successive flips driven by an radiofrequency  solenoid were performed within \SI{300}{s}. Assuming exponential attenuation of polarization, the average spin flipper efficiency was found to be $\epsilon_\text{flip} = 0.9872\pm 0.0001$\,\cite{PhysRevSTAB.18.020101}, corresponding to a lifetime of the continuously flipping spin of $\tau_\text{flip} = \SI{240}{s}$. With the radiofrequency  spin flipper turned off, the vertical polarization was found to have a much longer lifetime of $\tau_\text{p} = (2.7 \pm 0.8)\cdot\SI{e5}{s}$, indicating a close connection of the depolarization to the SF dynamics.

The hierarchy of frequencies given above (Table\,\ref{tab:hierarchy-frequencies}) allows one to pursue all aspects of the RF-driven spin dynamics within a unified Bogoliubov-Krylov averaging approach and paves the way to the first fully analytic and compact formalism for the detuned RF-driven parametric spin resonance taking into account the decoherence of the polarization. The present work extends earlier considerations considerably\,\cite{SpinTuneMapping,Silenko2017,JEDIphase,PhysRevAccelBeams.23.024601} and is intended as a satellite publication to the one describing the first experimental test of the Pilot Bunch concept\,\cite{Slim:2023lpd}, the corresponding numerical estimates are presented for the conditions of this experiment.
 
There is a strong need for such a description  because fitting the experimental data with multiple spin flips requires large number of calls of the spin evolution code, which can not be readily met by the numerical solution of the spin evolution for up to $\sim \num{e8}$ revolutions of the beam. To this end we emphasize that the above specified conditions are about typical for storage rings dedicated to the search for the charged particles electric dipole moments\,\cite{srEDM,YannisHybrid,AbusaifCYR}. We regard our formalism as a toolbox for the determination of the detuning parameter for individual fills of a machine,  and it may find applications in accelerator physics beyond the description of the pilot bunch regime. In the case of the pilot bunch, we point out tricky features of the  partial depolarization of the pilot bunch in the regime of incomplete masking (gating-out) the  RF of the spin rotator. We pay particular attention to the as yet unexplored role of the phase of the spin envelope of the horizontal polarization on the control of the stable performance of the RF-driven spin rotations, for which we provide a fully analytic description.
	
The following presentation is organized as follows. (The most important variables and parameters are collected in the glossary in Table\,\ref{tab:parameters}.) In 	Sec.\,\ref{sec:stroboscopic-off-resonance}, we present basics of the Bogoliubov-Krylov-averaging approach to continuous spin flips in a form best suited for the interpretation of experimental data in the regime of detuned resonances. Section\,\ref{sec:impact-of-detuning} contains an introduction to the main effects stemming from frequency detuning. Manifestations of detuning in the polarimetry of the in-plane polarization, most crucial for the pilot-bunch technique, are treated in Sec.\,\ref{sec:Polarimetry}. The impact of spin decoherence on spin flips is a subject treated in Sec.\,\ref{sec:decoherence}. In Sec.\,\ref{sec:Tomography}, we discuss spin-flip tomography along the bunch length and depolarization of the pilot bunch caused by incomplete gating-out of the radiofrequency  Wien filter. Implications of the derived formalism to the interpretation  of the precursor EDM search experiments are explored in Sec.\,\ref{sec:precursor}. In Sec.\,\ref{sec:summary-and-conclusions}, we summarize our main results. The phenomenology of the results of the pilot bunch experiment\,\cite{Slim:2023lpd} within the synchrotron oscillation-mediated spin-decoherence approach is presented in Appendix\,\ref{SOappendix}.
			
\begin{table*}[t]
	\caption{\label{tab:parameters}  Glossary of frequently used parameters and variables (auxiliary variables derived are omitted).  }
	\renewcommand{\arraystretch}{1.1} 
	\begin{ruledtabular}
	\begin{tabular}{lcr}
	Parameter/Variable 	& Notation  &  Defined in or near \\ \hline
	Turn number	& $n$	& Eq.\,(\ref{MasterEq}) \\
	Spin tune	& $\nu_\text{s}  $	& Eq.\,(\ref{MasterEq})  \\
	Spin phase increment per turn & $\theta_\text{s}  $			& Eq.\,(\ref{MasterEq}) \\
	Spin stable axis & $\vec{c}$	& Eqs.\,(\ref{MasterEq}), (\ref{eq:et-er-coordinate-system}) \\		  			
	Wien filter tune	& $\nu_\text{WF}  $			& Eq.\,(\ref{eq:WFprecession})  \\
	Wien filter side band 	&  $K$	& Eqs.\,(\ref{eq:WFprecession}), (\ref{WFmatrix})  \\
	Wien filter phase increment per turn & $\theta_\text{WF}$	& Eq.\,(\ref{eq:WFprecession}) \\
	Spin kick in the Wien filter& $\chi_\text{WF}$	& Eq.\,(\ref{eq:WFprecession})  \\
	Magnetic anomaly of a particle	& $G$	& Eq.\,(\ref{eq:WFprecession}) \\	  			
	Beam velocity in units of the speed of light & $\beta $	& Eq.\,(\ref{eq:WFprecession})  \\
	Relativistic factor & $\gamma$	& Eq.\,(\ref{eq:WFprecession}) \\
	Polarization vector		& $\vec{S}$	& Eqs.\,(\ref{MasterEq}), (\ref{Envelope})  \\
	Polarization envelope  	& $\vec{p}$	& Eq.\,(\ref{Envelope}) \\
	Spin-flip oscillation phase	& $x$	& Eqs.\,(\ref{EnvelopePhase}), (\ref{SF-phase}) \\
	Spin-flip tune on the exact spin resonance 		& $\nu_\text{SF}^0$	& Eq.\,(\ref{Etune}) \\
	Initial phase of the in-plane polarization 	& $\Phi_\text{in}$ & Eq.\,(\ref{InitialPhase}) \\	  			
	Spin precession vs. Wien filter frequency detuning parameter & $\delta$	& Eq.\,(\ref{eq:def-detuning}) \\
	Spin-flip tune off the  exact spin resonance & $\nu_\text{SF}$	& Eq.\,(\ref{eq:DetunedEnvelopeTune})\\		  			
	Angle of orientation of the spin envelope precession axis	& $\rho $& Eq.\,(\ref{eq:DetunedEnvelopeTune2}),(\ref{eq:vecm})  \\
	Shift of the spin-flip symmetric interval $x \in [\zeta, 2\pi+\zeta]$  & $\zeta$ & Eq.\,(\ref{zeta})\\
	In-plane polarization envelope phase during continuous spin flips	& $\phi(x)$	& Eq.\,(\ref{rt-phase})  \\
	Spin precession feedback period  & $ t_\text{fb}$   &  Eq.\,(\ref{eq:cosrhofb}) \\
	Spin precession phase walk during feedback period  & $ \sigma_\text{fb}$   &  Eq.\,(\ref{eq:cosrhofb}) \\
	In-plane polarization damping per turn in the exponential approximation & $ \Gamma$   &  Eqs.\,(\ref{Damping}), (\ref{eq:Gamma-damping}) \\
	Spin coherence time    & $\tau_\text{SCT} $            &  \text{Eqs.}  (\ref{eq:Gamma-damping}), (\ref{SOSCtime}) \\
	Fractional cyclotron phase of a particle in the bunch				& $\phi$			& Eq.\,(\ref{DampedEnvelopes1}) \\
	Slip factor 				& $\eta$  		& Eq.\,(\ref{SlipFactor1})  \\
	Gaussian rms width of the synchrotron oscillation amplitude distribution & $\sigma_\text{sy}$ & Eq.\,(\ref{BunchLength})  \\
	Amplitude of the synchrotron oscillations in  the spin precession phase & $\psi_\text{sy}$	& Eqs.\,(\ref{DeltaTheta})  \\
	Normalized synchrotron oscillation amplitude & $\xi$ & Eq.\,(\ref{DeltaTheta})  \\
	Synchrotron oscillation amplitude distribution function & $F(\xi)$ & Eq.\,(\ref{WeightF})\\  
	Parameter of the synchrotron oscillation driven slip of the Wien filter phase & $C_\text{WF}$ & Eq.\,(\ref{CWF}) \\
	Synchrotron oscillation strength in the spread of the spin-flip phase	& $Q_\text{sy}$	& Eq.\,(\ref{ApproxBessel1}), (\ref{ApproxBessel2}) \\
	Tilt of the spin stable axis by the electric dipole moment of a particle & $\xi^\text{EDM}$	& Eq.\,(\ref{IdealRing})  \\
	Gaussian rms length of the signal (s) bunch in the pilot (p) Bunch experiment & $\sigma_\text{s, p}$ & Appendix\,\ref{SOappendix} 	  			
	\end{tabular}
	\end{ruledtabular}
\end{table*}
			
\section{Stroboscopic spin evolution in the off-resonance regime}
\label{sec:stroboscopic-off-resonance}

\subsection{Master equation}
\label{sec:stroboscopic-master-equation}

In storage rings, the one-turn evolution of the spin $\vec{S}$ consists of the idle precession by an angle $\theta_\text{s} = 2\pi\nu_\text{s}$ about the spin stable axis ${\vec c}$, followed by the spin kick in the orbit-preserving radiofrequency   Wien filter, which is used as a spin flipper and is located in a straight section of the ring. Here $\nu_\text{s} = f_\text{s}/f_\text{c}$ denotes the spin tune, \textit{i.e.}, the number of spin precessions with respect to particle momentum per revolution. The length of the Wien filter is negligibly small compared to the ring circumference and it acts on the spin stroboscopically once per turn. As an introduction to the subject, in this section, we describe the radiofrequency excited	spin rotations in the SO(3) formalism\,\cite{PhysRevAccelBeams.23.024601} (for an alternative spinor formalism, see\,\cite{derbenev1971dynamics}, the textbook in Ref.\,\cite{SYLee}, and Ref.\,\cite{SpinTuneMapping}).
	 
The stroboscopic master equation for the spin vector $\vec{S}(n)$ as a function of the turn number $n$ is given by
\begin{equation}
	\vec{S}(n)= \matr{R}_\text{WF}(n) \matr{R}_\text{c}(\theta_\text{s}  ) \vec{S}(n-1)\,,
	\label{MasterEq}
\end{equation}
where $ \matr{R}_\text{c}(\theta_\text{s}  )$ and $\matr{R}_\text{WF}(n)$ are the ring and Wien filter spin transfer matrices, respectively. Alongside ${\vec c}$, we define the radial unit vector $\vec{e}_\text{r}$ and	the longitudinal unit vector $\vec{e}_\text{t}$ (tangential to the orbit). These three unit vectors form the orthogonal basis 
\begin{equation}
\begin{split}
  \vec{e}_\text{t}  & = \vec{e}_\text{r}\times\vec{c}\,, \\
  \vec{e}_\text{r} 	& = \vec{c}\times \vec{e}_\text{t}\,.
\end{split}
\label{eq:et-er-coordinate-system}
\end{equation}
 The vectors ${\vec e}_\text{r}$ and ${\vec	e}_\text{t}$ define the spin precession plane. Because of the  magnetic field imperfections in the ring lattice, the orientation of $\vec{c}$ differs slightly from $\vec{e}_y$, the normal one to the  storage ring plane, aka the $\{\vec{e}_x, \vec{e}_z\}$ momentum plane, and the spin precession plane is tilted with respect to the ring plane\,\cite{PhysRevAccelBeams.23.024601}. Wherever relevant, as will be the case in the discussion of the imperfection fields in Sec.\,\ref{sec:precursor}, we will distinguish between the spin and momentum bases, and our reference to $\vec{c}$ as the \textit{vertical} direction, and to the components of the spin in the spin precession plane as the \textit{horizontal} ones, should not cause any confusion.
	
We treat a particle on the reference orbit in the approximation of vanishing spin decoherence. Then the idle precession spin transfer matrix per turn is given by
\begin{equation}
		\matr{R}_\text{c}(\theta_\text{s})=
		\begin{pmatrix} \cos\theta_\text{s} & 0 &  \sin\theta_\text{s} \\
		0 &  1 & 0 \\
		-\sin\theta_\text{s}  & 0 &  \cos\theta_\text{s} 
		\label{IdleMatrix}
		\end{pmatrix}\, . 
\end{equation}
The Wien filter     axis $\vec w$ is along its magnetic field $ \vec{B}_\text{WF}$. The spin kick per pass of the Wien filter     of length $L_\text{WF}$ equals  
\begin{equation}
\chi(n) =      \chi_\text{WF} \cos( \theta_\text{WF}n) \label{WFkick}
\end{equation}
 with the amplitude
\begin{equation}
     \chi_\text{WF}= -\frac{q (1+G)B_\text{WF}L_\text{WF}}  {m \gamma^2 \beta} \, ,
\label{eq:WFprecession}
\end{equation}
where $q$,  $m$, $\beta$ and $G$ are the charge, mass, velocity, and magnetic anomaly of the orbiting particles. The Wien filter is operated at the frequency $f_\text{WF}$, the WF tune is given by $\nu_\text{WF}=f_\text{WF}/f_\text{c}$ and 
$\theta_\text{WF}=2\pi \nu_\text{WF}$. Evidently, the spin rotation in the WF is identical for all side bands $\nu_\text{WF} \Rightarrow \nu_\text{WF} +K, \quad K=0,\ \pm 1,\ \pm 2,\ \ldots$ Without loss of generality, we can focus the discussion on the so-called magnetic-dipole moment (MDM) mode, when  $\vec{w}=\vec{e}_\text{r}$ and $|\vec{c} \times \vec{w}|=1$. The spin transfer matrix for pass $n$ through the Wien filter equals 
\begin{equation}
\matr{R}_\text{WF}(n)=
\begin{pmatrix} 1 & 0 &  0 \\
0 &  \cos\chi(n) & -\sin \chi(n)  \\
0  & \sin \chi(n)  &  \cos \chi(n)  
\end{pmatrix} = 1 +\matr{W}(n)\,  . 
\label{WFmatrix}
\end{equation}
Note that the evolution of the experimentally observed polarization vector is identical to that of the quantum spin operator, and we retain $\vec{S}$ as notation for the polarization vector in what follows.

\subsection{Bogoliubov-Krylov averaging for exact spin resonance}
\label{BKaveraging}

The above outlined hierarchy of spin evolution frequencies (Table\,\ref{tab:hierarchy-frequencies}) dictates invoking the Bogoliubov-Krylov (BK) averaging \cite{BKaverage} as a tool for a solution of the master equation (\ref{MasterEq}). To give some background, we illustrate the main points of the case of exact resonance $\nu_\text{s}   = \nu_\text{WF}$ following the treatment in Ref.\,\cite{SpinTuneMapping}. The starting point is the interaction representation 
\begin{equation}
\vec{S}(n)= \left| \vec{S}(0) \right|\matr{R}_\text{c}( n\theta_\text{WF}  ) \vec{p}(n)\,, 
\label{Envelope}
\end{equation}
where $\vec{p}(n)$ is the spin envelope with initial condition $|\vec{p}(0)|=1$ defining the polarization as seen by a stationary observer in the co-rotating reference frame rotating about the axis $\vec{c}$ with frequency $f_\text{WF}$. Without loss of generality, in the following we set $|\vec{S}(0)|=1$.

A brief digression on this choice of the co-rotating frame is in order. The choice is dictated by the point that $f_\text{WF}$ is the only {\it known}  primary frequency in the problem. The spread of spin tunes in the bunch and the {\it unknown} walk of the spin precession frequency  necessitate a continuous  measurement of this {\it unknown} frequency in order to obtain a feedback for setting the Wien filter to another known frequency, etc. (In practice, of course, the beam interacts stroboscopically with the polarimeter target once per turn.) To the extent that intrabeam interactions are weak to depolarize the beam (see for instance Ref.\,\cite{PhysRevSTAB.18.020101} and the related discussion in Sec.\,\ref{sec:introduction}), the bunch can be treated as an ensemble of independent particles, so that we solve first the one-particle problem and then take the average over the ensemble.

To the laboratory-frame observer the idle precessing in-plane polarization  is described by 
\begin{equation}
\begin{split}
\vec{u}_\text{r} (n)& = \phantom{-}\vec{e}_\text{r} \cos(\theta_\text{WF} n) + \vec{e}_\text{t} \sin (\theta_\text{WF}n)\,,\\
\vec{u}_\text{t} (n)& = -\vec{e}_\text{r} \sin (\theta_\text{WF}  n) + \vec{e}_\text{t} \cos (\theta_\text{WF}n)\, . 
\label{IdleVectors}
\end{split}
\end{equation}
The master equation for the spin envelope takes the form 
\begin{equation}
\begin{split}
\vec{p}(n) = \matr{R}_\text{c}(-n\theta_\text{WF} )\matr{R}_\text{WF}(n)
\matr{R}_\text{c}(n\theta_\text{WF} )\vec{p}(n-1)\,.
\label{Master}
\end{split}
\end{equation}
In view of $ \chi_\text{WF} \ll 1$, the stroboscopic Eq.\,(\ref{Master}) can be cast in the differential form 
\begin{equation}
\frac{\vec{p}(n)}{\dd n}= \matr{R}_\text{c}(-n\theta_\text{WF})\matr{W}(n)
\matr{R}_\text{c}(n\theta_\text{WF})\vec{p}(n) \, . 
\label{DiffEq}
\end{equation} 
To the leading order in the small parameter  $ \chi_\text{WF}$ the BK averaging over the spin precession periods proceeds as follows:
\begin{widetext}
	\begin{equation}
	\begin{split}
	 \langle \matr{R}_\text{c}(-n\theta_\text{WF} n)\matr{W}(n) \matr{R}_\text{c}(n\theta_\text{WF} n) \rangle 
	& = \left\langle\begin{pmatrix}
	0 & -\chi(n)\sin(n\theta_\text{WF}) & 0 \\
	\chi(n)\sin(n\theta_\text{WF})&  0& -\chi(n)\cos(n\theta_\text{WF}) \\
	-0 & \chi(n)\cos(n\theta_\text{WF}) & 0
	\end{pmatrix} \right\rangle\\
	&=
	\begin{pmatrix}
	0 & 0 & 0\\
	0&  0 & -\frac{1}{2}     \chi_\text{WF} \\
	-0 & \frac{1}{2}     \chi_\text{WF}  & 0
	\end{pmatrix} 
	= 2\pi\nu_\text{SF} \begin{pmatrix}
	0 & 0 & 0 \\
	0&  0 & -1 \\
	0 & 1 & 0 
	\end{pmatrix}   =2\pi\nu_\text{SF} \matr{U} 
	\, , \label{matrU}
	\end{split}
	\end{equation}
	\end{widetext}
where we applied
\begin{equation}
\langle \cos^2 (\theta_\text{WF} n)\rangle  \to \frac{1}{2} \quad \text{and} \quad
\langle \cos (\theta_\text{WF} n) \sin (\theta_\text{WF} n)\rangle  \to 0\,.
\end{equation}
The solution of Eq.\,(\ref{DiffEq}) for the envelope will be 
\begin{equation}
\vec{p}(x) =\exp(2\pi\nu_\text{SF} n \matr{U})\vec{p}(0)=\matr{E}_0(x)\vec{p}(0)\, , \label{EnvEvol}
\end{equation}
where the subscript 0 stands for zero detuning. Making use of the recursive relations, 
\begin{equation}
\matr{U}^{2n+1} =(-1)^n \matr{U}\,, \quad	\matr{U}^{2n} =(-1)^{n-1} \matr{U}^2\,, \label{Recurs}
\end{equation}
we decompose the Taylor expansion of $\matr{E}_0(x)$ into sums of the odd and even powers of $\matr{U}$ with the result
	\begin{equation}
	\begin{split}
	\matr{E}_0(x)
	&=\sum_{k=0} \frac{(-1)^k x^{2k}}{(2k)!}
	\matr{U}^{2k} + \sum_{k=0} \frac{(-1)^k x^{2k+1}}{(2k+1)!}
	\matr{U}^{2k+1} \\
	&= \matr{1} + \sin x \matr{U} + (\cos x  -1
	)\matr{U}^2  \\
&=\begin{pmatrix}
1 & 0 & 0\\
0&  \cos x & -\sin x \\
0 & \sin x  & \cos x
\end{pmatrix} \, ,
	 \label{EnvelopeEvolution1}
\end{split}
\end{equation} 
where 
\begin{equation}
x=2\pi \nu_\text{SF}^0 n\
\label{EnvelopePhase}
\end{equation} 
is the SF phase with the SF tune 
\begin{equation}
\nu_\text{SF}^0= \frac{1}{4\pi}      \chi_\text{WF}  \left| \vec{c}\times 	\vec{w} \right|\, , 
\label{Etune}
\end{equation} 
which defines the SF frequency $f_\text{SF} = \nu_\text{SF}^0 f_\text{c}  $. The factor  $\left| \vec{c}\times 	\vec{w} \right|$ emerges  for generic orientation of the  Wien filter     axis $\vec{w}$\,
\cite{SpinTuneMapping,PhysRevAccelBeams.23.024601}. For instance, the so-called EDM mode corresponds to $\vec{w} \approx \vec{e}_y$. 

\begin{figure}[tb]
	\includegraphics[width=1.0\columnwidth]{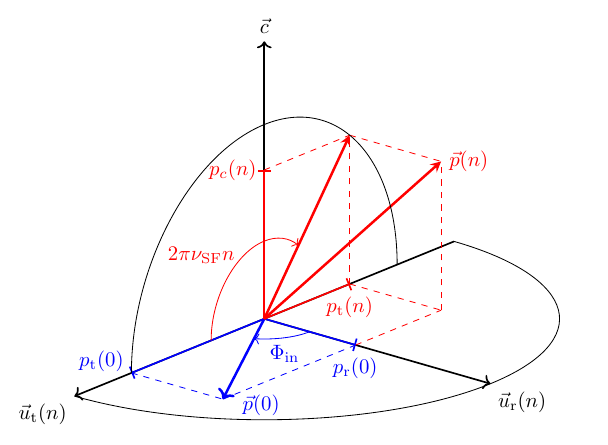}
	\caption{\label{fig:scotimesrf} 
		Evolution of the spin envelope in the reference frame, co-rotating at the idle spin precession frequency $f_\text{s}  $. The initial polarization $\vec{p}(0)$ is in the horizontal $\{\text{rt}\}$ ring plane. The spectator radial component $p_\text{r}(0) = p(0)\cos\Phi_\text{in}$ is immune to the radiofrequency   Wien filter and continues to precess unchanged. The active tangential component $p_\text{t}(0)=p(0)\sin\Phi_\text{in}$ starts rotations driven by the Wien filter in the vertical $\{\text{ct}\}$ plane with the spin-flip frequency $f_\text{SF} = \nu_\text{SF}f_\text{rev}$. To the observer in the co-rotating frame, the idly precessing unit vectors $\vec{u}_\text{r}(n)$ and  $\vec{u}_\text{t}(n)$ appear as being constant along the radial and tangential directions.}
\end{figure}

Note that SFs proceed via rotation of the vertical envelope to the tangential one with the frequency $f_\text{SF}$, while the radial envelope remains a spectator,
\begin{equation}
\begin{split}
p_\text{r}(x)&=p_\text{r}(0)\,, \\
p_\text{c}(x)&=p_\text{c}(0) \cos x -p_\text{t}(0) \sin x\, ,\\
p_\text{t}(x)&=p_\text{c}(0) \sin x -p_\text{t}(0) \cos x\, .\\
\end{split}
\end{equation}
The final result for the polarization is
\begin{equation}
\vec{S}(n) = \matr{R}_\text{c}(n\theta_\text{WF}) \matr{E}_{0}(x)\vec{S}(0)\,, \label{SpinRot}
\end{equation}
with the expansion
\begin{equation}
\begin{split}
\vec{S}(n) &= \matr{R}_\text{c}(n\theta_\text{WF}) \matr{E}_{0}(x)\vec{S}(0)\\
& = \left| \vec{S}(0) \right| \left\{ p_\text{r}(x) \vec{u}_\text{r} (n) + p_\text{c}(x) \vec{c}_\text{r} +  p_\text{t}(x) \vec{u}_\text{t} (n) \right\}\,.
\label{Decomposition}
\end{split}
\end{equation}
The generic initial condition is defined by the initial spin precession phase $\Phi_\text{in}$. Our convention is
\begin{equation}
\begin{split}
\vec{p}(0) &= 	p_\text{c}(0)\vec{c} +	p_\text{r}(0)\vec{e}_\text{r} +	p_\text{t}(0)\vec{e}_\text{t}\\
&=	p_\text{c}(0)\vec{c} +
p_\text{rt}(0)(\cos\Phi_\text{in}\vec{e}_\text{r}
+ \sin\Phi_\text{in}\vec{e}_\text{t})\, , 
\label{InitialPhase}
\end{split}
\end{equation}
where $p_\text{rt}=\sqrt{p_\text{r}^2 +p_\text{t}^2}$ denotes the modulus of the in-plane polarization. These features of the radiofrequency   driven polarization are shown in Fig.\,\ref{fig:scotimesrf}. For the pure in-plane initial polarization $p_\text{c}(0)=0$, the envelope of the vertical polarization evolves as $p_\text{c}(x)=-\sin\Phi_\text{in}\sin x$.

Unitarity features of the master equation (\ref{MasterEq}) are noteworthy. Here two unitary spin transfer matrices do describe sequential rotations with preservation of the magnitude of the polarization. Our final result in Eq.\,(\ref{SpinRot}) has precisely the same unitarity property. 

In order to estimate the higher-order corrections  to the SF tune, one must proceed in Eq.\,(\ref{matrU}) with the BK averaging of the exact expression $\sin\chi(n)\cos(n\theta_\text{WF})$, instead of the perturbative expression $\chi(n)\cos(n\theta_\text{WF})$, with the result 
\begin{equation}
\langle \sin\chi(n)\cos(n\theta_\text{WF}) \rangle = J_1 (     \chi_\text{WF})\, , 
\end{equation}
where
where $J_n(z)$ is the Bessel function,
\begin{equation}
J_\text{n}(z) = \left(\frac{z}{2}\right)^n \sum_{m=0}^{\infty} \frac{(-1)^m}{m!(m+n)!}\left(\frac{z^2}{4}\right)^m\, .
\end{equation} 
For conditions of the typical JEDI experiments with deuterons, we have an extremely small argument in the Bessel function,
\begin{equation}
\frac{     \chi_\text{WF}}{2} = 2\pi \nu_\text{SF}^0 = 2\pi \frac{f_\text{SF}}{f_\text{c}} \approx 10^{-6}\, ,
\end{equation}
and the correction to the linear approximation for the SF tune amounts to $\approx 10^{-12}$. This gives a time independent renormalization of the polarization and can safely be neglected, see the related discussion of Eq.\,(\ref{IdleSO}) in Sec.\,(\ref{sec:MasterEqSpinEnv}).

\subsection{Off-resonance spin rotations}
\label{sec:off-resonancerotations}

We have at our disposal two \textit{known} parameters: the Wien-filter frequency $f_\text{WF}$ and the Wien-filter strength $\chi_\text{WF}$ (spin kick). Detuning is parameterized in terms of the small angle 
\begin{equation}
	\delta =\theta_\text{s}   - \theta_\text{WF} = 2\pi (\nu_\text{s}-\nu_\text{WF} ) = 2\pi \frac{\Delta f_\text{s}}{f_\text{c}}\,.
	\label{eq:def-detuning}
\end{equation}
Correspondingly, we define the interaction representation in terms of the \textit{known} Wien filter frequency as in Eq.\,(\ref{Envelope}), and cast the spin evolution in Eq.\,(\ref{MasterEq}) in the form
\begin{equation}
	\begin{split}
	\matr{R}_\text{c}(n\theta_\text{WF}) \vec{p}(n)  
	=  \matr{R}_\text{WF}(n) \matr{R}_\text{c}(\delta)\matr{R}_\text{c}(n\theta_\text{WF} )
	\vec{p}(n-1)\,.  \label{MasterEqDetuned}
	\end{split}
\end{equation}
Following Eq.\,(\ref{WFmatrix}), we introduce the detuning corrected expression $\matr{W}(n) =  \matr{R}_\text{WF}(n) \matr{R}_\text{c}(\delta)-\matr{1}$ and proceed to 
the BK averaging of
\begin{widetext}
	\begin{equation}
	\begin{split}
	&  \matr{R}_\text{c}(-n\theta_\text{WF} n)\matr{W}(n) \matr{R}_\text{c}(n\theta_\text{WF} n)  
	=\begin{pmatrix}
	0 & -\chi(n)\sin(n\theta_\text{WF}) & \delta \\
	\chi(n)\sin(n\theta_\text{WF})&  0& -\chi(n)\cos(n\theta_\text{WF}) \\
	-\delta & \chi(n)\cos(n\theta_\text{WF}) & 0
	\end{pmatrix} 
	\, .  \label{Wdetuned}
	\end{split}
	\end{equation}
\end{widetext}
The  corresponding matrix $\matr{U}$ takes the form 
\begin{equation}
\begin{split}	
\matr{U} & =\begin{pmatrix}
0 & 0 & \cos \rho \\
0&  0 & -\sin\rho \\
-\cos\rho & \sin\rho  & 0 
\end{pmatrix}
\, .\label{matrUdetuned}
\end{split}
\end{equation}
The detuning modified SF tune equals
\begin{equation}
\begin{split}
	\nu_\text{SF} =  \frac{\sqrt{     \chi_\text{WF}^2 + 4\delta^2}}{4\pi} = 	\frac{\nu_\text{SF}^0}{\sin\rho}\,,
	 \label{eq:DetunedEnvelopeTune}
	\end{split}
\end{equation}
where we parameterize detuning in terms of the angle $\rho$ such that
\begin{equation}
\begin{split}
\sin\rho = \frac{     \chi_\text{WF}}{4\pi\nu_\text{SF}}\,, \quad 
\cos\rho = \frac{2\delta}{4\pi\nu_\text{SF}}\, . 
\label{eq:DetunedEnvelopeTune2}
\end{split}
\end{equation}	
We reiterate that in the  generic case the substitution
$     \chi_\text{WF} \Rightarrow \left| \vec{c}\times \vec{w}\right|     \chi_\text{WF}$ is in order, so that
\begin{equation}
\nu_\text{SF}^2= \frac{1}{16\pi^2}\left(      \chi_\text{WF}^2  \left| \vec{c}\times 	\vec{w} \right|^2 +4\delta^2\right)\, . \label{NuSquared}
\end{equation}

The above derived $\matr{U}$ satisfies the recursive relations from Eq.\,(\ref{Recurs}), so that application of the decomposition in Eq.\,(\ref{EnvelopeEvolution1}) yields  
\begin{widetext}
\begin{equation}
	\begin{split}
		\matr{E}(x)=\begin{pmatrix}
		E_\text{rr}(x)  &  E_\text{rc}(x) & E_\text{rt}(x)\\
		E_\text{cr}(x) & E_\text{cc}(x)  & E_\text{ct}(x) \\
		E_\text{tr}(x) & E_\text{tc}(x) & E_\text{tt}(x)
		\end{pmatrix} 
		=\begin{pmatrix}
		\sin^2 \rho+\cos^2 \rho \cos x  &  \cos \rho \sin\rho (1-\cos x) & \cos \rho \sin x \\
		\cos \rho \sin\rho (1-\cos x) & \cos^2 \rho+\sin^2 \rho\cos x &
  -\sin \rho \sin x \\
-\cos \rho \sin x &  \sin \rho \sin x  & \cos x 
\end{pmatrix}\, , \label{Envelopes}
\end{split}
\end{equation}
\end{widetext}
which describes the envelope rotations about the axis 
\begin{equation}
	\vec{m} = \sin\rho \, \vec{e}_\text{r} -\cos\rho \, \vec{c} \,,\label{NewAxis}
\end{equation} 
with the SF phase 
\begin{equation}
x=2\pi\nu_\text{SF}n =2\pi \nu_\text{SF}f_c t\,. 
\label{SF-phase}
\end{equation} (for generic SO(3) rotations, see Ref.\,\cite{SO3rotations}). In the subsequent discussion, the $x$-dependence and the time-dependence are interchangeable.
 
Within the spinor formalism, an early derivation of  Eq.\,(\ref{Envelopes}) was already presented in the 2017 JEDI publication\,\cite{SpinTuneMapping}, and the alternative and equivalent treatment of the same problem was reported in the follow-up JEDI publication in 2018\,\cite{JEDIphase}. The above outlined SO(3) formalism  will play a pivotal role in the subsequent incorporation of the spin-decoherence effects that will be discussed in Sec.\,\ref{sec:decoherence}.

\subsection{Radiofrequency solenoid as a spin rotator}
\label{Solenoid}
	
The above formalism is fully applicable as well to the orbit preserving radiofrequency   solenoid as a spin rotator. In that case, one needs to interchange $\vec{e}_\text{t} \Rightarrow \vec{e}_\text{r},\, \vec{e}_\text{r} \Rightarrow -\vec{e}_\text{t}$ and also the corresponding indices $\text{r} \Leftrightarrow \text{t}$ in the matrix elements of $\matr{E}$. The spin kick $     \chi_\text{WF}$ in the Wien filter     must be swapped for the spin kick in the solenoid $\chi_\text{sol}$,
\begin{equation}
	     \chi_\text{WF}\Rightarrow \chi_\text{sol} =
	-\frac{q (1+G)}{m v} \int dz B(z)\, ,
	\label{eq:Solprecession}
\end{equation}
where $B(z)$ is the longitudinal magnetic field in the solenoid.	In the co-rotating frame of reference, the spin envelope would precess about the axis 
\begin{equation}
	\vec{m} = -\sin\rho \, \vec{e}_\text{t} +\cos\rho \, \vec{c}\, . 
	\label{eq:vecm}
\end{equation} 
In the limit of vanishing detuning, $\cos\rho=0$, the spectator in-plane polarization will be directed along $\vec{e}_\text{t}$. In addition, the convention for the initial spin phase has to be modified such that $\Phi_\text{in} \to \Phi_\text{in} +\sfrac{\pi}{2}$.
	
\section{Impact of detuning on the vertical polarization}
\label{sec:impact-of-detuning}
	
\subsection{Evolution of vertical polarization}
\label{sec:evolution-of-py-from-py}
	
We start with the beam polarization stored along the spin stable axis $\vec{c}$, so that $p_\text{c}(0)=1$ and  $p_{r}(0)=p_{t}(0)=0$.  Note that the notion of an initial spin phase $\Phi_\text{in}$ is meaningful only for a non-vanishing precessing horizontal component of the polarization. With operating Wien filter, the vertical polarization will evolve as
\begin{equation}
	\begin{split}
	p_\text{c}(x)  = E_\text{cc}(x)p_\text{c}(0) 
	= ( \cos^2 \rho+\sin^2 \rho \cos x )p_\text{c}(0)\, . 
	\end{split}
	\label{cc}
\end{equation}

This result nicely illustrates the interplay of the detuning by $\delta$ [see Eqs.\,(\ref{eq:def-detuning}) and (\ref{eq:DetunedEnvelopeTune2})] and the spin kick $\chi_\text{WF}$ in the Wien filter:

\begin{enumerate}
	\item The envelope exhibits oscillations with amplitude $\sin^2\rho \leq 1 $ on top of the offset $\cos^2\rho$. 	
	
	\item In the regime of negligible detuning,  the offset	$\cos^2\rho \ll 1$ 	can be neglected and the vertical polarization will oscillate with full amplitude  $p_\text{c}(x)  = p_\text{c}(0) \cos x$. 
	
	\item As the detuning increases, the oscillation amplitude decreases, and at $\sin^2\rho < \sfrac{1}{2}$ the SF is incomplete: the offset term takes over and the vertical polarization no longer passes through zero. 
	
	\item At finite detuning, $\cos^2 \rho <\sfrac{1}{2}$,  the pure horizontal polarization is reached at the envelope phase
		\begin{equation}	
		   \cos x_0 = - \cot ^2\rho\,. 
		   \label{ZeroVertical}
		\end{equation}
	
	\item Conversely, to achieve the often-required $\sfrac{\pi}{2}$ rotation from the vertical to the horizontal spin orientation, usually performed on a time scale of approximately $\SI{1}{s}$ with the radiofrequency   solenoid\,\cite{morozov2003first}, the detuning needs to satisfy only the very liberal condition that 
		\begin{equation}
			\Delta f_\text{s} < \frac{1}{\sqrt{2}} f_\text{SF}\, .
		\end{equation}
	
	\item The detuning can be constrained by a comparison of the flipped, $S_\text{c}(\pi)$, and initial, $S_\text{c}(0)$, vertical polarizations,
		\begin{equation}
			2\cos^{2}\rho = 1 -  \frac{S_\text{c}(\pi)}{S_\text{c}(0)}\,.
		\end{equation} 
	The $\cos^{2}\rho$ thus determined must not be confused with the $\epsilon_\text{flip}$, which is  determined from the exponential attenuation of the vertical polarization\,\cite{PhysRevSTAB.18.020101}.
		
	\item In the limiting case of strong detuning, $\cos^2 \rho \to 1$, the amplitude of the oscillating term vanishes, the  rotation axis of the envelope becomes equal to the vertical axis, $\vec{m}=\vec{c}$, and the vertical polarization is preserved, $p_\text{c}(x)=p_\text{c}(0)$.
		
	\item The phase locking of spin precession with the radiofrequency   Wien filter     developed by the JEDI collaboration requires continuous feedback. In practice, continuous means stepwise, since one must collect statistics for $t_\text{fb} =  5 - \SI{10}{s}$ to measure the spin precession frequency with sufficient accuracy. The implications of the emerging  detuning with changing sign of Eq.\,(\ref{FeedBackDetuning}) will be discussed in Sec.\,\ref{sec:feedback}.
		
	\end{enumerate}

\subsection{Build-up of vertical polarization from in-plane polarization}
	
In this case, the initial conditions are $p_\text{c}(0)=0$ and  $p_\text{rt}(0)=1 $, and the initial in-plane polarization can be parameterized in terms of the initial spin phase $\Phi_\text{in}$, as given in Eq.\,(\ref{InitialPhase}).
	
Reading $E_\text{cr}(x)$ and $E_\text{ct}(x)$ from the envelope evolution matrix $\matr{E}(x)$ of Eq.\,(\ref{Envelopes}), we find
\begin{equation}
	\begin{split}
	p_\text{c}(x)  & = E_\text{cr}(x)p_\text{r}(0) +  E_\text{ct}(x)p_\text{t}(0) \\
	&= \sin\rho
	 \left(\cos \rho \cos \Phi_\text{in} (1-\cos x) -\sin\Phi_\text{in} \sin x\right) \\
	&=q(\Phi_\text{in},\rho)\sin\rho  
	 \sin\left(\frac{x}{2}\right) \sin\left(\frac{x}{2} -\zeta\right)\,,
	\end{split}	
	\label{HorToVert}
\end{equation}
where
\begin{equation}
	\begin{split}
	q(\Phi_\text{in},\rho) & =\sqrt{\sin^2\Phi_\text{in} + \cos^2\rho
		\cos^2\Phi_\text{in}}\,,\\
	\sin\zeta & = \frac{\sin\Phi_\text{in}}{\sqrt{\sin^2\Phi_\text{in} +
			\cos^2\rho \cos^2\Phi_\text{in}}}\,, \text{and} \\
	\cos\zeta & = \frac{\cos\rho\cos\Phi_\text{in}}{\sqrt{\sin^2\Phi_\text{in} +
			\cos^2\rho \cos^2\Phi_\text{in}}}\,. \label{zeta}
		\end{split}	
\end{equation}
In the case of $\zeta=0$, the vertical polarization is invariant under the interchange $x \Leftrightarrow 2\pi -x$ within the symmetric period interval $[0,2\pi]$, while for finite $\zeta$ the related invariance under $x-\zeta \Leftrightarrow 2\pi -(x-\zeta)$ persists in the shifted symmetric interval $[\zeta,2\pi+\zeta]$.
	
It is noteworthy that in the case exactly on resonance, $\cos\rho=0$,  and
\begin{equation}
	p_\text{c}(x) =  -p_\text{t}(0) \sin x = -\sin\Phi_\text{in} \sin x\, , \label{PlaneToVertical}
\end{equation}
so that only the initial tangential polarization is the active one, while the radial component of the horizontal  polarization remains a spectator component and does not contribute at all to the build-up of the vertical polarization.
	
\section{Polarimetry of the in-plane polarization}
\label{sec:Polarimetry}
	
\subsection{Amplitude and phase conventions}
In the generic case, the polarization components  are given by Eq.\,(\ref{Decomposition})
\begin{equation}
	\begin{split}
	S_\text{r}(x,n) & = \phantom{-}p_\text{r}(x)\cos (n\theta_\text{WF})
	+p_\text{t}(x)\sin (n\theta_\text{WF}) \, ,\\
	S_\text{c}(x,n) & = \phantom{-}p_\text{c}(x)\,, \\
	S_\text{t}(x,n) & = - p_\text{r}(x)\sin (n \theta_\text{WF})  +
	p_\text{t}(x)\cos (n \theta_\text{WF})\,. \label{Polarizations}
		\end{split}
\end{equation}
The running envelope $\vec{p}(x)$ is given by Eq.\,(\ref{EnvEvol})  with the $\rho$- and $x$-dependent evolution matrix of Eq.\,(\ref{Envelopes}), subject to the $\Phi_\text{in}$-dependent initial envelope $\vec{p}(0)$ of Eq.\,(\ref{InitialPhase}). The spin-flip phase $x$ and the turn number $n$ are related by Eq.\,(\ref{SF-phase}), we kept both on purpose to distinguish spin-flip rotations of the envelopes from the idle spin precession. Because of parity conservation in strong interactions, the tangential (longitudinal) polarization at the polarimeter $S_\text{t}(x,n)$ is not measurable. The up-down asymmetry in the polarimeter measures the radial (transverse) polarization $S_\text{r}(x)$. This measurement takes place stroboscopically once per revolution of the beam. The polarimeter signal as a function of turn number $n$ is Fourier-analyzed bin by bin, with a bin duration corresponding to about \num{e6} turns in the machine, but still sufficiently short so that the variation of the spin-flip phase $x$ and the walk of the in-plane-polarization envelopes $p_\text{r}(x)$ and $p_\text{t}(x)$ can be neglected.
  
A cartoon of the  Fourier analysis boils down to the evaluation of
\begin{equation}
  	\begin{split}
  		p_\text{r}(x)&= \frac{2}{N}\sum_{k=1}^N S_\text{r}(x,k) \cos k\xi_\text{WF}\,,\\
  		p_\text{t}(x)&= \frac{2}{N}\sum_{k=1}^N S_\text{t}(x,k) \sin k\xi_\text{WF}\,.
  	\end{split}
\end{equation}
where $k$ is the turn number of the corresponding event in the polarimeter, and $N$ is a total number of events in the bin. These definitions are supported by the least squares analysis, and both $ p_\text{r}(x)$ and $p_\text{t}(x)$ take their maximal magnitudes  at $\xi_\text{WF} = \pm \theta_\text{WF}$. Because only one component of the rotating spin vector $\vec{S}(x,k)$ is observed, there is a non-essential sign ambiguity in $p_\text{t}(x)$.
  
The orientation of $\vec{p}_\text{rt}$ is given by the phase $0 < \psi(x)<  2\pi$, specified in terms of 
\begin{equation}
  \begin{split}
  \sin\psi(x)&= \frac{p_\text{r}(x)}{\sqrt{p_\text{r}^2(x) +p_\text{t}^2(x)}}\, ,\\
  \cos\psi(x)&= \frac{p_\text{t}(x)}{\sqrt{p_\text{r}^2(x) +p_\text{t}^2(x)}}\, . 
  \label{psi}
  \end{split}
\end{equation}
The full-fledged four-quadrant determination of $\psi(x)$ is well possible, but without any loss of information, it is convenient to map the phase $\psi(x)$ onto the band $0< \phi(x)<\pi$, where 
\begin{equation} 
\begin{split}
\phi(x)  & = \arccos \left(\frac{p_\text{t}(x)}{\sqrt{p_\text{r}^2(x) +p_\text{t}^2(x)}}\right)\\
& = \arccos\left[(\cos\psi(x)\right]\,. 
\label{rt-phase}
\end{split}
\end{equation}
It terms of the four-quadrant definition, this amounts to assigning to the radial polarization  its modulus,
\begin{equation} 
  \begin{split}
  |p_\text{r}(x)|  = p_\text{rt}(x)|\sin \psi(x)| = p_\text{rt}(x)\sin \phi(x) \, .
  \end{split}
\end{equation} 

A comment on the statistical limitations is in order. With limited statistics, the magnitude $p_\text{rt}(x)$ of the in-plane component of the close-to-vertical polarization can only be measured to a certain accuracy $\Delta p_\text{rt}$, and  the accuracy of determination of the phase of $p_\text{rt}(x)$ deteriorates for small in-plane polarization, $\Delta \phi(x) \propto  \Delta p_\text{rt} / p_\text{rt}$. 

\subsection{Continuous spin rotation by the WF: build-up of pure initial  in-plane polarization }
\label{sec:continuous-WF}

We find it instructive to illustrate the RF-driven spin dynamics on the special case of \textit{continuous} spin rotations by the Wien filter. In terms of the generic three-stage process, outlined in Sec.\,\ref{sec:introduction}, in stage I, instead of making use of the the
 radiofrequency     solenoid, the spins are rotated by the Wien filter. Stage II is skipped altogether and stage III begin at the instant when the vanishing vertical polarization has been reached in stage I.	While in the generic three-stage process the detuning of the Wien filter     in stage III can be different from the detuning of the radiofrequency  -solenoid,  in stage I, due to the tuning of  the Wien filter     to the spin precession frequency, measured in stage II, in the regime of  \textit{continuous} Wien filter     operation the detuning angle $\rho$ is kept constant from stage I to stage III on.
 
Now we treat the  spin evolution starting  with the  initial polarizations $p_\text{c}(0)=1$ and	$p_\text{r}(0)=p_\text{t}(0)=0$.
The envelope rotation phase $x=0$ corresponds  to the time at which the spin rotator is switched on. The radial and tangential polarization envelopes are given by
\begin{equation}
	\begin{split}
	p_\text{r}(x) = E_\text{rc}(x) p_\text{c}(0) & =\cos \rho \sin\rho \, (1-\cos x )
	\, , \\
	p_\text{t}(x) = E_\text{tc}(x) p_\text{c}(0) & =\sin\rho \sin x \, p_\text{c}(0)
	\,. \label{VertToPlane}
	\end{split}
\end{equation}
It is interesting to note that although $p_\text{r}(x)$ is zero at $\cos x = 1$, in this regime it does not change its sign at any value of $x$. The positively defined envelope $p_\text{rt}(x)$ of the in-plane polarization equals
\begin{equation} 
	\begin{split}
	p_\text{rt}(x) & = \sqrt{p_\text{r}^2(x)+p_\text{t}^2(x)} \\
	& =2|\sin \rho| \cdot |\sin \frac{x}{2}|\sqrt{\cos^2 \frac{x}{2}+ \cos^2\rho \sin^2 \frac{x}{2}}  \,. \label{Prt}
	\end{split}
\end{equation}
	
\subsection{Cross talk of vertical, tangential and radial polarizations}
Special features of the case exactly on resonance ($\cos \rho=0$) are noteworthy. Although mathematically exact resonance is a special case, we will always come across its special properties, and it is still instructive. In this case the envelope rotation axis $\vec{m}$ of Eq.\,(\ref{NewAxis}) is a purely radial one. Viewed in the co-rotating frame, the vertical polarization can not rotate into the radial one along the rotation axis. Indeed, according to Eq.\,(\ref{VertToPlane}), in this case $p_\text{r}(x)$ would vanish. In other words, the spectator radial polarization decouples from the  vertical one, while the active tangential envelope will oscillate with the full amplitude $p_\text{c}(0)$. Similarly, the tangential polarization cannot rotate into the radial one. Alternatively formulated, the polarization along the rotation axis $\vec{m}$ is \textit{immune} to the RF-driven rotations and is preserved.
	
This decoupling of both the vertical component from the spectator in-plane component and the active component from the spectator in-plane component is lifted once $\cos\rho \neq 0$. In the former case, this is  clear from Eq.\,(\ref{VertToPlane}). In the latter case, the cross talk of radial and tangential polarizations is given by the matrix elements $E_\text{rt}  (x) = -E_\text{tr} (x)$ in Eq.\,(\ref{Envelopes}). For instance, if $ p_\text{c}(0) = p_\text{r}(0)=0$ and $p_\text{t}(0)=1$, then
\begin{equation}
		p_\text{r}(x) = \cos\rho \sin x  \,p_\text{t}(0)\,.
\end{equation}
Vice versa, at $ p_\text{c}(0)=p_\text{t}(0)=0$ and $p_\text{r}(0)=1$, we find
\begin{equation}
	p_\text{t}(x) = -\cos\rho \sin x \, p_\text{r}(0)\,.
\end{equation}
This cross talk is a natural consequence of the vertical component $\cos\rho \vec{c}$ of the rotation axis $\vec{m}$ of the envelope.
\begin{figure*}[htb]
	\includegraphics[width=0.9\textwidth]{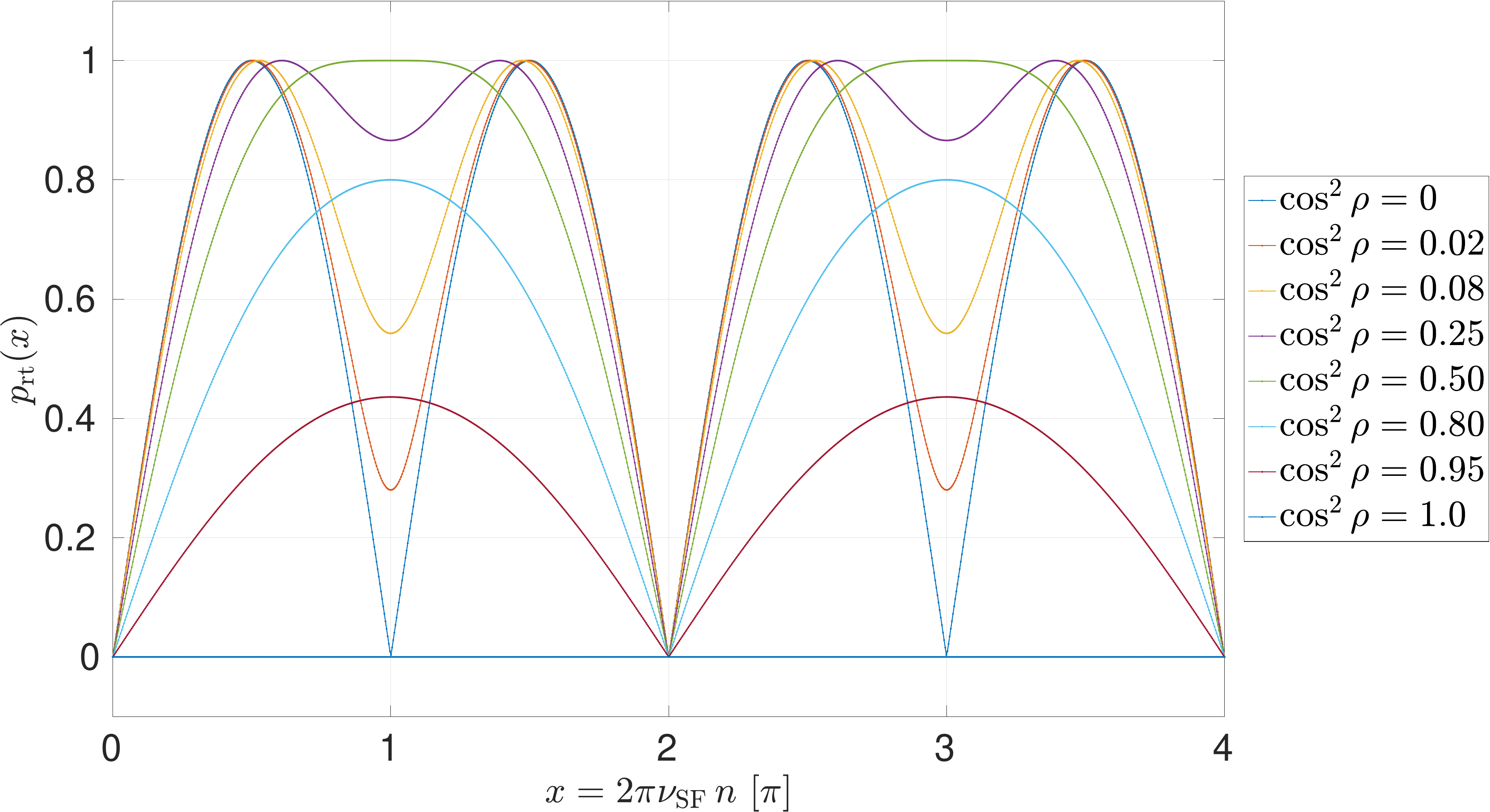}
	\caption{\label{fig:ContinuousInPlaneSpin} Pattern of the time dependence of the envelope of the horizontal polarization, which evolves from the pure vertical initial polarization $p_\text{c}(0)=1$, under the RF-driven continuous full or partial spin flips for different detunings, as given by Eq.\,(\ref{Prt}). Note that the central zero of $p_\text{rt}(x)$ at $x = \pi$ and $x = 3\pi$ (full spin flip) occurs exclusively at zero detuning, \textit{i.e.}, for $\delta=0$ or $\cos^2\rho = 0$. Within each period, the double hump structure with hump height $p_\text{rt}=1$ persists for $\cos^2\rho < \sfrac{1}{2}$. At even greater detuning, for $\cos^2\rho \geq \sfrac{1}{2}$, $p_\text{rt}(x)$ exhibits a single hump whose height vanishes in the limit $\rho \to 0$.}
\end{figure*}	
\subsection{Continuous spin rotation by the Wien filter    and envelope of in-plane polarization}

The result for $p_\text{rt}(x)$ has already been given in Eq.\,(\ref{Prt}). {The predicted dependence of the spin envelope on the detuning is depicted in Fig.\,\ref{fig:ContinuousInPlaneSpin} for $\cos\rho \ge 0$. As a function of the  phase $x$, the envelope $p_\text{rt}(x)$ is a periodic function with a period of $2\pi$, but in order to better demonstrate the periodicity properties of the in-plane polarization, we show the results for $x \in [0,4\pi]$. We start with the special case of vanishing detuning, \textit{i.e.}, with $\cos\rho=0$ and $\sin\rho=1$, when we recover the second line of Eq.\,(\ref{VertToPlane}),
\begin{equation}
	p_\text{rt}(x) = | 2 \left| p_\text{c}(0)\sin \left(\frac{x}{2}\right) \cos \left(\frac{x}{2}\right) \right|\,. \label{CosRhoZero}
\end{equation}
 In the interval $[0,\,2\pi]$ the envelope  has two end-point zeros at $x_1=0$ and $x_2= 2\pi$, stemming from $\sin(x/2) =0$. There is still another zero at midpoint $x_3=\pi$, stemming from $\cos(x/2)=0$. There are two maxima at $x_4 = \pi/2$ and $x_5= \pi/2 +\pi$, stemming from $p_\text{rt}'(x) =|p_\text{c}(0)| \cos x=0$. The change of the sign of $\sin\rho \Leftrightarrow$ corresponds to the change $\phi(x) \Leftrightarrow \pi-\phi(x)$.
 
 \begin{figure*}[htb]
	\includegraphics[width=0.9\textwidth]{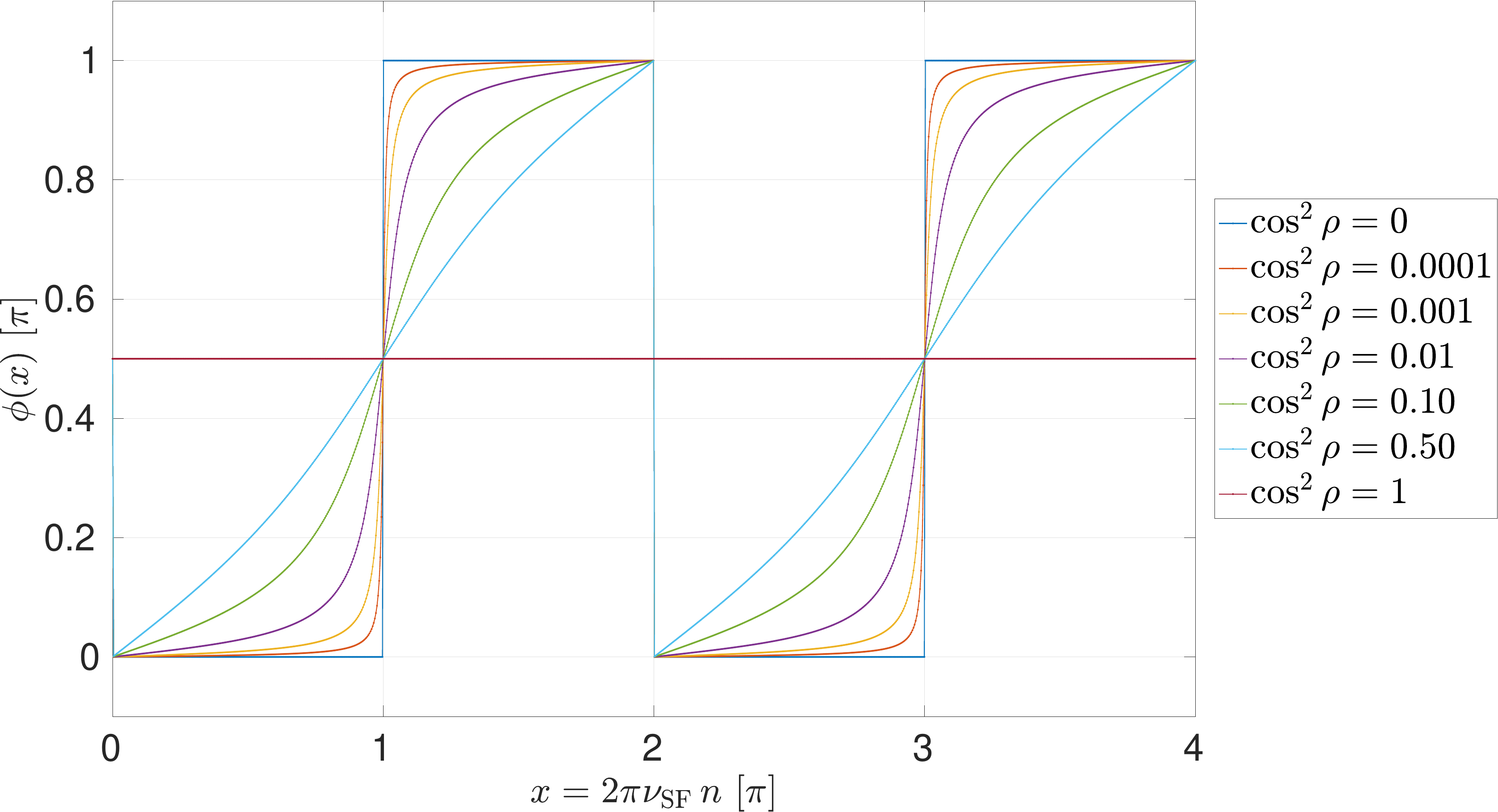}
	\caption{\label{fig:Phasemotion} Phase motion of the horizontal polarization envelope during the RF-driven continuous spin flips for different detunings, as predicted  by Eq.\,(\ref{PTRphase1}) for $\cos\rho \in [0,1]$. The phase exhibits a jump by $-\pi$  from $x=2\pi-0$ to $2\pi+0$, which repeats itself periodically at any $x=2\pi M,$\, where $M = 0,\ 1,\ 2,\ ,3,..$. In the vicinity of the phase jump, the slope $\phi'(2\pi-0) =\phi'(2\pi+0)= \frac{1}{2}\cos\rho$. Yet another jump by $+\pi$ develops at $x= \pi + 2\pi M$, where the slope, $\phi'(x=\pi) = 2/\cos\rho$, of the phase becomes singular for $\cos^2\rho \to 0$.}
\end{figure*}
	
The walk of these zeros and extrema with $\rho$ is as follows. The 	functional form in Eq.\,(\ref{Prt}) retains the end-point zeros at $\sin (x/2) =0$, {\it i.e.,} the $\rho$-independent  $x_1=0$ and $x_2=2\pi$. However, as soon as $\cos\rho \neq 0$, the midpoint zero disappears, and one has to look for zeros of the derivative $S'_\text{rt}=0$, which are roots of the equation
\begin{equation}
		\cos\left(\frac{x}{2}\right) \left[(1-2\sin^2\rho +2\sin^2\rho \cos^2 \left(\frac{x}{2}\right) \right] = 0\, . 
\end{equation} 
Here $\cos (x/2) = 0$ gives the mid-point extremum at $x_3=\pi$, where 
\begin{equation}
		p_\text{rt}(x_3) = \left| p_\text{c}(0)\sin2\rho \right| \,. \label{Midpoint} 
\end{equation}
The two other extrema are roots of the equation
\begin{equation}
		\cos^2\left(\frac{x}{2}\right) = 1- \frac{1}{2\sin^2\rho}\,,
\end{equation}
which has solutions only at $\sin^2\rho\geq \sfrac{1}{2}$,
\begin{equation}
	\begin{split}
		x_{4,5}(\rho)=\pi \pm 2\arcsin\sqrt{1- \frac{1}{2\sin^2\rho}}\,.
	\label{Roots}
	\end{split}
\end{equation}
The separation of these two roots,
\begin{equation}
	x_5(\rho)-x_4(\rho) = 4\arcsin\sqrt{1- \frac{1}{2\sin^2\rho}}
\end{equation} 
starts at $\pi$ at $\sin^2\rho=1$ and vanishes at $\sin^2\rho =\sfrac{1}{2}$, when the  roots $x_4$ and $x_5$ merge with $x_3=\pi$. Note that prior to this merger, the minimum of the envelope $p_\text{rt}(x) =|p_\text{c}(0)\sin2\rho| <|p_\text{c}(0)|$ will be  sandwiched between the maxima $p_\text{rt}(x_{4,5}) = |p_\text{c}(0)|$, while at still smaller $\sin^2\rho < \sfrac{1}{2}$, the envelope will exhibit a single bump with height $p_\text{rt}(x) =|p_\text{c}(0)\sin2\rho|$.
	
\subsection{Continuous spin rotation by the Wien filter     and phase of  in-plane polarization}
\label{continuous-from-vertical}
	
The expected phase motion for $\cos\rho >0$ is depicted in Fig.\,\ref{fig:Phasemotion} for several values of $\rho$. According to Eq.\,(\ref{VertToPlane}), in the considered case the radial envelope does not change its sign at all, \textit{i.e.}, $\sgn(p_\text{r}(x))=+1$, while $p_t(x)$ changes the sign at $x=\pi$. Still, at $x\neq \pi$ the phase remains well defined. Making use of $p_\text{t}(x)$ from Eq.\,(\ref{VertToPlane}) and $p_\text{rt}(x)$ from Eq.\,(\ref{Prt}), we obtain		  
\begin{equation}
	  \phi(x)=\arccos\left(\frac{\sgn(\sin x)\sgn(\sin\rho)}{ \sqrt{1+ \cos^2\rho \tan^2\frac{x}{2}}}\right)  \,.
		\label{PTRphase1}
\end{equation}
Evidently, the change of the sign, $\sin\rho \Leftrightarrow -\sin\rho$,  entails the change of phase $\phi(x) \Leftrightarrow \pi-\phi(x)$.
We predict $\cos\phi(x)=0$ and $\phi(x) = \pi/2$ at $x\to \pi$,  \textit{regardless} of the detuning angle $\rho$. The approach to $\phi(x) = \pi/2$ is singular in a sense that for $\cos^2\rho \ll 1$, it takes place in the very narrow range of $x$ in the vicinity of $x=\pi$, which is best seen from
\begin{equation}
	\cot \psi(x) = \frac{1}{\cos\rho}\cot \frac{x}{2}\, . 
\end{equation}
One readily finds that at $x=\pi$, the derivative of the phase equals $\phi'(x)=2/\cos\rho$ which is singular at $\cos\rho \to 0$, thus the phase motion degenerates into the step function. Still more singular is the case of $x=2\pi$, when 
\begin{equation}
\cos\phi(x)= \sgn(\sin x)\sgn(\sin\rho) 
\label{Jump}
\end{equation}
 and changes sign from $-1$ for $x=2\pi-0$ to $+1$ for $x=2\pi+0$, \textit{i.e.,} the envelope phase  has a phase jump by $-\pi$ irrespective of the detuning. Finally, Eq.\,(\ref{PTRphase1}) predicts the slope at $x=+0$ and $x=2\pi-0$,
\begin{equation}
\phi'(+0) = \phi'(2\pi-0)=\phi'(2\pi+0)=\frac{1}{2}|\cos\rho|\, . \label{Slope1}
\end{equation}

\subsection{Interplay of detuning and initial phase in the generic three-stage regime}
\label{subsec:two-stage}
		
In spin physics experiments on tests of fundamental symmetries such as the search for parity and time-reversal-invariance violating permanent charged particle electric dipole moments\,\cite{srEDM,YannisHybrid,AbusaifCYR}, of major interest is the signal of  spin rotations during stage III, where we make use of the radiofrequency       Wien filter     starting with in-plane polarization. In principle, alongside the measured spin precession frequency, the polarimetry of the idle spin precession during stage II gives access also to the orientation of the in-plane polarization at the activation of the Wien filter in stage III. The JEDI collaboration demonstrated the continuous retention of the corresponding phase $\Phi_\text{in}$ to an accuracy of $\SI{0.21}{\radian}$\,\cite{PhaseLock}. While the proof of principle for the pilot bunch concept consists in the mere observation that the radiofrequency       Wien filter     does not affect the pilot bunch spins, in the detailed treatment the initial spin phase $\Phi_\text{in}$ becomes another free parameter that has to be determined by fitting the experimental data. The clocks for the in-plane precession phase gain on top of $\Phi_\text{in}$ and the spin envelope phase $x$ [Eq.\,(\ref{EnvelopePhase})] begins to count when the Wien filter     is switched on. The generic solution for the vertical polarization as a function of $\Phi_\text{in}$ is given by Eq.\,(\ref{HorToVert}). 
	
In the evolution of the horizontal polarization, the dependence on $\Phi_\text{in}$ is much more subtle and deserves a dedicated analysis.
	
\subsubsection{Envelope of in-plane polarization}
\label{two-stage-envelope}
\begin{figure*}[htb]
	\includegraphics[width=0.9\textwidth]{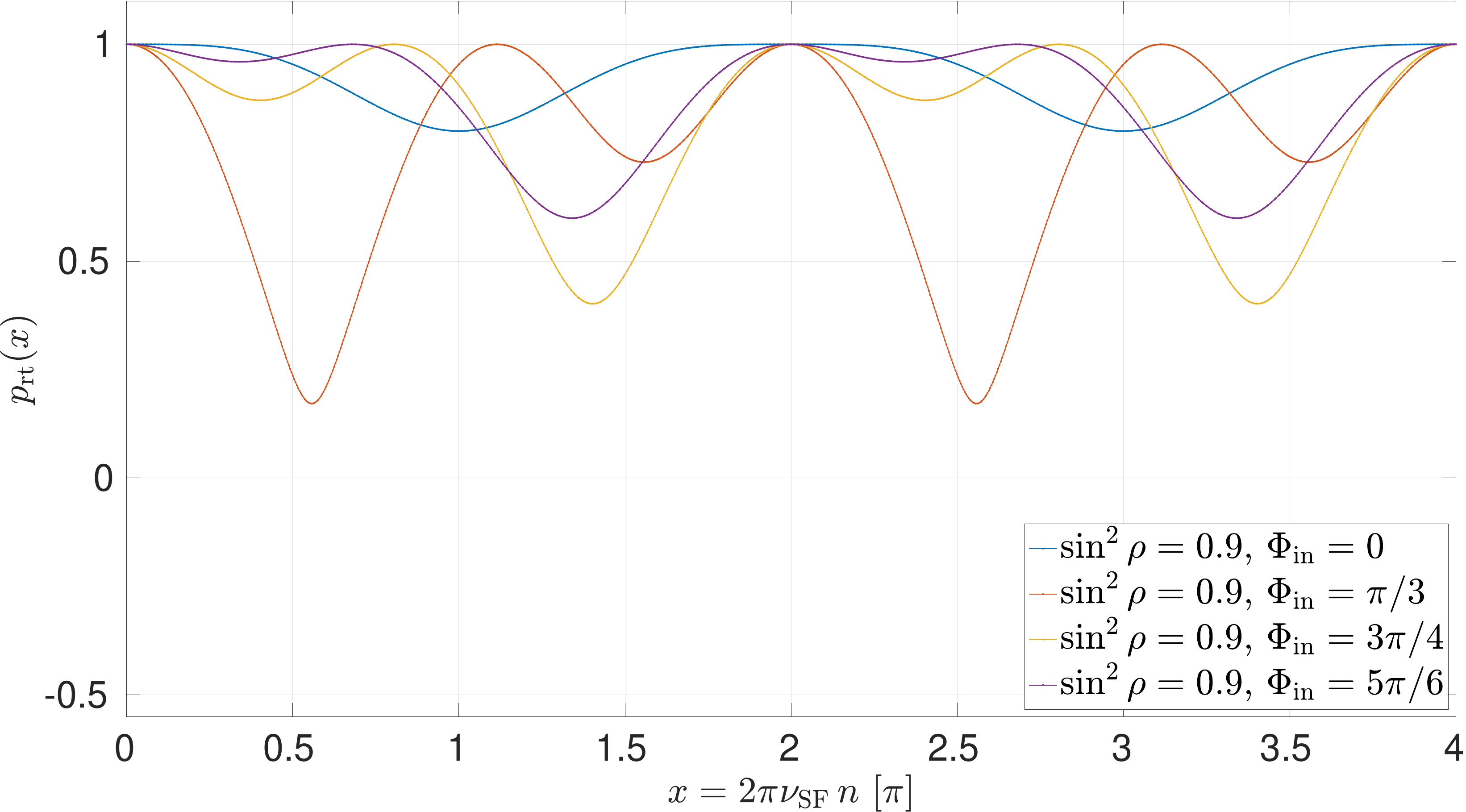}
	\caption{\label{fig:Prt-for-diff-Phiin} Pattern of the $x$-dependence of the horizontal polarization envelope $p_\text{rt}$, which evolves from the initial horizontal polarization with different initial spin precession phases $\Phi_\text{in}$. Within the interval $[0,2\pi]$, the left-right symmetry of the envelope polarization at $\Phi_\text{in}=0,\,\pi$ is broken at $0< \Phi_\text{in} <\pi$ [see Eq.\,(\ref{SrtAsym})]. However, the left-right symmetry is recovered within the symmetric period $[\zeta,2\pi+\zeta]$, see the discussion of symmetry properties of Eq.\,(\ref{HorToVert}).}
\end{figure*}

Resorting to the envelope evolution matrix $\matr{E}(x)$ of	Eq.\,(\ref{Envelopes}), we obtain 
\begin{widetext}
\begin{equation}
			\begin{split}
			p_\text{r}(x) & = E_\text{rr}(x) \cos\Phi_\text{in} + 			E_\text{rt}(x)\sin\Phi_\text{in}\\
			& =(\sin^2\rho+\cos^2\rho \cos x)\cos\Phi_\text{in}+\cos\rho \sin\Phi_\text{in}\sin x 
			 = \sin^2 \rho \cos\Phi_\text{in} + q(\Phi_\text{in},\rho) \cos\rho\cos y\,,\\
			p_\text{t}(x) & =   E_\text{rr}(x) \cos\Phi_\text{in} + E_\text{rt}(x)\sin\Phi_\text{in} =
			-\cos\rho \cos\Phi_\text{in}\sin x+\sin\Phi_\text{in}\cos x
			= -q(\Phi_\text{in},\rho)\sin y\,, \\ 
			\end{split}			
			\label{HorToHor}
\end{equation}
\end{widetext}
where $y= x-\zeta$ [see also Eq.\,(\ref{zeta})]. The predicted dependence of $p_\text{rt}(x)$ on the initial spin precession phase $\Phi_\text{in}$ is shown in Fig.\,\ref{fig:Prt-for-diff-Phiin}. It is instructive to start the discussion exactly on resonance, \textit{i.e.}, when $\cos\rho=0$ and $\sin\rho=1$. Under these conditions, we have  
\begin{equation}
	\begin{split}
		p_\text{r}(x) &=   \cos\Phi_\text{in} \, ,\\
		p_\text{t}(x) &=    \sin\Phi_\text{in} \cos x \, , \\
		p_\text{rt}(x) &= \sqrt{\cos^2\Phi_\text{in}
			+\sin^2\Phi_\text{in} \cos^2 x }\,.
	\end{split}
\end{equation}

This result nicely illustrates the emergence of the spectator radial polarization $p_\text{r}$, which is immune to the radiofrequency-driven rotations, and the active tangential polarization $p_\text{t}$, which is a partner to the vertical polarization. The distinctive appearance of a spectator polarization component is an exclusive feature of the case of vanishing detuning with $\cos\rho=0$. The envelope $p_\text{rt}(x)$  is a smooth function of $x$ with minima  at $x_4=\pi/2$ and $x_5=\pi/2 +\pi$,  and the maxima, $p_\text{rt}=1$,  at $x_3 =\pi$ and  at the end-points  $x_1=0$ and $x_2= 2\pi$. These features are evident from Fig.\,\ref{fig:ContinuousInPlaneSpin}, since $p_\text{rt} = (1-p_\text{c}^2)^{1/2}$.

We recall that the result from Eq.\,(\ref{Prt}) for the continuous operation of the Wien filter     beginning  with pure vertical polarization, shown in Fig.\,\ref{fig:ContinuousInPlaneSpin}, is symmetric with respect to the substitution  $x \Leftrightarrow  2\pi -x$. This symmetry is manifestly broken for non-vanishing values of $\Phi_\text{in}$ and $\cos\rho$ [see Eq.\,(\ref{HorToVert})], and we obtain
\begin{equation}
	\begin{split}
		 & p^2_\text{rt}(x)  -  p^2_\text{rt}(2\pi -x) = p^2_\text{c}(2\pi -x)-p^2_\text{c}(x)\\
		& = 4 \sin\Phi_\text{in} \cos\Phi_\text{in} \sin^2\rho \cos\rho \sin x\ (1-\cos x)\, . 
		\label{SrtAsym}
	\end{split}
\end{equation}
For finite $\xi(\Phi_\text{in},\rho)$, one rather has an invariance of $p_\text{rt}(x)$ with respect to the interchange $x-\zeta \Leftrightarrow 2\pi-(x-\zeta)$ within the shifted symmetric interval $[\zeta,2\pi+\zeta]$ [see the related discussion of Eq.\,(\ref{HorToVert})].

\subsubsection{Phase of in-plane polarization envelope for pure radial and longitudinal initial polarizations}
\label{two-stage-phase}
		
The motion of the phase $\phi(x)$ of the envelope $\vec{p}_\text{rt}$ is quite  sensitive to the initial phase $\Phi_\text{in}$ and the detuning angle $\rho$. It is sufficient to treat the case $\cos\rho \geq 0$, an extension of the results to $\cos\rho <0$ is straightforward. 
	
We start from Eq.\,(\ref{HorToHor}) with the pure radial initial polarization case of $\Phi_\text{in}=0$, when $p_\text{r}(x)= \sin^2 \rho + \cos^2 \rho \cos x $ and $p_\text{t}(x) = -\cos \rho \sin x $. The results are shown in Fig.\,\ref{fig:PhiPi0}. First of all, $\phi(x)$ is antisymmetric with respect to $x \Leftrightarrow 2\pi-x$. Second, for all detuning angles we find $\phi(x) = \pi/2$ at $x=0,\ \pi, \ 2\pi,..$. Third,  $\cos\phi(x_1)= -\sgn(\cos\rho \sin x_1) =\pm 1$, \textit{i.e.,} $\phi_1= 0,\ \pi$, can be reached only if $p_\text{r}(x_1)=0$, \textit{i.e.,}, at  
\begin{equation}
			\cos x_1 = - \tan^2\rho\, ,
\end{equation}
which is only possible for $\cos^2\rho \geq \sfrac{1}{2}$. 

The phase motion about the pointed tips at $|\cos\phi(x_1)|=1$ can be understood as follows. In the vicinity of $x_1$, we have  $p_r(x)= -\cos^2\rho \sin x_1 \cdot(x-x_1)$ and 
\begin{equation}
	\begin{split}
	|\cos\phi(x)| &= 1 - \frac{1}{2} [\phi(x)-\phi_1]^2\\
	&=\frac{1}{\sqrt{1+\cos^2\rho \ (x-x_1)^2}} \\
	&= 1 -\frac{1}{2}\cos^2\rho\ (x-x_1)^2\, ,
	\label{TipPhiZero}
	\end{split}
\end{equation}
which yields the slope
\begin{equation}
	\begin{split}	
		\phi(x)-\phi_1 &= \pm |\cos\rho|\ |x-x_1|\,. \label{SlopePhiZero}
	\end{split}
\end{equation}
Note that the magnitude of the slope at the tip, $|\cos\rho|$, varies from $ 1/\sqrt{2}$ to 1.
\begin{figure*}[htb]
	\includegraphics[width=0.9\textwidth]{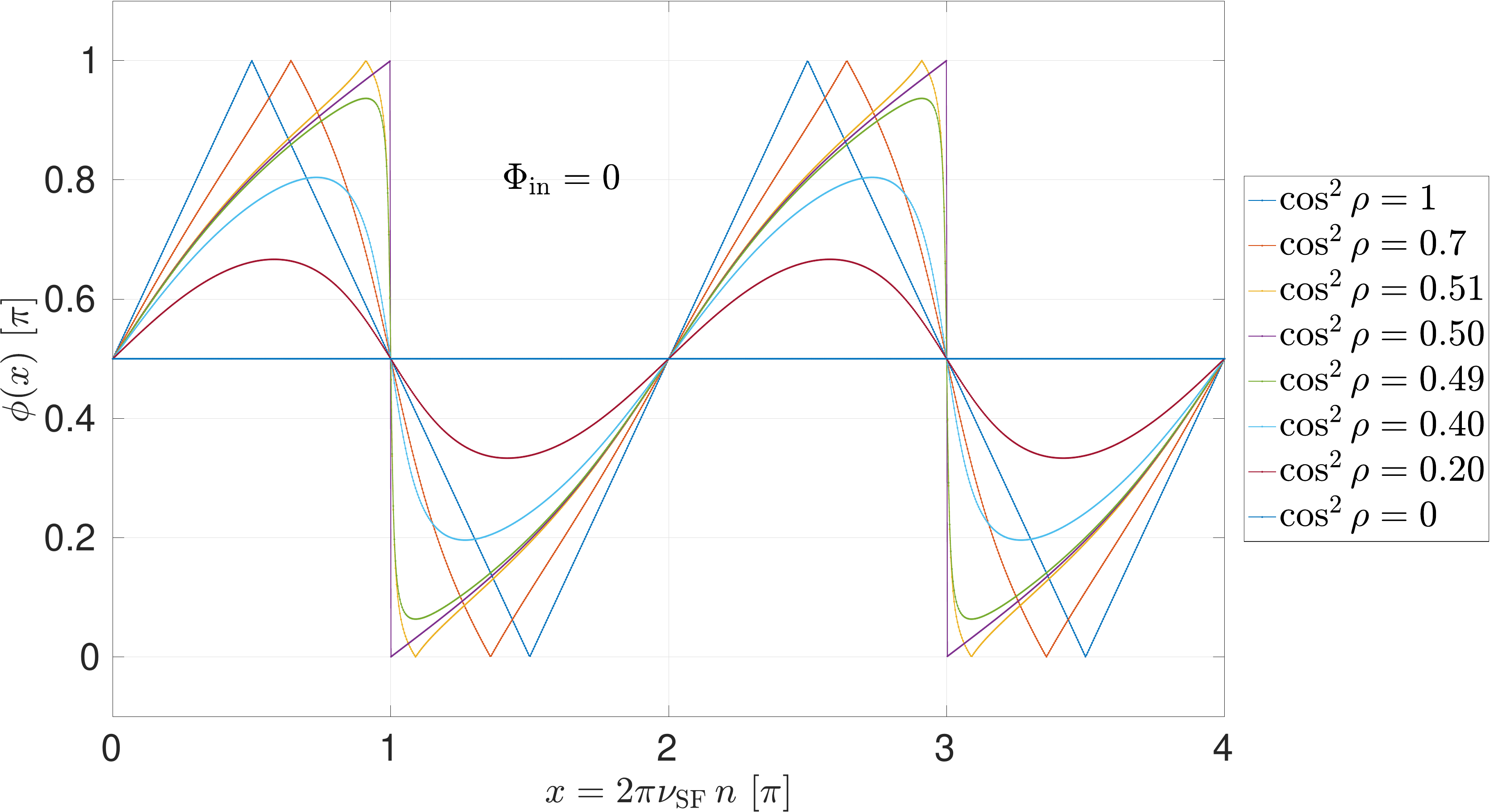}
	\caption{\label{fig:PhiPi0} Phase motion of the horizontal polarization envelope for $\Phi_\text{in}=0$ as predicted by Eq.\,(\ref{rt-phase}). The full phase swing of $\phi_\text{max}-\phi_\text{min}=\pi$ is reached only for $\cos^2\rho\geq \sfrac{1}{2}$, when $\phi(x)$ exhibits a pointed tip with the slope $\pm |\cos\rho|$. The phase motion evolves into the phase jump for the transition detuning, $\cos^2\rho \to \sfrac{1}{2}$.} 
\end{figure*}

In the opposite case of $\cos^2\rho < \sfrac{1}{2}$, the envelope phase span is less than $\pi$. Indeed, at $|\cos\rho|\ll 1$ we have 
\begin{equation}
	\begin{split} 
		\cos\phi(x) &\approx -\cos\rho \sin x \,, \quad \text{and} \\
 		\phi(x) &\approx \frac{3\pi}{2} +\cos\rho \sin x\,, 
	\end{split}
\end{equation}
with a phase span of $\phi_\text{max} - \phi_\text{min} \approx 2|\cos\rho|$. For generic  $\cos\rho < 1/2$,  the  extremal values of $\phi(x)$ come from the equation $(\cos \psi(x))' = 0$, which takes the form
\begin{equation}
		\cos x + \sin^2\rho \cos^2\rho\ (1-\cos x)^2 =0\, , 
		\label{yMaxPhi0}
\end{equation}
and yields the  root  $\cos x = -\cot^2 \rho$. The resulting phase span equals 
\begin{equation}
		 \phi_\text{max} - \phi_\text{min} = 2 \arccos |\cot \rho|\, .
\end{equation}
Finally, note how with approach to the boundary of the two regimes, $\cos^2\rho \to \sfrac{1}{2}$,  the phase motion evolves into the phase jump.

The next interesting case we would like to discuss is the pure tangential initial polarization, characterized by 
\begin{equation}
	\begin{split}
	\Phi_\text{in} & = \pi/2 \,, \\ 
	p_\text{r}(x)  & = \cos\rho \sin x \,, \\  
	p_\text{t}(x)  & = \cos x\,,  \\
	p_\text{rt}    & = \sqrt{\cos^2\rho \sin^2 x +\cos^2 x}\,,
	\end{split}
\end{equation}
so that
\begin{equation}
	\begin{split}
		\cos\phi(x) = \frac{\sgn(\cos x)}{\sqrt{1+\cos^2\rho \tan^2 x}}\,. \label{TipPi2}
	\end{split}
\end{equation}
	
The corresponding results are presented in Fig.\,\ref{fig:PhiPi2}. The phase $\phi(x)$ is symmetric with respect to $x \Leftrightarrow 2\pi-x$ and the phase swing $\phi_\text{max}- \phi_\text{min} = \pi$ for all $\rho$. It exhibits pointed tips at $x=x_1$, when $\tan^2 x_1=0$, \textit{i.e.,} when $\phi(x_1)=0$ for $x_1=0,\ 2\pi,...$ and when $\phi(x_1)=\pi$ for $x_1=\pi,\ 3\pi,...$  In the vicinity of the pointed tip at $x=x_1$,  the phase motion is given by
	\begin{equation}
	\begin{split}	
		\phi_1-\phi(x) = \pm|\cos\rho|\cdot |x-x_1|\,, \label{SlopePhiPi2}
	\end{split}
\end{equation}
yielding exactly the same slope as in Eq. (\ref{SlopePhiZero}). The only distinction to the case of $\Phi_\text{in}=0$ is that here $|\cos\rho| \leq 1/\sqrt{2}$. Note that $\phi(\pi/2) = \pi/2$, and at $|\cos \rho|\ll 1$, the phase $\phi(x)$ passes ${\pi}/{2}$ steeply in the narrow range of  $|x-\pi/2| < |\cos \rho|$. This steep variation of $\phi(x)$ about $x=\pi/2$ tends to a step function as $|\cos\rho|\to 0$.
\begin{figure*}[htb]
		\includegraphics[width=0.9\textwidth]{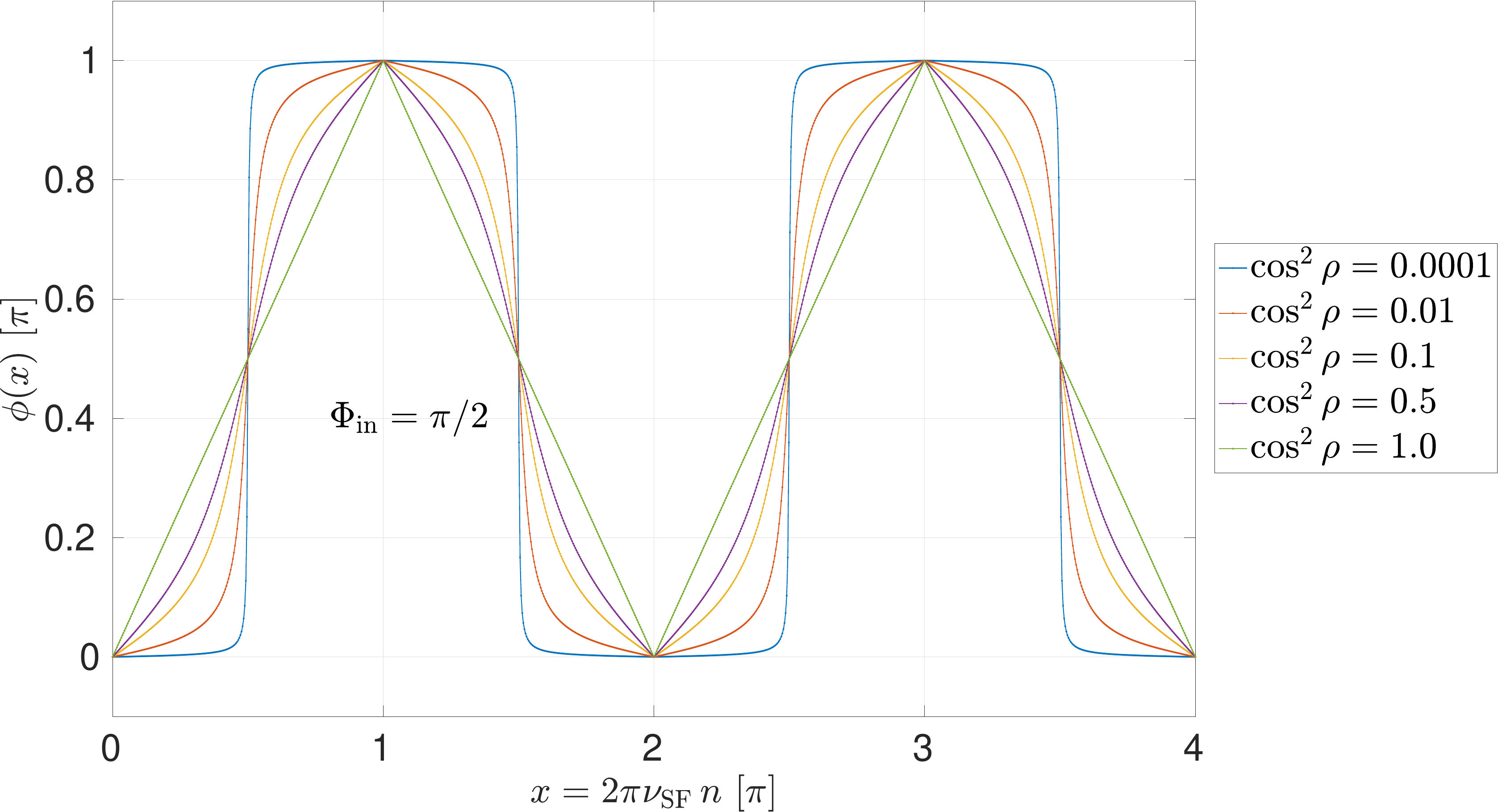}
		\caption{\label{fig:PhiPi2} Phase motion of the horizontal polarization envelope for $\Phi_\text{in}=\pi/2$ as predicted by Eq.\,(\ref{rt-phase}). In the limit of $\cos\rho \to 0$, the phase motion evolves into the phase jumps and the central bumps at $x = \pi$ and $3\pi$ exhibit a rectangular shape.}
\end{figure*}
\begin{figure*}[htb]
		\includegraphics[width=0.9\textwidth]{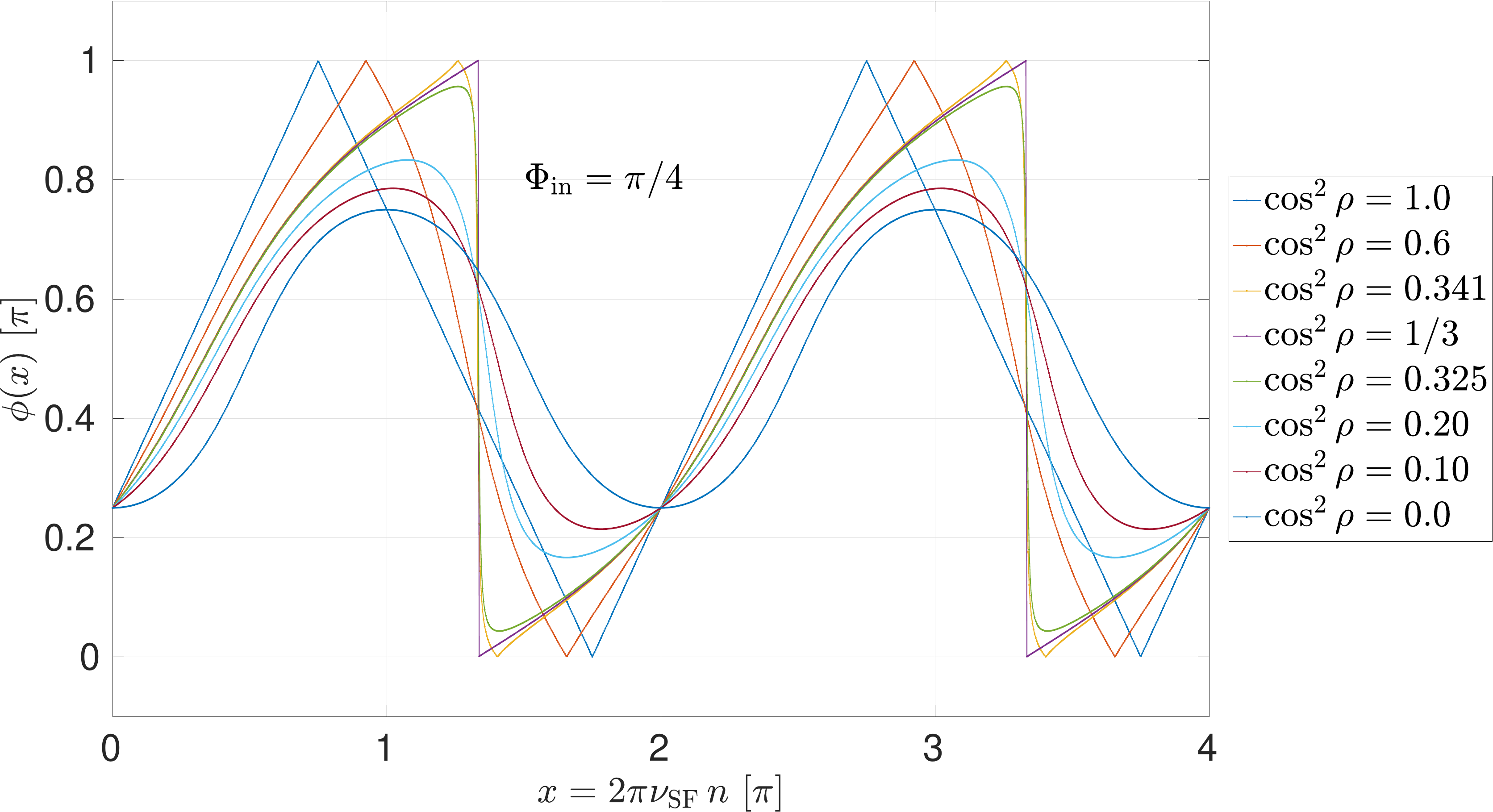}
		\caption{\label{fig:PhiPi4} Phase motion of the horizontal polarization envelope for $\Phi_\text{in}=\pi/4$, as predicted by Eq.\,(\ref{rt-phase}). For $\cos^2\rho \geq \cos^2\rho_m = \sfrac{1}{3}$ [see Eq.\,(\ref{CosRhoSmall})], the pattern of the phase motion resembles that for $\Phi_\text{in}=0$, depicted in Fig.\,\ref{fig:PhiPi0}. The phase jump for $\cos^2\rho = \cos^2\rho_m$ is located at $x=x_m = 4\pi/3$, as predicted by Eq.\,(\ref{Phi-xm}). In contrast to the case of $\Phi_\text{in}=0$ in Fig.\,\ref{fig:PhiPi0}, the phase motion for $\cos^2\rho <\cos^2\rho_m$  has no symmetry center.}
\end{figure*}

\subsubsection{Evolution of the phase of in-plane polarization envelope for generic orientation of the initial polarization}
\label{arbitrary-phi}

The analysis is based on Eqs.\,(\ref{HorToHor}) and (\ref{rt-phase}). The salient features of $\phi(x)$ for generic $\Phi_\text{in}$ are illustrated in Fig.\,\ref{fig:PhiPi4} for the example that  $\Phi_\text{in}=\pi/4$. To start with, at $x=0$ and $x=2\pi$,  Eq.\,(\ref{HorToHor}) implies that
\begin{equation}
	\phi(0)=\phi(2\pi)=\frac{\pi}{2} -\Phi_\text{in}\, , \label{phiZero} 
\end{equation}
\textit{independent} of the detuning parameter $\rho$. 
	
The subsequent analytic discussion is most conveniently  performed in terms of the variables $y= x-\zeta(\Phi_\text{in},\rho)$ and $q(\Phi_\text{in},\rho)$ [see Eqs.\,(\ref{zeta}) and\,(\ref{HorToHor})]. A major finding is that the same universal slope at the tip, $\pm |\cos\rho|$, persists for all $\Phi_\text{in}$. Indeed, according to Eq.\,(\ref{HorToHor}), we have $p_\text{r}(x)=0$ at
\begin{equation}
	\begin{split}
	\cos y_1 = -\frac{\sin^2 \rho \cos\Phi_\text{in}}{q(\Phi_\text{in},\rho) \cos\rho}\,. \label{cosy1}
	\end{split}
\end{equation}
This solution is only possible if
\begin{equation}
	\cos^2\rho \geq \cos^2\rho_\text{m}=\frac{\cos^2\Phi_\text{in}}{1+\cos^2\Phi_\text{in} }\, , \label{CosRhoLarge}
\end{equation}
where $\rho_m$ denotes the boundary  detuning angle for which the solution (\ref{cosy1}) does still exist.

In close similarity to the case $\Phi_\text{in}=0$, shown in Fig.\,\ref{fig:PhiPi0}, the phase $\phi(x)$ exhibits pointed tips $x_1 = y_1+\zeta$.  In the vicinity of the tips  we have 
\begin{equation}
	\begin{split}
	p_\text{r}(x)& =- q(\Phi_\text{in},\rho) \cos\rho \sin y_1 \cdot(x-x_1)\\
	& = p_\text{t}(x_1)\cos\rho \sin y_1 \cdot(x-x_1)\,,
	\end{split}
\end{equation}
which entails
\begin{equation}
	\begin{split}
	\cos\phi(x) = \frac{\sgn(\cos x)}{\sqrt{1+\cos^2\rho \ (x-x_1)^2}}\,,
	\label{SlopePhiGen}
	\end{split}
\end{equation}	
and we recovered Eq.\,(\ref{TipPhiZero}) and the familiar slope $ \pm \cos\rho$ at the pointed tips.

In the evaluation of the phase span at 
\begin{equation}
	\cos^2\rho \leq \cos^2\rho_\text{m}=\frac{\cos^2\Phi_\text{in}}{1+\cos^2\Phi_\text{in}}\,, \label{CosRhoSmall}
\end{equation} 
we follow the procedure developed for the case of $\Phi_\text{in} =0$. The phase extrema are roots of the equation $(\cos \phi(x_m))' =0$, which takes the form [here below $q=q(\Phi_\text{in},\rho)$]
\begin{equation}
	\cos^2 y +2w \cos y+1=0\,, \label{PhiCosy}
\end{equation}
with the roots
\begin{equation}
	\cos y_\pm = w \pm \sqrt{w^2-1}\,, 
	\label{PhiRoots}
\end{equation}
where 
\begin{equation}
	w = \frac{\sin^2\rho (q^2+\cos^2\rho\cos^2\Phi_\text{in})-1}{2 q \sin^2\rho \cos\rho\cos\Phi_\text{in}}\,.
\end{equation}
The solutions exist for $w^2 \geq 1$. It is easy to check that the boundary case,  $w=1$, corresponds to the exact equality in the condition (\ref{CosRhoSmall}). Subject to the constraint $|\cos y_\pm|\leq 1$, the admissible roots  are $\cos y_-$ at $w\geq 1 $, and $\cos y_+$ at $w\leq -1$, and the two branches are related by
\begin{equation}
	\cos y_-(w) = -\cos y_+(-w)\, . \label{TwoBranches}
\end{equation}
The limit of $w^2 \gg 1$ corresponds to $|\cos\rho|\ll |\tan \Phi_\text{in}|$, when 
\begin{equation}
	\begin{split}
	q^2 		& \to \sin^2 \Phi_\text{in}\,,\\
	\cos y_\pm 	& \to \frac{1}{2w} = -\frac{|\sin\Phi_\text{in}|}{\cos\Phi_\text{in}}\cos\rho\,,\\
	\zeta 		& \to \frac{\pi}{2}\sgn(\sin \Phi_\text{in})\,.
	\end{split}
\end{equation} 
\begin{figure*}[htb]
		\includegraphics[width=0.9\textwidth]{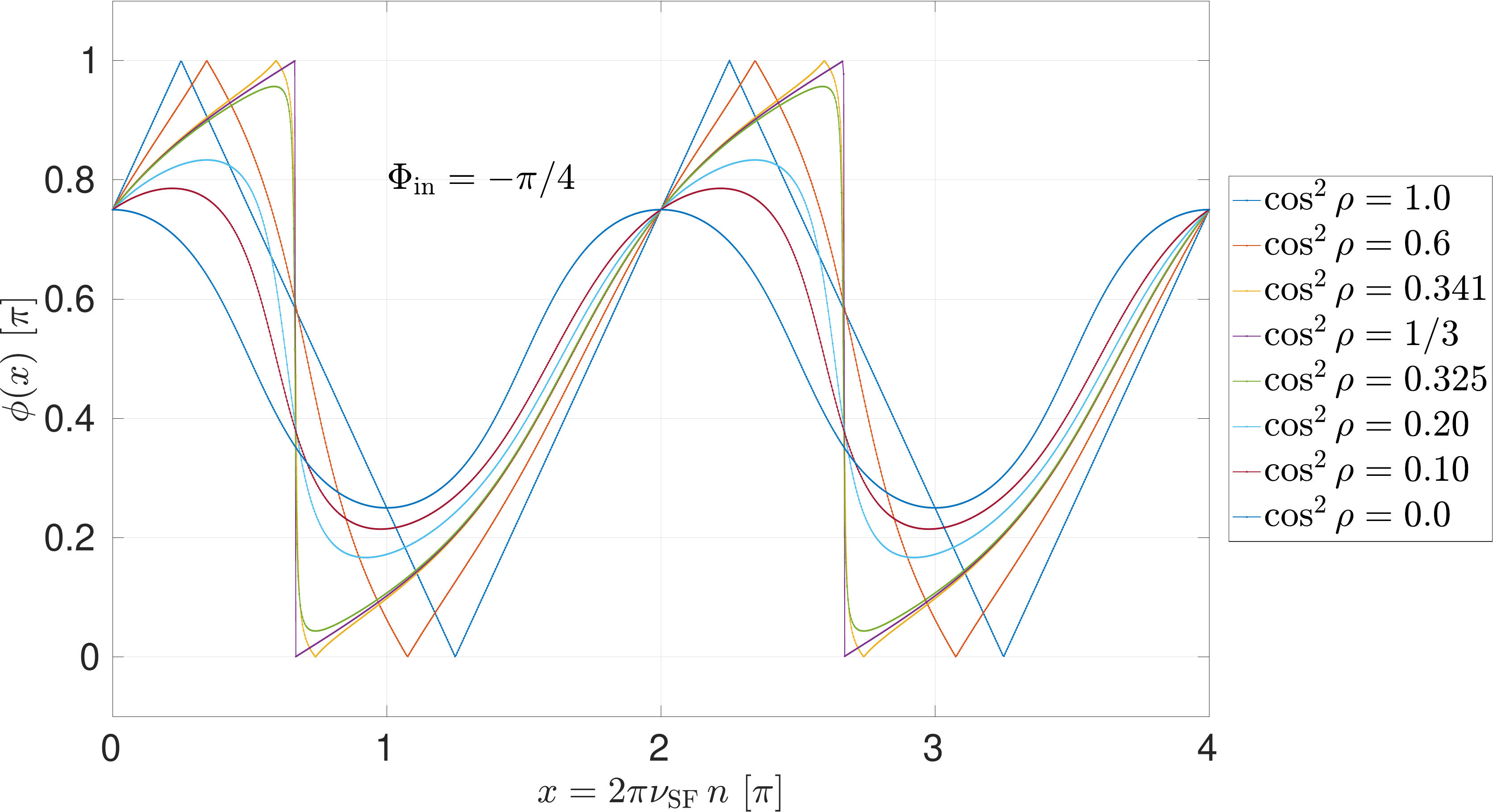}
		\caption{\label{fig:Phi-Pi4} Phase motion for $\Phi_\text{in}=-\pi/4$ as predicted by Eq.\,(\ref{rt-phase}). For $\cos^2\rho \geq \cos^2\rho_m = \sfrac{1}{3}$ [see Eq.\,(\ref{CosRhoSmall})] the pattern of the phase motion resembles that for $\Phi_\text{in}=0$ depicted in Fig.\,\ref{fig:PhiPi0}. The phase jump for $\cos^2\rho = \cos^2\rho_m$ is located at $x=x_m=\sfrac{4\pi}{3}$ as predicted by Eq.\,(\ref{Phi-xm}). In contrast to the case of $\Phi_\text{in}=0$ in Fig.\,\ref{fig:PhiPi0}, the phase motion for $\cos^2\rho <\cos^2\rho_m$  has no symmetry center.}
\end{figure*}

Now we focus on the boundary case $\cos\rho =\cos\rho_\text{m}>0$. According to Eq.\,(\ref{HorToHor}), $p_\text{t}(x)$ changes the sign at $y=\pi$, and we encounter the by now familiar phase jump  depicted in Fig.\,(\ref{fig:PhiPi0}). Upon some algebra, we find
\begin{equation}
	\cos\zeta(\Phi_\text{in},\rho_m) =  \cos^2\Phi_\text{in}\,, \label{cos-zeta}
\end{equation}
which in our  case $\Phi_\text{in}= \pi/4$ entails $\zeta(\Phi_\text{in},\rho_m) = \pi/3$, and we predict
\begin{equation}
	x_m = \pi + \arccos \left( \cos^2(\Phi_\text{in}) \right)= \frac{4}{3}\pi \,, \label{Phi-xm}
\end{equation}
in perfect agreement with the numerical results shown in Fig.\,\ref{fig:PhiPi4}. 

As we observed in Sec.\,\ref{subsec:two-stage}, a finite initial phase $\Phi_\text{in}$ introduces an asymmetry with respect to $x\Leftrightarrow 2\pi -x$. The symmetry is restored in the exceptional case of $\cos\rho=0$ [see Eq.\,(\ref{SrtAsym})], when we predict $\phi(x = \pi)= 3\pi/4$ in agreement with the numerical results shown in Fig.\,\ref{fig:PhiPi4}. 

Finally, we consider the case of $\Phi_\text{in}= -\pi/4$. The corresponding phase motion is shown in Fig.\,\ref{fig:Phi-Pi4}. First, according to Eq.\,(\ref{phiZero}), we get 
\begin{equation}
	\phi(0)=\phi(2\pi)=\frac{\pi}{2} -\Phi_\text{in} = \frac{3}{4}\pi 
\end{equation}
Second, according to Eq.\,(\ref{zeta}), now we must take a branch $\zeta =-\arccos \left( \cos^2(\Phi_\text{in})\right)$. As far as the $x$-dependence of the phase $\phi(x)$ is concerned, a chain of substitutions   
\begin{equation}
	\begin{split}
		y=x-\zeta|_{\pi/2} &\Rightarrow x-\zeta|_{-\pi/2} = x+\zeta|_{\pi/2} \\ 
		&\Rightarrow \tilde{y} = - [(-x)- \zeta|_{\pi/2}]\,,
	\end{split}
\end{equation} 
amounts to the inversion of the $x$-axis accompanied by the shift by $2\pi$, and simultaneous phase inversion $\phi(x) \Rightarrow \pi -\phi(x)$.

We found a very rich pattern of the in-plane-envelope phase motion depending on the detuning and the initial spin phase. Still, there are certain universal features of the graphs shown in Figs.\,\ref{fig:PhiPi0}, \ref{fig:PhiPi2}, \ref{fig:PhiPi4} and \ref{fig:Phi-Pi4} which are worth of emphasis. Irrespective of $\Phi_\text{in}$, in all graphs the envelope phase exhibits the phase jump by $\pi$ with the known $\Phi_\text{in}$-dependence of the location of the jump. The same is true for the continuous spin rotation [see Fig.\,\ref{fig:Phasemotion}], although this case has certain exceptional features to be discussed below. For non-vanishing detuning, $\phi(x)$ exhibits pointed tips with a universal slope equal to $\pm \cos\rho$ at the tip, irrespective of the initial spin phase, while in Fig.\,\ref{fig:Phasemotion}, the rated slope equals $|\cos\rho|/2$. Finally, the phase continuity condition $\phi(x=0)=\phi(x=2\pi)$ holds for all $\Phi_\text{in}$ with the detuning-independent $\phi(0)$, again with the exception of Fig.\,\ref{fig:Phasemotion}. Regarding the pointed tips, according to Eq.\,(\ref{CosRhoLarge}), they persist for a finite range of detunings, apart from the exceptional cases $\Phi_\text{in}= \pm \pi/2$, when the tips for all $\rho$ share identical locations at $x=0,\pi, 2\pi,..$.   
 
The WF-driven \textit{continuous} evolution from the pure vertical initial polarization is distinct from the generic three-stage evolution used in actual JEDI experiments. As explained in Sec.\,\ref{sec:continuous-WF}, here Wien filter     operates in the capacity of the spin rotator in stage I and continuous on to stage III at one and the same detuning angle $\rho$.	
	 Specifically, the rotation of the polarization into the horizontal plane happens at $\cos x_0(\rho)= -\cot^2\rho$ [see  Eq.\,(\ref{ZeroVertical})]. In the spirit of generic three-stage process, this instant can be viewed as a start of stage III with the initial phase $\Phi_\text{in}$ defined by
\begin{equation}
	\begin{split}
		 \cos\Phi_\text{in}&= p_\text{r}(x_0)=\cot\rho\, ,\\
		 \sin\Phi_\text{in}&=p_\text{t}(x_0)=\sgn(\sin \rho)\sqrt{1-\cot^2\rho}\,.    \label{phi-cintinuous}
	\end{split}
\end{equation} 
Our convention for stage III is that the envelope evolution phase starts with $x=0$. Evidently, the further evolution of $p_\text{r,t}(x)$ will be still  described by Eq.\,(\ref{VertToPlane}) subject to the trivial substitution $x\to x+x_0(\rho)$.
This way in Fig.\,\ref{fig:Phasemotion} we lumped together the detuning dependence of $\phi(x)$ for a  very special subset of initial phases $\Phi_\text{in}(\rho)$ as opposed to the $\rho$-independent initial phase in other cases. This distinctive feature of continuous evolution is behind the $\rho$-independent phase jump at $0,$ $2\pi,$ $4\pi,\ldots$ , and the degeneracy of the tip and jump locations, and  a phase slope at the tip, $\frac{1}{2}\cos\rho$, which is half of that in the generic case.

The above analysis suggests that the phase of the envelope of the horizontal polarization has a great potential for the diagnostics of the RF-driven spin dynamics (see also early considerations in Ref.\,\cite{JEDIphase}). We demonstrated a remarkably strong sensitivity of the phase motion to the initial phase of the horizontal spins and to the detuning of the spin precession frequency. This phase remained the as yet unexplored feature of the RF-driven spin dynamics in storage rings and we make a point that variations of the dependence of this phase with respect to time may prove as a good indicator of the stability of the detuning during the cycle, or as an indicator for the lack or presence of unwanted phase walks.

\section{Spin decoherence incorporated}
\label{sec:decoherence}
\subsection{Decoherence through feedback to compensate for spin precession walk}
\label{sec:feedback}

As mentioned in Sec.\,\ref{sec:introduction}, the observed idle spin precession phase walk during the feedback (fb) time interval $t_\text{fb} = (5-10)\,\si{s}$ on the scale of $\sigma_\text{fb} \approx 0.2\,\text{rad}$ corresponds to a detuning of the spin precession on the rms scale of  $\Delta f^{(\text{fb})}_\text{s} =\cos\rho_\text{fb} f_\text{SF}\approx \SI{5}{mHz}$, where the perturbative  parameter in the problem is
\begin{equation}
\cos \rho_\text{fb} = \frac{\sigma_\text{fb}}{2\pi f_\text{SF}t_\text{fb}} \, .	
\label{eq:cosrhofb}	
\end{equation} 
When the ring instabilities are slow on the time scale $t_\text{fb}$, the smooth spin phase walk can be approximated by constant detuning. Then the spin envelope evolution can be approximated by Eq.\,(\ref{Envelopes}) with the spin-flip tune of Eq.\,(\ref{eq:DetunedEnvelopeTune}): 
\begin{equation}
\begin{split}
\nu_\text{SF}&=\nu_\text{SF}^0 \left(1+ \frac{1}{2}\cos^2\rho_\text{fb}\right)\, .
\label{EnchancedSpinFlip}
\end{split} 
\end{equation}
To set the ballpark, for $\sigma_\text{fb} = 0.2$, $t_\text{fb} = \SI{10}{s}$ and $f_\text{SF} = \SI {80}{mHz}$ as in the pilot bunch experiment\,\cite{Slim:2023lpd}, we obtain $\cos^2 \rho_\text{fb} = 0.0016$, but this parameter becomes as large as 0.1 for $f_\text{SF} = \SI{10}{mHz}$.

Qualitatively, the feedback follows the windshield-wiper pattern, which can be cast into a toy model of consecutive spin envelope rotations,
\begin{equation}
\matr{E}^{\text{(fb)}}(2x_\text{fb}) = \matr{E}(-\cos\rho_\text{fb},x_\text{fb}) \matr{E}(\cos\rho_\text{fb},x_\text{fb})\, 
\end{equation}
where $x_\text{fb}$  is the SF phase acquired per feedback period $t_\text{fb}$, and we show explicitly the dependence on $\cos\rho$ in the SF matrix of Eq.\,(\ref{Envelopes}).  Here, the  first envelope transfer matrix $\matr{E}(\cos\rho_\text{fb},x_\text{fb})$ parameterizes the experimentally measured spin phase walk in terms of the detuning $\Delta f^{(\text{fb})}_\text{s} =\cos\rho_\text{fb} f_\text{SF}$. In order to compensate the acquired relative phase walk during text next period $t_\text{fb}$, the Wien filter     is operated at a frequency corrected by $2\Delta f^{(\text{fb})}_\text{s}$, {\it i.e.}, with the  flipped sign of the detuning, which is modeled by $\matr{E}(-\cos\rho_\text{fb},x_\text{fb})$.  In the limit of vanishing spin walk $
\matr{E}^{\text{(fb)}}(2x_\text{fb}) = \matr{E}(0,2x_\text{fb})$, and we define the feedback matrix   
\begin{equation}
\matr{R}_{\text{fb}} = \matr{E}^\text{(fb)}(2x_\text{fb}) \matr{E}^{-1}(0,x_\text{fb}) \,. 
\end{equation}
The corresponding feedback-corrected envelope evolution matrix takes the familiar stroboscopic form
\begin{equation}
\vec{p}(2(k+1)x_\text{fb}) = \matr{R}_\text{fb}\matr{E}(0,x_\text{fb})\vec{p}(2k x_\text{fb}) \,.
\end{equation}
Note that in our toy model, this matrix $\matr{R}_\text{fb}$ is time independent. We skip  the lengthy derivation of $\matr{E}^{\text{(fb)}}(2x_\text{fb})$ and the corresponding BK averaging and give the behavior of the resulting SF matrix for large $k$,
\begin{widetext}
	\begin{equation}
	\begin{split}
	\matr{E}^{(\text{fb})}(x)&=\begin{pmatrix}
	\exp(-2\Gamma_\text{fb} x) & 0 & 0 \\
	0 & \exp(-\Gamma_\text{fb} x) \cos x & -\exp(-\Gamma_\text{fb} x)\sin x \\
	0 & \exp(-\Gamma_\text{fb} x)\sin x & \exp(-\Gamma_\text{fb} x)\cos x 
	\end{pmatrix} \, ,
	\label{EnvelopesF}
	\end{split}
	\end{equation}
\end{widetext}
which supports the spectator radial polarization. 

The spin precession walk depolarizes the vertical polarization with the lifetime $\tau_\text{fb}$ given by
\begin{equation}
\frac{1}{\tau_\text{fb}} = 2\pi \Gamma_\text{fb}  f_\text{SF} = \frac{\cos \rho_\text{fb}^2  (1- \cos x_\text{fb} )^2}{t_\text{fb}}\,,
\label{eq:lifetime-taufb}
\end{equation}
while the spectator radial in-plane polarization depolarizes twice faster.  The spin decoherence time for the active in-plane polarization, $\tau_\text{SCT}$,  is equal to $\tau_\text{fb}$. Indeed, the detuning of the spin precession does not lead to a depolarization of the vertically oriented spins (see the related discussion below in Sec.\,\ref{ExponAnsatz}).  The spin-flip tune acquires two corrections: 
\begin{equation}
\nu_\text{SF} = \nu_\text{SF}^0  \left[1 + \frac{1}{2} \cos \rho_\text{fb}^2 + \frac{\sin x_\text{fb}}{2 x_\text{fb}} \cos^2 \rho_\text{fb} (2-\cos \rho_\text{fb} )\right]
\end{equation}
The first correction stems from Eq.\,(\ref{EnchancedSpinFlip}), while the second one derives from spin-flip rotations during the feedback periods. The corresponding SF phase is given by  $x = 2\pi\nu_\text{SF}n$. The above toy-model corrections to the spin tune, as well as the rate of depolarization, must be regarded as gross estimations. Nevertheless, they are a good example of how the feedback to maintain phase locking between the spin precession and Wien-filter     phases has a non-vanishing influence on the spin-flip dynamics. For instance, if taken at face value, for the conditions of the pilot bunch experiment and the above-given feedback parameters ($\sigma_\text{fb} = 0.2$, $t_\text{fb} = \SI{10}{s}$, $f_\text{SF} = \SI {80}{mHz}$), Eq.\,(\ref{eq:lifetime-taufb}) predicts $\tau_\text{fb} \approx \SI{e4}{s}$, while at $x_\text{fb} < 1$, it predicts
\begin{equation}
  \tau_\text{fb} = \frac{t_\text{fb}}{{\sigma_\text{fb}}^2} \approx \SI{600}{s}\,.
\end{equation}

\subsection{Recovering the spectator polarization}
\label{uncovering-spectator}
		
As a prelude to further discussion of the spin decoherence effects, we observe that the envelope evolution matrix in Eq.\,(\ref{Envelopes}) can be cast in the form
\begin{widetext}
\begin{equation}
	\begin{split}
			\matr{E}(x)&=\begin{pmatrix}
			\sin^2 \rho+\cos^2 \rho \cos x  &  \cos \rho \sin\rho (1-\cos x ) & \cos \rho
			\sin x \\
			\cos \rho \sin\rho (1-\cos x ) & \cos^2 \rho+\sin^2 \rho \cos x    & -\sin \rho
			\sin x  \\
			-\cos \rho \sin x  &  \sin \rho \sin x   & \cos x 
			\end{pmatrix} \\
			& =\begin{pmatrix}
			\sin\rho  & -\cos \rho & 0 \\
			\cos\rho  & \sin \rho & 0 \\
			0 & 0  & 1 
			\end{pmatrix} 
			\cdot 
			\begin{pmatrix}
			1 & 0 & 0 \\
			0 &  \cos x & -\sin x \\
			0 & \sin x & \cos x 
			\end{pmatrix} 
			\cdot
			\begin{pmatrix}
			\sin\rho  & \cos \rho & 0 \\
			-\cos\rho  & \sin \rho & 0 \\
			0 & 0  & 1 
			\end{pmatrix}\, ,
	\label{Envelopes2}
	\end{split}
\end{equation}
\end{widetext}
which amounts to the rotation of coordinates such that the vector $\vec{m}$ of 	Eq.\,(\ref{NewAxis}) plays now the role of $\vec{c}$ in the case of idle precessions. In this new reference frame, the matrix in Eq.\,(\ref{Envelopes}) stems from the initial block-diagonal matrix $\matr{E}_0(x)$ of Eq.\,(\ref{EnvelopeEvolution1}), which features the spectator polarization. This observation serves as crucial guidance to link spin evolution to decoherence effects.
		
As a matter of fact, the presence of the hidden spectator component could have	been directly guessed from the original envelope rotation matrix of Eq.\,(\ref{Envelopes}). Indeed, besides the manifestly RF-driven terms $\propto \sin x $ and $\propto \cos x $, the four matrix elements of $\matr{E}(x)$ do contain the non-rotating components: $\sin^2 \rho$ in $E_\text{rr}(x)$, $\cos^2 \rho$ in $E_\text{cc}(x)$, and $\cos\rho\sin\rho$ in $E_\text{rc}(x)$  and $E_\text{cr}(x)$. 
		
\subsection{Ansatz of exponential decoherence of the in-plane polarization} 
\label{ExponAnsatz}	

\subsubsection{Damped spin rotations}
		
The  JEDI studies of spin decoherence have revealed\,\cite{SCT1000sJEDI} an enhancement of the spin-coherence time to the fine tuning of families of sextupole magnets to zero chromaticity to reduce the spread of spin tunes in the beam caused by orbit lengthening due to betatron oscillations\,\cite{KoopShatunov}. In the spirit of the Bloch approach\,\cite{bloch1946nuclear}, we present here the ad hoc treatment of the residual spin decoherence in terms of the exponential attenuation of the in-plane polarization and preservation of the vertical polarization in the idle precession regime.

Correspondingly, the master equation (\ref{MasterEq}) will be modified to yield 
\begin{equation}
\vec{S}(n)=    \matr{R}_\text{WF}(n)  
\matr{R}_\Gamma
\matr{R}_\text{c}(\theta_\text{WF})\vec{S}(n-1) \,,
\label{MasterEq2}
\end{equation}
where 
\begin{equation}
 \matr{R}_{\Gamma} =\begin{pmatrix}
1-\Gamma  & 0 & 0 \\
0  & 1 & 0 \\
0 & 0  & 1-\Gamma 
\end{pmatrix}  =\matr{1} + \matr{W}_{\Gamma}
\label{Damping}
\end{equation}
describes the attenuation per turn, where in terms of the spin coherence time $\tau_\text{SCT}$, $\Gamma$ is given by
\begin{equation}
\Gamma = \frac{1}{f_\text{c} \tau_\text{SCT}}\,.
\label{eq:Gamma-damping}
\end{equation} 
We shall also use the small decoherence parameter, 
\begin{equation}
Q=\frac{\Gamma}{4\pi \nu_\text{SF}}\,,  \label{Q}
\end{equation} 
which is defined such that $\Gamma n = 2Qx$. 

\subsubsection{Sequential Bogoliubov-Krylov averaging}

Anticipating the sequential BK averaging, we seek for a solution of the master equation (\ref{MasterEq2}) of the form
\begin{equation}
\vec{S}(n)=    \matr{R}_\text{c}(n\theta_\text{WF})\matr{E}_0(n) \vec{g}(n-1)  \,,
\label{Solution2}
\end{equation}
so that $\vec{g}(n)$ embodies the impact of the spin decoherence on the earlier defined spin envelope: $\vec{p}(n) =  \matr{E}_0(n) \vec{g}(n)$. Then, the master equation for $\vec{g}(n)$ reads
\begin{widetext}
\begin{equation}
\begin{split}
\vec{g}(n)=  \matr{E}_0^{-1}(n) \matr{R}^{-1}_\text{c}(n\theta_\text{WF})\matr{R}_\text{WF}(n)\matr{R}_\text{c}(n\theta_\text{WF})  \matr{R}_{\Gamma}  \matr{E}_0 (n-1)    \vec{g}(n-1)\,.
	\label{MasterEqGamma}	
	\end{split}	
\end{equation}	
The first stage of the BK averaging over spin precession yields
\begin{equation}
\left\langle \matr{R}^{-1}_\text{c}(n\theta_\text{WF})\matr{R}_\text{WF}(n) \matr{R}_\text{c}(n\theta_\text{WF})\right \rangle =\matr{E}_0(1) \,. \label{BKstage1}
\end{equation} 
Next we perform the BK averaging over spin flips which are fast compared to the spin damping,
\begin{equation}
\begin{split}
\matr{U}_\Gamma	 =  \left\langle \matr{E}_0^{-1}(n-1)  \matr{W}_\Gamma \matr{E}_0 (n-1)\right\rangle 
 = \Gamma \left\langle \begin{pmatrix}
1 & 0 & 0 \\
0 &  \sin^2 x & 0 \\
0 & 0 & \cos^2 x 
\end{pmatrix}\right\rangle =   
-\Gamma \begin{pmatrix}
1 & 0 & 0 \\
0 &  \frac{1}{2} & 0 \\
0 & 0 & \frac{1}{2}
\end{pmatrix} \, . \\
\label{BKstage2}
\end{split}
\end{equation}
\end{widetext}
The  corresponding solution of Eq.\,(\ref{MasterEqGamma}) is given by
\begin{equation}
\vec{g}(n)  = \matr{E}_\Gamma (n)\vec{p}(0) =  \exp(\matr{U}_\Gamma n) \vec{p}(0)\, ,
\end{equation}	
with 
\begin{equation}
\begin{split}
\matr{E}_\Gamma (x)=\begin{pmatrix}
\exp(-2Qx) & 0 & 0 \\
0 & \exp(-Qx) & 0 \\
0 & 0 & \exp(-Qx) 
\end{pmatrix}\, . \label{BKdamping}	
\end{split}
\end{equation}

While the idly precessing spectator component decoheres  $\propto \exp(-2Qx)$, the vertical  and the in-plane active polarizations decohere at half this rate,  $\propto \exp(-Qx)$. Indeed, the polarization decoheres when it is in the $rt$-plane, while the attenuation of the upward or downward polarization is negligibly weak on the time scale of $\tau_\text{SCT}$\,\cite{MachineDevelopmentSF2}, see the related discussion of Eq. (\ref{EnvelopesF}) in Sec.\,\ref{sec:feedback}. The corresponding damped envelope evolution reads $\vec{p}(x)=\matr{E}_\text{D}(x)\vec{p}(0)$ with the SF matrix\\
\begin{widetext}
\begin{equation}
\begin{split}
\matr{E}_\text{D}(x) = \matr{E}_0(x)\matr{E}_\Gamma(x)	 
		            & = \begin{pmatrix}
			\exp(-2Qx) & 0 & 0 \\
			0 &  \exp(-Qx) \cos x & - \exp(-Qx)\sin x \\
			0 &  \exp(-Qx)\sin x &  \exp(-Qx)\cos x 
			\end{pmatrix}\, ,
\label{Eexpon}
\end{split}
\end{equation}
which replaces $\matr{E}_0(x)$ in Eq.\,(\ref{Envelopes2}) with the result
\begin{equation}
			\begin{split}
			\matr{E}_\text{exp}(x)
			=\begin{pmatrix}
			e^{-2Qx}\sin^2 \rho+e^{-Qx}\cos^2 \rho \cos x  &  \cos \rho \sin\rho
			(e^{-2Qx}-e^{-Qx}\cos x) & e^{-Qx}\cos \rho \sin x \\
			-\cos \rho \sin\rho (e^{-2Qx}-e^{-Qx}\cos x) & e^{-2Qx}\cos^2 \rho+e^{-Qx}\sin^2 \rho \cos x
		   & -e^{-Qx}\sin \rho \sin x \\
			-e^{-Qx}\cos \rho \sin x &  e^{-Qx}\sin \rho \sin x & e^{-Qx}\cos x
			\end{pmatrix} \, .
			\label{DampedEnvelopes1}
			\end{split}
\end{equation}
\end{widetext}
In this purely phenomenological approach, the attenuation does not affect the SF tune\,\footnote{Recall the classic example of the frequency shift of the damped harmonic oscillator: $z(t)= A\exp(-\gamma t)\sin(\sqrt{\omega_0^2-\gamma^2}t +\phi_0)$.}. A treatment within this exponential decoherence model of the experimental results of the pilot bunch experiment is reported in ref.\,\cite{Slim:2023lpd}. 

\subsection{Spin decoherence by synchrotron motion}

\subsubsection{Spread of synchrotron oscillation amplitudes}

So far, we considered only central particles in the bunch. The synchrotron oscillations (SO) with frequency $f_\text{sy}$ modulate the particle momentum and the spin tune, and are endemic in storage rings. The emerging oscillating detuning between Wien filter     and spin  precession is a well defined dynamical mechanism of spin decoherence, and here we treat it as the leading one,  supposing that the betatron oscillation effects have been taken care of by fine tuning of the sextupole families. We follow the technique of an earlier study\,\cite{NonExpon} and extend these considerations. 

The oscillations of the particles around the center of the bunch can be evaluated using the time distribution of the events recorded in the internal polarimeter. Following Ref.\,\cite{Slim:2023lpd}, it is convenient to represent the longitudinal profile of the bunch in terms of a fractional cyclotron phase $\phi = \phi_\text{c} - 2\pi n$ such that $\phi \in [0,2\pi]$. In the further discussion, the synchrotron motion for an individual particle is defined with respect to a center of the bunch, $\phi= a \cos(2\pi \nu_\text{sy} f_\text{c}t +\lambda) $, where $\nu_\text{sy} =f_\text{sy}/f_\text{c}$ is the synchrotron tune and $\lambda \in [0.2\pi]$ is the individual particle's random phase. 

The one-particle contribution to the longitudinal density of the bunch $N(\phi)$ is inversely proportional to the SO velocity, and the one-particle density of the bunch 
\begin{equation}
	N(\phi) = \frac{1}{\pi} \int_\phi^\infty \frac{\dd a F(a)}{\sqrt{a^2 -\phi^2} }\, ,
	\label{eq:Nz}
\end{equation}
Clearly, for large-$\phi$ the bunch density receives contributions only from particles with synchrotron amplitudes $a > \phi$.
Now we observe that Eq.\,(\ref{eq:Nz}) assumes the form of the Abel transform with the solution for the synchrotron amplitude distribution
\begin{equation}
	F(a)= -2a \int_{a}^\infty  \frac{\dd\phi N'(\phi)}{\sqrt{\phi^2-a^2}} \, . 
	\label{eq:Fz}
\end{equation}
Using the Gaussian approximation,
\begin{equation}
N(\phi) \propto \exp(-\phi^2/2\sigma_\text{sy}^2)\, ,\label{Gaussian}
\end{equation}
which represents well the experimentally observed longitudinal profile of the bunch\,\cite{Slim:2023lpd}, one easily finds 
\begin{equation}
F(a)=\frac{a}{\sigma_\text{sy}^2}\exp\left(-\frac{a^2}{2\sigma_\text{sy}^2}\right)\,. \label{eq:analytic}
\end{equation}
Different functional form of $N(\phi)$ and $F(a)$ stems from the fact, that the small-$\phi$ central section of the bunch receives as well contributions from particles with large synchrotron amplitudes.}
		
The synchrotron modulation of the particle momentum $\Delta p(n)$ and the revolution period $\Delta T(n)$ are related by the slip factor $\eta$,
\begin{equation}
	\frac{\Delta T }{T} = \frac{\Delta \phi(n)}{2\pi}=\eta \cdot \frac{\Delta p(n)}{p} \,,\\
	\label{SlipFactor1}
\end{equation}
where $\eta$ 
\begin{equation}
	\eta = \frac{1}{\gamma^2}-\frac{1}{\gamma_{tr}^2}\,,
	\label{SlipFactor2}
\end{equation} 
and  $\gamma_\text{tr}$ is the transition gamma-factor. In Eq.\,(\ref{SlipFactor1}) we introduced $\Delta \phi(n)$, an angular advance (retardation) of a particle per revolution $n$  oscillating with time $\propto \cos(2\pi \nu_\text{sy} f_\text{c}t)$. These one-turn synchrotron phase shifts sum precisely to the $\phi$ defined above with an amplitude larger by the large factor $(2\pi\nu_z )^{-1}$  than that of $\Delta\phi(n)$. Averaging over the ensemble of particles yields the simple relationship
\begin{equation}
	\sigma_\text{sy} = \langle \phi^2\rangle^{1/2} = \frac{\eta}{\nu_\text{sy}}  \Big\langle \frac{\Delta p^2}{p^2}\Big\rangle^{1/2}\,. 
	\label{BunchLength}
\end{equation} 
The corresponding phenomenology of the experimental results from the pilot bunch experiment will be presented in the Appendix\,\ref{SOappendix}. The SOs generate a shift of the spin precession phase, $\Delta\theta_\text{s}(n)=\theta_\text{s}(n) -\theta_\text{s} n $, which is a sum of shifts per turn,
\begin{equation}
\begin{split}
	&\delta\theta_\text{s}(n) = 2\pi G\delta\gamma =2\pi G\gamma \beta^2 \frac{\Delta p(n)}{p}\, , \\
	&\Delta\theta_\text{s}(n) = \xi \psi_\text{sy} \sin(2\pi\nu_z n +\lambda)\, ,\\
	&\psi_\text{sy}   = \sqrt{2}G\gamma \beta^2\frac{\sigma_\text{sy}}{|\eta|}\,, 
	\label{DeltaTheta}
\end{split}
\end{equation}
where  $\xi$ is a convenient  phase-slip relative amplitude with the distribution function, 
\begin{equation}
		F(\xi)=2\xi \exp(-\xi^2)\, ,  
		\label{WeightF}
\end{equation}
and normalization $\langle \xi^2\rangle =1 $ (\textit{cf.} Eq.\,(\ref{eq:analytic})). 
		
The modulation $\Delta T$ of the revolution time results in the corresponding SO-driven slip of the Wien filter phase,
\begin{equation}
\begin{split}
	\Delta\theta_\text{WF}(n) &= \frac{f_\text{WF}}{f_\text{s}  } \cdot \frac{\eta}{\beta^2}\Delta\theta_\text{s}   = C_\text{WF}\Delta\theta_\text{s}  (n)  \,,\\
	C_\text{WF} & = 1 + \frac{K}{G\gamma}\, ,
	 \label{CWF}
\end{split}
\end{equation}
which will show up in the spin-flip dynamics\,\cite{LLMNR:SCT}.

\subsubsection{Master equation for spin envelope}
\label{sec:MasterEqSpinEnv}

It suffices to consider the case of the exact resonance for the central particle,  $f_\text{WF} = f_\text{s}$ , {\it i.e.,} $\theta_\text{s}=\theta_\text{WF}$\,\footnote{A treatment of the side band resonances at $f_\text{WF}= f_\text{s} + K f_\text{sy}$ (with $ K=\pm 1,\pm 2,\ldots$)\,\cite{derbenev1971dynamics}, is beyond the scope of this paper.}. The SO-modified one-turn spin transfer will be given by
\begin{equation}
\vec{S}(n)= \matr{R}_\text{WF}(n) \matr{R}_\text{c}(\theta_\text{s}   +\delta\theta_\text{s}(n)) \vec{S}(n-1)\,.
\label{MasterEqSO}
\end{equation}
Bearing in mind the subsequent Fourier analysis of the in-plane polarization, we stick to the definition of the spin envelope via
Eq.\,(\ref{Envelope}), {\it i.e.,}  we define the envelopes in the reference frame co-rotating with the fixed angular velocity $\omega_\text{WF}$.

Simple rotations in (\ref{MasterEqSO}) do preserve the magnitude of the polarization of individual particles. However, experimentally one measures the average polarization of an 
ensemble of particles with a typical observation time that is much longer than the SO period. This averaging over the ensemble leads to spin decoherence and depolarization.

As an exercise, we first treat the simplest case of the pure idle precession of the in-plane polarization. Here the determination of the envelope $p_\text{rt}$ by the Fourier analysis amounts to the projection of the polarization on the unit vector rotating with fixed frequency $f_\text{WF}$. For an individual particle, the average over the SO period equals
\begin{equation} 
p_\text{rt}(\xi) =  \langle \exp(i\Delta \theta_\text{s}(n))\rangle = J_0(\xi\psi_\text{sy}) \,,
\end{equation}
and the average over the ensemble of particles in the bunch  is
\begin{equation}
\begin{split}
p_\text{rt}&=\int_0^\infty  2\xi \exp(-\xi^2) J_0(\xi\psi_\text{sy})  d\xi\\
&  = \exp(- \frac{1}{4}\psi^2_\text{sy}) \approx 1- \frac{1}{4}\psi^2_\text{sy}\, . \label{IdleSO}
\end{split}
\end{equation} 
This slight attenuation is independent of time. It is of rather academic value, because an instantaneous injection of the horizontal polarization is technically impossible. Equally impossible is a polarimetry with sufficient statistics at times shorter than the SO period. Consequently, in practice the attenuation in Eq.\,(\ref{IdleSO}) is reabsorbed in the definition of the magnitude of the initial in-plane polarization, as  determined experimentally prior to switching the RF spin rotator on. 

Now we proceed to the WF-driven oscillations. The corresponding master equation for the envelope takes the form
\begin{equation}
\begin{split}
\vec{p}(n) = \matr{R}_\text{c}(-n\theta_\text{WF} )\matr{R}_\text{WF}(n)\matr{R}_\text{c}(\delta\theta_\text{s}(n))
\matr{R}_\text{c}(n\theta_\text{WF} )\vec{p}(n-1)\,.
\label{MasterSO}
\end{split}
\end{equation}
It is reminiscent of the master equation (\ref{MasterEqDetuned}), but with oscillating instantaneous running flip of the spin phase per turn, $\delta\theta_\text{s}(n)$, and with much larger slip of the Wien filter     phase $\Delta\theta_{WF}(n)$. In the Fourier analysis,  one is bound to sample trains of turns much longer than the SO period, so that the detuning per se averages out to zero,
$\langle\delta\theta_\text{s}(n)\rangle = 0$, but we have already seen the non-vanishing SO effect even in the case of idle precession, see Eq.\,(\ref{IdleSO}).
 
In the BK averaging over rapid spin precessions  of the corresponding counterpart of the matrix in Eq.\,(\ref{Wdetuned}), we encounter
\begin{widetext}
\begin{equation}
\begin{split}
&\langle \cos (\theta_\text{WF} n )\cos (\theta_\text{WF} n + C_\text{WF}\Delta\theta_\text{s}  (n))\rangle \Rightarrow \frac{1}{2}\cos ( C_\text{WF} \Delta\theta_\text{s}  (n))\, ,\\
&\langle \sin (\theta_\text{WF} n )\cos (\theta_\text{WF} n + C_\text{WF}\Delta\theta_\text{s}  (n))\rangle \Rightarrow -\frac{1}{2}\sin ( C_\text{WF}\Delta\theta_\text{s}  (n))\, , \label{BK-SO-1}
\end{split}
\end{equation}
and obtain 
\begin{equation}
\begin{split}
\matr{U}_\text{SO}(n) 
=\begin{pmatrix}
0 & -\frac{1}{2}        \chi_\text{WF}\sin( C_\text{WF}\Delta\theta_\text{s}  (n)) & \delta_\text{s}\theta(n) \\
\frac{1}{2}        \chi_\text{WF}\sin ( C_\text{WF}\Delta\theta_\text{s}  (n))   &  0 & -\frac{1}{2}        \chi_\text{WF} \cos( C_\text{WF}\Delta\theta_\text{s}  (n))   \\
-\delta_\text{s}\theta(n) & \frac{1}{2}        \chi_\text{WF}\cos ( C_\text{WF}\Delta\theta_\text{s}  (n)) & 0 \end{pmatrix}  \, . \label{USO}
\end{split}
\end{equation}
\end{widetext}
Next stage is BK averaging over the period of  SOs that are much faster than the envelope rotations:
\begin{equation}
\label{BKSO}
\begin{split}\langle \cos ( C_\text{WF}\Delta\theta_\text{s}  (n))\rangle &= \langle \cos( \xi C_\text{WF} \psi_\text{sy} \sin(2\pi\nu_\text{sy} k +\lambda))\rangle \\ 
&=J_0(\xi C_\text{WF} \psi_\text{sy})  \, ,\\
\langle \sin ( C_\text{WF}\Delta\theta_\text{s}  (n))\rangle &=0\, ,\\
\langle\delta\theta_\text{s}(n)\rangle &= 0\, , 
\end{split}
\end{equation}
so that we recover the familiar 
\begin{equation}
\langle \matr{U}_\text{SO}(n) \rangle = \frac{1}{2}        \chi_\text{WF}J_0(\xi C_\text{WF} \psi_\text{sy}) \matr{U}\, .
\end{equation}

Compared to a discussion in Sec. II-B, the principal change is the SO dependent renormalization of the SF tune 
\begin{equation}
\begin{split}
&\nu_\text{SF}\Rightarrow \nu_\text{SF}(\xi)=\nu_\text{SF} J_0(\xi C_\text{WF} \psi_\text{sy})\,.
\label{EtuneSpread}
\end{split}
\end{equation}
In the case of weak to moderate SO effects, we can approximate  
\begin{equation}
1-J_0(\xi C_\text{WF} \psi_\text{sy}) \approx Q_\text{sy}\xi^2\, , 
\label{ApproxBessel1}
\end{equation}
where 
\begin{equation}
Q_\text{sy} = \frac{1}{4} C_\text{WF}^2 \psi_\text{sy}^2= \frac{1}{2}(K+G\gamma)^2 \sigma_\text{sy}^2\, .
\label{ApproxBessel2}
\end{equation}
Note the strong dependence of $Q_\text{sy}$ on the angular length of the bunch and the Wien filter sideband $K$, which is an important feature of the SO mechanism.  

\subsubsection{Evaluation of synchrotron oscillation-driven spin decoherence of the bunch polarization}

The above defined $Q_\text{sy}$ is the principal parameter which defines the SO driven spread of the spin-flip tune (\ref{EtuneSpread}) and the spin-flip phase,
\begin{equation}
		x \Rightarrow x(\xi) = x J_0(\xi C_\text{WF} \psi_\text{sy}) \approx x - Q_\text{sy}\xi^2 x\,.
\label{xRenormalization}
\end{equation}
The SO-driven decoherence is quantified by the expectation value over the ensemble of particles in the bunch,  $\langle \matr{E}(x(\xi))\rangle_\xi$, with the weight function $F(\xi)$ of Eq.\,(\ref{WeightF}). We need to evaluate 
\begin{equation}
\begin{split}
	&\langle \exp(i x(\xi)) \rangle_\xi \\
	&
	=\exp(ix)\int_0^\infty d\xi F(\xi)\exp(-i Q_\text{sy}\xi^2 x)\\
	&= \exp(ix) D(x)\exp(-i\varphi_\text{sy}(x))   \, . \label{Expectation}
\end{split}
\end{equation}
The corresponding envelope rotation matrix takes the form
\begin{equation}
\begin{split}
	\matr{E}_\text{sy}(x_\text{sy})	 = \begin{pmatrix}
	1 & 0 & 0 \\
	0 &  D(x) \cos x_{sy} & - D(x) \sin x_\text{sy} \\
	0 &   D(x) \sin x_\text{sy}  &  D(x) \cos x_\text{sy} 
	\end{pmatrix}\, ,
\label{ESO}
\end{split}
\end{equation}
where 
\begin{equation}
x_\text{sy}= x - \varphi_\text{sy}(x). \label{sy-phase}
\end{equation}
To the approximation in Eq.\,(\ref{ApproxBessel1}), we obtain
\begin{equation}
\begin{split}
	&\langle \exp(i x(\xi)) \rangle_\xi =
	\frac{\exp(i x)}{1+iQ_\text{sy}x}\, ,
\label{SOexponential}
\end{split}
\end{equation}
yielding
\begin{equation}
\begin{split}
	&D(x) =\frac{1}{\sqrt{1+Q_\text{sy}^2 x^2}}\, ,\\
	&\varphi_\text{sy}(x) = \arctan( Q_\text{sy} x) \,
\label{SOdamping}
\end{split}
\end{equation}

The synchrotron oscillation mediated matrix $\matr{E}_\text{sy}(x_\text{sy})$ differs from the exponential-model matrix $\matr{E}_\text{exp}(x_\text{sy})$ in several  aspects. 	
In the SO mechanism, the time dependent spin decoherence takes place only in the spin-flip process. In contrast to the exponential attenuation Ansatz of Sec.\,\ref{ExponAnsatz},  see  Eq.\,(\ref{Eexpon}), in the SO mechanism the  idly precessing spectator radial polarization doesn't decohere, see also the discussion of  Eq.\,(\ref{IdleSO}).  The SO damping factor starts  as $D(x)\approx 1 -\frac{1}{2} Q_\text{sy}^2 x^2$ at  $Q_\text{sy}x \ll 1$ in contrast to $ \exp(-Qx) \approx 1- Qx$ for the exponential Ansatz, while the large-time attenuation $D(x) \approx 1/(Q_\text{sy}x)$  is slower than the exponential one. A signature of the SO dominated spin coherence time is that its scale is set by $Q_\text{sy}x \sim 1$ and exhibits strong dependence on the SF frequency:
\begin{equation}
	\tau_\text{SCT} \sim \frac{1}{2\pi f_\text{SF} Q_\text{sy}}\,. 
	\label{SOSCtime}
\end{equation}
In the above derivation, the exact spin resonance was assumed for the central particles in the bunch. 
Finally, the synchrotron oscillations entail a nonlinear spin-flip phase walk $\varphi_\text{sy}(x)$. It  is an indispensable feature of the SO mechanism of spin decoherence, and it cannot be eliminated by the feedback process targeting the vanishing detuning. This phase walk $\varphi_\text{sy}(x)$  entails the running SF tune
\begin{equation}
	\nu_\text{SF}^\text{(sy)}(x) = \nu_\text{SF}^\text{(sy)} \frac{d x_\text{sy}(x)}{dx}= \nu_\text{SF}^\text{(sy)}\left(1-\frac{Q_\text{sy}}{1+ Q_\text{sy}^2 x^2}\right)\,, \label{SOSFtune}
\end{equation}
where $\nu_\text{SF}^\text{(sy)}$ is the constant spin-flip tune which defines the principal spin-flip phase $x$ and is given by Eqs.\,(\ref{eq:DetunedEnvelopeTune},\ref{NuSquared}) [see further Sec.\,\ref{sec:precursor}].

\subsubsection{Excursion on not compensated betatron oscillation effects}

A strong enhancement of the spin coherence time by tuning the chromaticity, which suppresses orbit lengthening effects caused by betatron oscillations (BO), is well demonstrated experimentally\,\cite{KoopShatunov,SCT1000sJEDI2,SCTchromaticityJEDI}.
Here we comment on the possibility that the residual spin decoherence is an artifact of under-compensated BO effects. BO tunes are large, for example in COSY $\nu_{x,y} \approx 3.6$, some 4 orders of magnitude larger than the SO tune, yet the above treatment of SO effects can be extended to BOs as well. In fact, the prolongation of the orbit by BOs can be considered as a time-independent feature of individual particles. Its effect on the spin tune is proportional to the square of the BO amplitude, 
\begin{equation}
\nu_\text{s}  (\xi) = (1-Q_\text{sy}\xi^2)\nu_\text{s}\, , 
\label{BOspintune}
\end{equation}
which is equivalent to a finite detuning of
\begin{equation}
\delta(\xi) = 2\pi\nu_\text{WF}Q_\text{sy}\xi^2\, , \label{BOdetuning}
\end{equation}
where $\xi$ is the relative amplitude of the BOs with the distribution function $F(\xi)$ of Eq.\,(\ref{WeightF}). According to Refs.\,\cite{KoopShatunov,SCT1000sJEDI2,SCTchromaticityJEDI}, by fine tuning the chromaticity the BO parameter $Q_\text{sy}$ could ideally be brought to zero. 

We abstract from the dynamical considerations and comment here on the phenomenological consequences of the under-compensated BO effects. The most important point is a BO-dependent spread of the detuning, which results in a spread of SF tune. The small-$\delta$ expansion of the SF tune of Eq.\,(\ref{eq:DetunedEnvelopeTune}) gives
\begin{equation}
\begin{split}
\nu_\text{SF}(\xi) & = \nu_\text{SF}^0 (1+\frac{1}{2}Q_\beta \xi^4)\,, \quad \text{where} \\
Q_\beta & = Q_\text{sy}^2 \left(\frac{\nu_\text{WF}}{\nu_\text{SF}^0}\right)^2\,. 
\label{BOSFtune}
\end{split}
\end{equation}
The BO correction to the SF tune starts with a term $\propto \xi^4$ compared to the $\propto \xi^2$ term in the SO Eq.\,(\ref{EtuneSpread}), while the qualitative features are preserved.

Indeed, for the average over the ensemble,  the BO-driven spread of the SF phase factor yields 
\begin{equation}
\begin{split}
\int_0^\infty d\xi F(\xi)\exp[i\frac{x}{2}Q_\beta \xi^4 ]
& = \frac{1}{\sqrt{1-i2Q_\beta x \rho_\beta (x)}} \\& =D_\beta(x)\exp(i\varphi_\beta(x))\,,  
\label{BOexpectation}
\end{split}
\end{equation}
with 
\begin{equation}
\begin{split}
 D_\beta(x) & = \left\{1+ 4 Q_\beta^2 x^2 \rho_\beta^2(x)\right\}^{1/4} \\
 \varphi_\beta(x) & = \frac{1}{2} \arctan\left[2Q_\beta x \rho_\beta(x)\right]\\
\rho_\beta(x) & \approx \frac{1+\pi^{-1} Q_\beta^2 x^2}{1+ Q_\beta^2x^2}\, ,
\end{split}
\end{equation} 
where $\rho_f(x)$ interpolates the damping factor  from  $D_\beta(x)\approx 1$  for $Q_\beta x <1$ to 
\begin{equation}
 D_\beta(x)\approx \sqrt{\frac{\pi}{2Q_\beta x }}  
\end{equation} 
 for $Q_\beta x \gg 1$. 

For $Q_\beta x \gg 1$, the phase $\phi_\beta(x)$  saturates at $\sfrac{\pi}{4}$ compared to $\sfrac{\pi}{2}$ in the case of $ \phi_\text{sol}(x)$.  For  $Q_\beta x <1$ the interpolation function $\rho_\beta(x)  \approx 1$, while for $Q_\beta x \gg 1$, it only controls small details of saturation at $\sfrac{\pi}{4}$, so that the corresponding running spin tune can be approximated by
\begin{equation}
\nu_\text{SF}^{\beta}(x) \approx \nu_\text{SF}\left(1-\frac{Q_\beta}{1+ 4 Q_\beta^2 x^2}\right)\,. \label{BOtune}
\end{equation} 
Here $ \nu_\text{SF}$ is the SF tune defined by Eqs.\,(\ref{eq:DetunedEnvelopeTune},\ref{NuSquared}).  
In summary, despite the very different hierarchy of frequencies involved, the  synchrotron and betatron oscillations have quite a similar impact on the SF dynamics.
 
\section{Spin tomography of synchrotron oscillations}
\label{sec:Tomography}
The remarkable feature of the SF tune, given in Eq.\,(\ref{EtuneSpread}), is its dependence on the SO amplitude, which can  be tested experimentally tagging events in the polarimeter by their angular coordinate $\phi$.	The first look at this effect  was undertaken in the pilot bunch experiment\,\cite{Slim:2023lpd}, where the full data sample of $\phi \in [-\xi_\text{max},\xi_\text{max}]\sigma_\text{sy}=[-2,2]\sigma_\text{sy}$ was split into the central set\,I (with $\phi \in [-\xi_\text{med},\xi_\text{med}]\sigma_\text{sy} = [-0.6,0.6]\sigma_\text{sy}$), and set\,II (with $\xi \in [\xi_\text{med},\xi_\text{max}]$), to be referred to as the head and tail set). The median $\xi_\text{med} = 0.6$  was chosen to have about the same number of recorded events in the sets I and II.

Particles in the bunch do perpetually oscillate from the head to tail and vice versa, crossing back and forth the central region $\xi\leq \xi_\text{med}|$, and a fraction of  the time they spend at $ |\phi_\text{med}| <|\phi|< |\phi_\text{max}|$ is given by the duty cycle
\begin{equation}
	\mathcal{D}(\xi_\text{max},\xi_\text{med}, \xi^2)
		=\frac{2}{\pi}\left[\arccos\left(\frac{\xi_\text{med}}{\xi}\right)  -\arccos\left(\frac{\xi_\text{max}}{\xi}\right) \right] \ \, . 
\label{XiDutyCycle}
\end{equation}
For arbitrary domain $\mathcal{R}$, the expectation value of the phase factor is given by
\begin{equation}
\begin{split}
	\langle \exp(i x(\xi)) \rangle_\xi = \frac{\int_{\mathcal{R}} d\xi F(\xi)\mathcal{D}(\mathcal{R},\xi^2) \exp(i x(\xi))}{\int_{\mathcal{R}} d\xi F(\xi)\mathcal{D}(\mathcal{R},\xi^2)}\,. \label{Set-R}
\end{split} 
\end{equation}
The integrand in Eq.\,(\ref{Set-R}) has remarkable factorization properties. Consider the set ${\mathcal{R}}$ of $\xi \geq \xi_\text{m}$. In terms of the convenient new  variable $\zeta_\text{sy}= \xi^2-\xi_\text{m}^2$, the expansion of Eq.\,(\ref{ApproxBessel1}) gives  $J_0(\xi C_\text{WF} \psi_\text{sy})\approx J_0(\xi_\text{m} C_\text{WF} \psi_\text{sy}) - Q_\text{sy}\zeta_\text{sy}$, so that the phase factor in the integrand factorizes. A similar factorization works for the Gaussian factor in $F(\xi)$, and we obtain
\begin{equation}
\begin{split}
		&\langle \exp(i x(\xi))\rangle_\xi = \exp(ix (\xi_\text{m}))\\
		&\times \frac{\int_{\mathcal{R}} d\zeta \mathcal{D}(\mathcal{R}, \xi_\text{m}^2+\zeta_\text{sy}) \exp(-(1+iQ_\text{sy} x)\zeta_\text{sy})}
		{\int_{\mathcal{R}} 
			d\zeta \mathcal{D}(\mathcal{R}, \xi_\text{m}^2+\zeta_\text{sy}) \exp(-\zeta_\text{sy} )}\,.  \label{XiDutyCycle2}
\end{split} 
\end{equation}
In the generic case, the duty cycle prevents an analytic integration.  For the sake of illustration,  consider the domain  $\mathcal{R} = [\infty,\xi_\text{m}]$. For sufficiently large $\xi_\text{m}> 1$ one can use the approximation $\mathcal{D}(\infty,\xi_\text{m},\xi^2)\approx \sqrt{\zeta_\text{sy}/\xi_\text{m}^2}$. Then the integrals in Eq.\,(\ref{XiDutyCycle2}) reduce to the Euler gamma-functions with the result
\begin{equation}
	\langle \exp(i x(\xi)) \rangle_\xi \approx \frac{\exp(ix (\xi_\text{m}))}{1+i Q_\text{sy}(\xi_\text{m})x}\,, \label{Tomography}
\end{equation}
where $Q_\text{sy}(x_\text{m})= C(\xi_\text{m})  Q_\text{sy}(x_\text{m})$, and  $C(\xi_m \gg 1 )= 3/2$, while for $\xi_\text{m}=0$, Eq.\,(\ref{SOdamping}) corresponds to $C(0)=1$.
Hence we predict a more rapid depolarization of the head and tale portions of the bunch, 
	\begin{equation}
	\frac{S_\text{c}(\infty,\xi_\text{m})}{S_\text{c}(\infty,0)} \approx \sqrt{\frac{1+ Q_\text{sy}^2 x^2}{1+ C^2(\xi_\text{m})Q_\text{sy}^2 x^2} }\,, \label{RelativeAttenuation}
	\end{equation}
		
As another case of spin-flip tomography, we comment on the thought experiment with incomplete masking (gating-out) of the pilot bunch, in which the head and tail particles of the pilot bunch are subjected to spin-flips by the RF field of the WF, while the central body of the bunch is shielded from the RF field of the WF. The interplay between the finite time duration of the gate and the bunch length is as follows. At each turn, the head of the bunch with $\phi >\xi_\text{m}\sigma$ crosses the Wien filter     still in operation, and the spins in the bunch are subjected to the spin flip kicks. The main part of the bunch traverses the already switched-off WF. 
In terms of SF, this masking can be considered as operation of the Wien filter with $\chi_{WF}=0$. 
Since these particles spend part of the time in the central region of the bunch, their depolarization will mimic a partial depolarization of the central part of the bunch . We do not further discuss this effect, which can be easily quantified within the framework of the formalism presented above and will be taken up again elsewhere. 

The above discussion can also be extended to transverse spin tomography of beam bunches. The transverse profile of the polarization was previously studied at RHIC, where a significant variation of the transverse polarization from the core to the skin particles in the beam was observed\,\cite{Nakagawa:2008zzd}. In this case, the skin is populated by particles having large betatron amplitudes, while alongside the particles with small betatron amplitudes also large-amplitude particles spend part of their time in the core region. 

\section{Implications for spin-flip tune mapping}
\label{sec:precursor}

Here we explore implications of detuning and spin decoherence on the search for the EDM of charged particles in all magnetic storage rings with emphasis on the activity of the JEDI collaboration. 


The signal for an EDM is the spin rotation of particles spin in an electric field. In the co-moving frame in a magnetic field, the spins of charged particles  are subject to the electric field generated by the Lorentz transformation. The familiar Frenkel-Thomas-BMT result for the angular velocity of the idle spin precession with respect to the particle momentum in a homogeneous magnetic field reads \cite{BMT,FukuyamaEDM}
\begin{widetext}
\begin{equation}
\begin{split}
\vec{\Omega} = - \frac{q}{m} \left[
G{\vec{B}} +\left(\frac{1}{\beta^2} -1 -G\right) \vec{\beta}\times\vec{E} 
+ 
\frac{1}{2} \eta_\text{EDM}(\vec{E}+[\vec{\beta}\times\vec{B}])
\right] \, ,\label{eq:BMT}
\end{split}
\end{equation}
\end{widetext}
where $\eta_\text{EDM}$ defines the EDM in units of the nuclear magneton via $ d = \eta_\text{EDM} q/(2m)$.
In an ideal purely magnetic ring, the EDM tilts the spin stable axis $\vec c$ according to, 
\begin{equation}
\begin{split}
&\xi^{\text{EDM}}=\arctan \left(\frac{\eta_\text{EDM}}{2G\beta} \right)\, ,\\
&\vec{c} = \sin\xi^{\text{EDM}} \vec{e}_\text{r}+ \cos\xi^{\text{EDM}}\vec{e}_y\, .
\label{IdealRing}
\end{split}
\end{equation}
If the Wien filter     axis were aligned perpendicular to the momentum plane\,\footnote{We refer to this orientation as the EDM mode.}, $\vec{w} =  \vec{e}_y$, Eq.\,(\ref{Etune}) would yield
\begin{equation}
\begin{split}
\left| \vec{c}\times \vec{w} \right|= \sin\xi^{\text{EDM}} \,\, \text{and} \,\, \nu_\text{SF}= \frac{1}{4\pi}\nu_\text{WF }\sin\xi^{\text{EDM}}\,, \label{IdeaWFaxis}
\end{split}
\end{equation}	
and the experimental measurement of the SF tune $\nu_\text{SF} $ would amount to the measurement of the EDM of the particle\,\cite{SpinTuneMapping,MuonEDM}. However, since the spin stable axis is also tilted by imperfection magnetic fields, tangential $a^\text{MDM}_z$  and radial $a^\text{MDM}_x$, which are endemic in all-magnetic rings like COSY, so that 
\begin{equation}
\vec{c} = \vec{c}_y + \sin\xi^{\text{EDM}}\vec{e}_x + a^\text{MDM}_x\vec{e}_\text{r} +a^\text{MDM}_z \vec{e}_z\, . 
\end{equation}
The interaction of the magnetic dipole moment (MDM) of the stored particles with imperfection fields will overwhelm the EDM effect in the SF tune $\nu_\text{SF}$. 

Nevertheless, one can resort to an active compensation of the intrinsic imperfections by two artificial imperfections -- this approach was suggested in\,\cite{SpinTuneMapping,PhysRevAccelBeams.23.024601} and has been used in the recent JEDI experiment with deuterons stored in COSY ring\,\cite{JEDI-precursor-expt}. Specifically, what matters in the cross product $\left| \vec{c}\times \vec{w} \right|$ is the relative orientation of $\vec{c}$ and $\vec{w}$. The spin stable axis $\vec{c}$ is tilted by the static magnetic field of the Siberian snake in the straight section opposite the Wien filter     which rotates the spins around the $z$-axis by an angle $\chi^\text{sol}$, while the magnetic field axis $\vec{w}$ of the Wien filter     is tilted around the $z$ axis by an angle $\phi^\text{WF}$. Since the solenoid fields affect the idle spin precession tune, the Wien filter     frequency has to be corrected accordingly.  

In the case of the exact resonance, one finds
\begin{widetext}
	\begin{equation}
	\nu_\text{SF}  =   \frac{\chi_\text{WF}}{4\pi}\left| \vec{c}\times \vec{w} \right|
	=  \frac{\chi_\text{WF}}{4\pi} \left[\left(\xi^\text{MDM}+a_x^\text{MDM}-\phi^\text{WF} \right)^2 
	+ \left( a_z^\text{MDM} + \frac{1}{2\sin \pi\nu_\text{s}} \chi^\text{sol} \right)^2 \right]^{1/2} \, . 
	\label{SpinFlipMap}
	\end{equation}
\end{widetext}
As a function of the artificial imperfection parameters, $\phi^\text{WF}$ and  $\chi^\text{sol}$, the SF tune $\nu_\text{SF}$ describes  an elliptic cone.  The accuracy with which the location of the cone apex at $\nu_\text{SF}^0$ can be determined defines the best accuracy with which $\xi_\text{EDM}$ can be determined using the described technique\,\cite{JEDI-precursor-expt}. Barring accidental cancellations, one can reinterpret this accuracy as a tentative upper bound for $\xi^\text{EDM}$. 

At finite detuning, the observed SF tune will be modified according to Eq.\,(\ref{NuSquared})
\begin{widetext}
	\begin{equation}
	\nu_\text{SF} =   \frac{1}{4\pi} \left\{ \chi_{\text WF}^2 \left[ \left( \xi^\text{MDM}+a_x^\text{MDM}-\phi^\text{WF} \right)^2 + \left( a_z^\text{MDM} + \frac{1}{2\sin \pi\nu_\text{s}} \chi^\text{sol} \right)^2 \right]+ \frac{1}{4}\delta^2 \right\}^{1/2} . \label{SpinFlipMap2}
	\end{equation}
\end{widetext}

As far as the detuning is relatively weak, it should not affect the location of the cone apex. To this end, we emphasize that the detuning parameter $\delta$ is not a free parameter as the detuning angle $\rho$ can be determined \textit{independently} from the combined analysis of the evolution of the vertical and horizontal polarizations. However, one should be wary of the effects of the feedback effect described in Sect.  -- here one needs more experimental input from the spin-precession phase walk studies.


In the exclusive regime of exact spin resonance and vanishing spin decoherence, the  SF tune $\nu_\text{SF}$ defines the slope of the time dependence of the SF phase,
\begin{equation}
		\frac{\dd p_\text{c}(x)}{\dd t}\Big\vert_\text{t=0} =  -\sin\Phi_\text{in} \frac{\dd x}{\dd t}  = -2 \pi  f_\text{c}    \sin\Phi_\text{in}\nu_\text{SF} \, . 
		\label{IntialSlope1}
\end{equation}
For instance, this is the case in the exponential decoherence model.
In the case of spin decoherence dominated by synchrotron oscillation, the phase response $\varphi_\text{sy}(x)$ must be taken into account [see Eq.\,(\ref{SOSFtune})]. As far as the experimental data were taken in the regime of $Q_\text{sy} x <1$, as suggested by the analysis given in Appendix\,\ref{SOappendix}, the net effect is a minor renormalization of the visible spin flip tune 
\begin{equation}
		\nu_\text{SF}^{(\text{exp})} \approx \nu_\text{SF}^{(\text{sy})} (1-Q_\text{sy})\, . \label{SF-phase-renormalization}
\end{equation}
Here  $\nu_\text{SF}^{(\text{exp})}$ is the spin tune which  one will get if the spin-flip data were treated within the exponential model, where it is given by Eqs.\,(\ref{eq:DetunedEnvelopeTune},\ref{NuSquared}). In the regime of $Q_\text{sy} x <1$, Eq.\,(\ref{SF-phase-renormalization}) entails simple overall rescaling of the spin-flip tune without affecting the location of the apex of the map in Eq.\,(\ref{SpinFlipMap}). However,  were  $Q_\text{sy} x \sim 1$, then it would have been necessary  to directly use the nonlinear $\varphi_\text{sy}(x)$ in the extraction of $\nu_\text{SF}^{(\text{sy})}$ from the experimental spin flip data. The same point refers to the spin decoherence controlled by betatron oscillations. Here we reiterate that neither $\phi_\beta(x)$ nor $\varphi_\text{sy}(x)$ can be eliminated by the feedback set to maintain the phase locking between Wien filter     and spin precession as accurately as possible. 

\section{Summary and Conclusions}
\label{sec:summary-and-conclusions}

Inspired by the JEDI studies of high-precision spin dynamics in storage rings, we have developed a theoretical description of RF-driven spin rotations that accounts for detuning with respect to the exact spin resonance.   Such a description serves in part as the theoretical basis for the first search for the EDM of deuterons and for tests of the pilot-bunch approach to co-magnetometry recently performed at COSY. The fully analytical description of the multiple spin flips, complemented by in-plane polarization precession and various spin depolarization mechanisms, is essential for data analysis down to the smallest detail, since fitting the experimental data requires multiple calls to the spin rotation and depolarization codes. 

As part of our generic approach to RF-driven spin rotations, we have presented results for three different mechanisms of spin decoherence. We found great similarities between synchrotron oscillations and betatron oscillations as driving spin decoherence, with detuned spin precession being a common denominator. Interestingly, in the presence of ring instabilities, detuning is an integral part of the feedback mechanism to maintain the most accurate phase locking between the RF Wien filter     and the spin precession. 

Parameters common to the two spin-decoherence mechanisms considered include the magnitude and orientation of the stored initial polarization, the detuning, and the spin-decoherence parameter.  It has been shown that different spin-decoherence models result in different patterns of depolarization of different components of the continuously flipping polarization. We emphasized the importance of a concurrent analysis of vertical and in-plane precessing polarization components, in particular the previously unexplored phase of the in-plane polarization envelope, as an insight into RF-driven spin dynamics in storage rings.

The synchrotron oscillation mechanism of decoherence is shown to be governed by the bunch length and we suggest a spin-flip based tomography of the synchrotron oscillation-driven spin dynamics. Within the statistical accuracy currently achieved, the main results of the JEDI pilot bunch experiment are consistent with the quantitative expectations of the synchrotron oscillation model, and we commented on the possibility of improving the sensitivity of spin-flip tomography.  

\begin{acknowledgments}
The work presented here has been performed in the framework of the JEDI collaboration and was supported by an ERC Advanced Grant of the European Union (proposal No.\,694340: Search for electric dipole moments using storage rings) and by the Shota Rustaveli National Science Foundation of the Republic of Georgia (SRNSFG Grant No.\,DI-18-298: High precision polarimetry for charged particle EDM searches in storage rings). This research is part of a project that has received funding from the European Union’s Horizon 2020 research and innovation program under grant agreement STRONG-2020, No.\,824093. The work of A.\,Aksentev, A.\,Melnikov and  N.\,Nikolaev on the topic was supported by the Russian Science Foundation (Grant No.\,22-42-04419). Thanks are due to A.\,Zelenski for useful discussions.
\end{acknowledgments}

\appendix

\section{Phenomenology of spin decoherence driven by synchrotron oscillations}
\label{SOappendix}

\begin{figure}[htb]
	\includegraphics[width=\columnwidth]{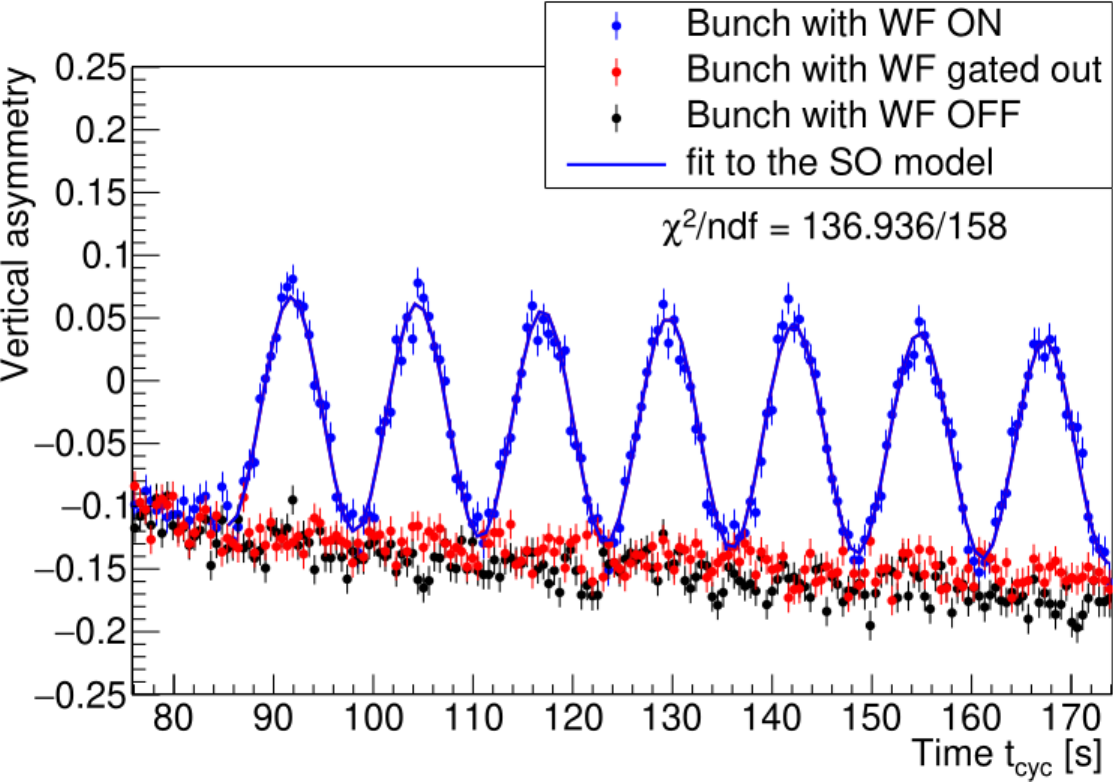}
	\caption{\label{fig:2sigma-inclusive} 
		Measured WF induced vertical oscillation of the signal bunch  polarization in terms of the left-right asymmetry in the polarimeter (not normalized for the dC analyzing power) 
		for a cycle with two bunches stored  in the machine. The RF Wien filter is switched ON at $t_0 = \SI{85.55}{s}$. The blue points indicate the vertical polarization asymmetry for events within the $\pm 2\sigma_s$ boundary of the signal bunch. The results of corresponding fits within the synchrotron oscillation model are presented in Table\,\ref{tab:SO-analysis}. The red points reflect the case for the pilot bunch, \textit{i.e.}, when the RF of the Wien filter is gated out. The black points indicate the situation when, during a different cycle, the WF is completely switched OFF. The blue line indicates a fit with Eq.\,(\ref{SO-fit}), using events from within the $\pm 2\sigma_\text{s}$ boundary of the signal bunch distribution, it is practically indistinguishable from the exponential decoherence  fit shown in Ref.\,\cite{Slim:2023lpd}, see also the discussion in the text. }
\end{figure} 

Here we present a brief phenomenology of the experimental results of the pilot bunch experiment\,\cite{Slim:2023lpd} in the framework of the model of spin decoherence mediated by synchrotron oscillations. The main parameters of the COSY ring are listed in Table\,\ref{tab:machine-conditions}.  

Within the model, the main source of spin decoherence is the longitudinal momentum spread, which is related to the angular length of the bunch by Eq.\,(\ref{BunchLength}). With the momentum spread $\Delta p/p$ and the slip factor $\eta$ from Table \,\ref{tab:machine-conditions}, we obtain $\sigma_\text{sy}=0.177 \pm 0.018$, which agrees with the RMS value $\sigma_\text{s}$ of the Gaussian approximation to the longitudinal density of the signal bunch, varying from $\sigma_\text{s}=0. 11$ at the beginning of the measurement cycle after the cooling was turned off, through $\sigma_\text{s}=0.18$ in the middle of the cycle to $\sigma_\text{s}=0.20$ at the end of the cycle\,\cite{Slim:2023lpd}. The Wien filter was operated in the $K=-1$ sideband, and from Eq.\,(\ref{ApproxBessel2}), we expect to find 
\begin{equation}
Q_\text{sy} \approx 0.0211 \pm 0.0043 \,. 
\label{Qsy}
\end{equation}  
\begin{figure}[htb]
	\includegraphics[width=\columnwidth]{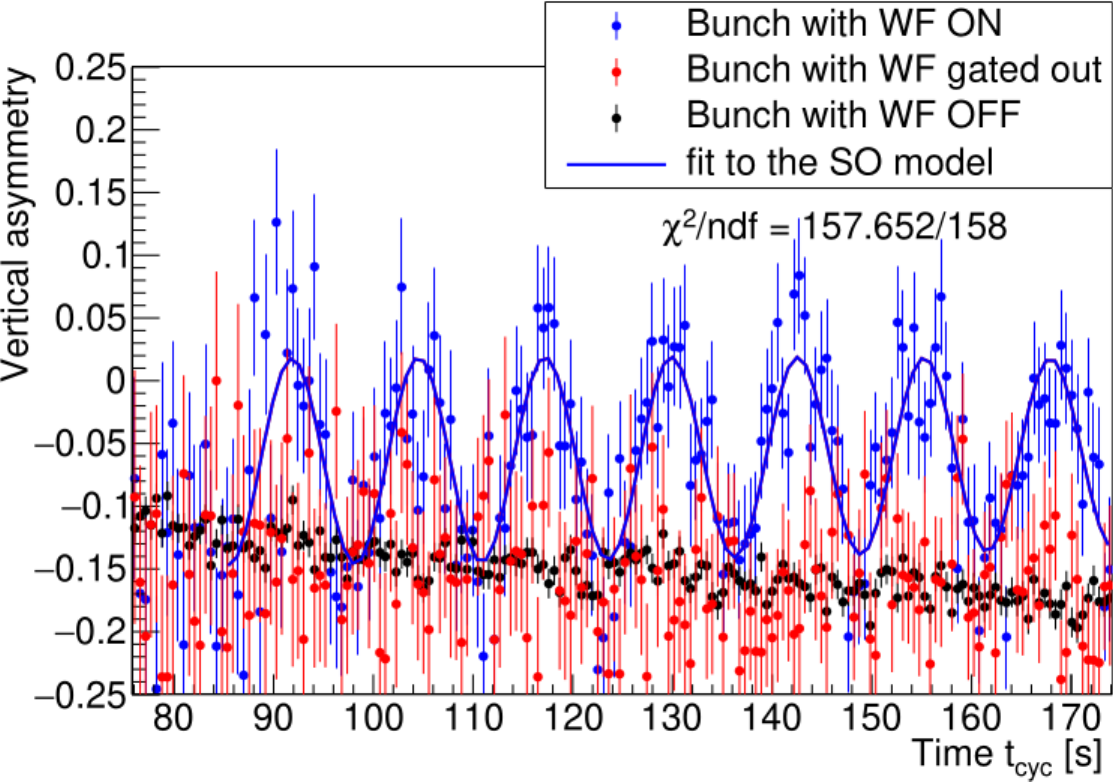}
	\caption{\label{fig:Head-and-Tail} 
		The same graph as shown in Fig.\,\ref{fig:2sigma-inclusive}, but here for particles with synchrotron oscillation  amplitudes outside of the $\pm 2 \sigma_\text{s}$ cut on the longitudinal bunch distribution.}
\end{figure}
The obvious feature of the synchrotron oscillation mechanism is that the head and tail particles have larger synchrotron oscillation amplitudes, entailing stronger spin decoherence, and here we focus on the determination of $Q_\text{sy}$ from the pilot-bunch experimental data. In Fig.\,\ref{fig:2sigma-inclusive}, we show the polarization left-right  asymmetry with a $\pm 2 \sigma_\text{s}$ cut on the signal bunch distribution, so that the experimental data are exactly the same as shown in Fig.\,2 of Ref.\,\cite{Slim:2023lpd}. For the purposes of our discussion, it is not necessary to convert the polarization asymmetry to the actual polarization, as this only adds an overall normalization uncertainty from the dC analyzing power to all data points. 

\begin{table*}[htb]
	\caption{\label{tab:machine-conditions} Parameters of the deuteron kinematics, the COSY ring and the synchrotron motion in the pilot bunch experiment.}
	\begin{ruledtabular}
		\renewcommand{\arraystretch}{1.2}
		\begin{tabular}{lll}
			Parameter 								& Symbol [Unit] 		& Value \\ \hline
			Deuteron momentum (lab)					& $P$ [MeV/c]			& \num{970.000} \\
			Lorentz factor 							& $\gamma$ [1] 			& 1.126 \\
			Beam velocity 							& $\beta$  [c] 			& 0.460  \\
			
			Nominal COSY orbit circumference 		& $\ell_\text{COSY}$ [m] 			& \num{183.572}  \\
			Revolution frequency 					& $f_\text{c}$ [Hz] 	& \num{750602.6} \\
			Spin precession frequency    			& $f_s$ [Hz] 			& \num{-120860.5} \\

			Slip factor								& $\eta$ [1]			& 0.6545 \\
			Momentum spread	in middle of cycle		& $\Delta p/p$ [1]		& \num{7.397e-5} \\
			Synchrotron oscillation frequency       & $f_\text{sy}$ [Hz]    & $205 \pm 21$
		\end{tabular}
	\end{ruledtabular}
\end{table*}

A fit to the asymmetry with the formula describing the synchrotron oscillations,
\begin{widetext}
\begin{equation}
\begin{split}
A_\text{sy}(t) = a (t-t_0) + b
+ \frac{c}{\sqrt{1+ \left[ 2\pi Q_\text{sy} f_\text{SF} (t-t_0) \right] ^2}}
\times\cos \left[ 2\pi f_\text{SF} (t-t_0)  - \arctan(2\pi Q_\text{sy} f_\text{SF} (t-t_0) ) \right]\, ,
\label{SO-fit}
\end{split}
\end{equation}
resulted in $Q_\text{sy}(\pm 2 \sigma_\text{s})= (0.0077 \pm 0.0036)$, which is in the ballpark of the model expectation of Eq.\,(\ref{Qsy}). In this fit, we kept fixed $t_0 = \SI{85.5}{s}$, as determined in Ref.\,\cite{Slim:2023lpd}, and where the same data were fitted to the exponential decoherence formula, given by
\begin{equation}
\begin{split}
	A_\text{exp}(t) = a (t-t_0) + b
	+ c\exp \left[-\Gamma(t-t_0)\right]	\times\cos \left[ 2\pi f_\text{SF} (t-t_0 )\right]\, .
\label{SEXP-fit}
\end{split}
\end{equation}
\end{widetext} 

The quality of the synchrotron oscillation model fit, $\chi^2/\text{ndf} = 136.936/158=0.867$, is basically identical to  $\chi^2/\text{ndf} = 136.071/157=0.867$ for the exponential attenuation model, applied in Ref.\,\cite{Slim:2023lpd}, and for all practical purposes, the synchrotron oscillation model in Fig.\,\ref{fig:2sigma-inclusive} is indistinguishable from the exponential-decoherence model. Indeed, in view of the  weak signal of attenuation, the two parametrizations can not be discriminated with the present accuracy of the experimental data. In order to not confuse the two formula-wise different fits, we changed the color code of the fit curve and of the related data points, so that the blue curve in in Fig.\,\ref{fig:2sigma-inclusive}  must be compared to the red curve in  Fig.\,2 of Ref.\,\cite{Slim:2023lpd}.


According to the discussion in Sec.\,\ref{sec:Tomography}, for the head and tail particles, we expect an enhancement of the parameter $Q_\text{sy}$ by a factor up to $\approx 9/4$. As a subsample of events with the largest attainable synchrotron oscillations, we considered separately the  head and tail particles outside of the  $\pm 2 \sigma_\text{s}$ cut. The experimental results for the corresponding polarization asymmetry are shown in Fig.\,\ref{fig:Head-and-Tail}.  With low statistics in the head-and-tail sample, a fit to the data using Eq.\,(\ref{SO-fit}) yields $Q_\text{sy}(|\phi_\text{s}|> 2\sigma_\text{s}) = 0.0098 \pm 0.0108$, which is consistent with the estimate given in Eq.\,(\ref{Qsy}).  \\

As a further check of the synchrotron oscillation model, following Ref.\,\cite{Slim:2023lpd}, we considered still grouping of signal bunch  events within the  $\pm 2 \sigma_\text{s}$ cut into set I and set II,  shown in Table\,\ref{tab:Binning}. The boundary of $0.6\sigma_\text{s}$ between the two sets was chosen as to have about equal number of events in each of the sets. It should be noted that the two sets are not entirely  statistically independent, as particles from set II spend part of their time in set I. Again, within the present experimental accuracy, the  corresponding results for $Q_\text{sy}$ of our interest from fits to the parametrization of synchrotron oscillations, given in Eq.\,(\ref{SO-fit}), are in the ballpark of our estimate, given in Eq.\,(\ref{Qsy}).

\begin{table*}[htb]
	\caption{\label{tab:SO-analysis} Parameters obtained from fits of the asymmetry oscillation pattern of the signal bunch with the synchrotron oscillations model described by Eq.\,(\ref{SO-fit}) to two different sets of events, shown in shown in Figs.\,\ref{fig:2sigma-inclusive} and\,\ref{fig:Head-and-Tail}. }
	\begin{ruledtabular}
		\renewcommand{\arraystretch}{1.2} 
		\begin{tabular}{crll}
			Parameter 	   	& Central events inside the $ [-2,2]\sigma_\text{s}$ cut 			& Head and tail events outside of the $ [-2,2]\sigma_\text{s}$ cut			& Unit 						\\\hline
			$a$       	   	& \num{-4.01}$\pm$ \num{0.38}      & \num{0.79} $\pm$ \num{1.55}	    & \SI{e-4}{\per \second} 	\\
			$b$				& \num{-0.02967}$\pm$ \num{0.00191}  	& -\num{0.06490}$\pm$ \num{0.00849}  	& 1							\\	
			$c$ 			& \num{-0.092419}$\pm$ \num{0.002046}		&  -\num{0.082409}	 $\pm$ \num{0.007904}& 1					\\
			$Q_\text{sy}$	& \num{0.007728} $\pm$ \num{0.003602}	& \num{0.009837}$\pm$ \num{0.010764}	& \si{\second}				\\
			$f_\text{SF}$  	& \num{0.079984} $\pm$ \num{0.000278}	& \num{0.079617} $\pm$  \num{0.000825} 	& \si{\hertz}				\\
		\end{tabular}
	\end{ruledtabular}
\end{table*}	

\begin{table*}[htb]
	\caption{\label{tab:Binning} Synchrotron oscillation parameters $Q_\text{sy}$, deduced from fits to the vertical asymmetry with Eq.\,(\ref{SO-fit}) of the signal bunch for different sets of events of the longitudinal bunch distribution.}
	\renewcommand{\arraystretch}{1.2}
		\begin{ruledtabular}
		\begin{tabular}{ccccc}
			Set  & Cut on beam distribution 																																							& $Q_\text{sy}$	& $\chi^2/\text{ndf}$	\\\hline
			I    & $\phi_\text{s} \in [-0.6,0.6]\sigma_\text{s}$   	& $\num{0.0092} \pm \num{0.0046}$ &  $179.821/158 = \num{1.138}$ \\
			II   & $\phi_\text{s} \in [- 2\sigma_\text{s}, -0.6] \sigma_\text{s} \, \wedge \, \phi_\text{s} \in [+0.6, +2] \sigma_\text{s}$	& $\num{0.0037}    \pm \num{0.0102}$ 		& $132.685/158 = \num{0.840}$\\
		\end{tabular}
	\end{ruledtabular}
\end{table*}

Some comments on the interpretation of results for the spin-flip frequency are in order. In the ad hoc phenomenological model of exponential attenuation, the spin-flip phase motion is decoupled from the strength of the attenuation. Within this model, fits to the spin-flip pattern of events within the $\pm 2 \sigma_\text{s}$ boundary, observed in the pilot-bunch experiment yielded the spin-flip frequency $f_\text{SF}^\text{(exp)} $ to about one per mille accuracy, $f_\text{SF}^\text{(exp)} (\pm 2\sigma_\text{s})= 0.079442 \pm 0.000096\,$Hz\,\cite{Slim:2023lpd}. In contrast to that, the synchrotron oscillation dominance is a dynamical model with a well-defined correlation between spin decoherence and spin-flip phase motion. Here we capture on the point that in spite of  7 full spin-flip periods observed, the pilot-bunch experimental data still correspond to the regime of  small $Q_\text{sy} x <1$. Then we can invoke the approximation of  Eq.\,(\ref{SF-phase-renormalization}) to relate $f_\text{SF}^\text{(sy)}$ to $f_\text{SF}^\text{(exp)}$. Specifically, with entry for $Q_\text{sy}$ in Table\,\ref{tab:SO-analysis},  we find
\begin{equation}
f_\text{SF}^{(\text{sy})} \approx \frac{f_\text{SF}^\text{(exp)}}{1-Q_\text{sy}} = 0.080067\pm 0.000304, \label{SO-vs-Exp}
\end{equation}
which agrees with  the fit result for $f_\text{SF}^{(\text{sy})}$ in Table  \ref{tab:SO-analysis}.  Evidently, it is the present uncertainty of $\Delta Q_\text{sy} \approx \num{3.6e-3}$ which entails the about \SI{4}{\text{\textperthousand}} uncertainty in the determination of the 
 $f_\text{SF}^{(\text{sy})}(\pm 2\sigma_\text{s})$ in Table\,\ref{tab:SO-analysis}. 

The achieved precision of the JEDI pilot-bunch experiment is close to, but does not yet allow a decisive test of the discussed spin tomography of the longitudinal structure of the bunch.  We point out again that for more systematic studies it is advisable to increase the synchrotron oscillation parameter $Q_\text{sy}$ at the expense of either larger $\Delta p/p$ and correspondingly longer bunches, or to run the Wien filter at sidebands $K=\pm 2$ or at still larger $K$.
	
	
\bibliographystyle{apsrev4-2}
\bibliography{PilotBunch_18.06.2023}	

\end{document}